\documentclass[iop]{emulateapj}
\usepackage{graphicx}
\usepackage{subfigure}
\usepackage{multirow}
\usepackage{amssymb}
\usepackage{natbib}
\usepackage{rotating}
\usepackage[bookmarks=false, colorlinks=true, citecolor=blue, backref=true]{hyperref}
\usepackage{apjfonts}
\shorttitle{\textit{Spitzer IRS} spectroscopy of H{\sc i}-outflow radio galaxies}
\shortauthors{P. Guillard et al.}

\begin{document}


\title{Strong molecular hydrogen emission and kinematics of the multiphase gas in radio galaxies with fast jet-driven outflows}



 \author{P. Guillard\altaffilmark{1}}, \author{P. M. Ogle\altaffilmark{1}}, \author{B. H. C. Emonts\altaffilmark{2}}, 
\author{P. N. Appleton\altaffilmark{3}}, \author{R. Morganti\altaffilmark{4,5}}, \author{C. Tadhunter\altaffilmark{6}}, 
\author{T. Oosterloo\altaffilmark{4,5}}, \author{D. A. Evans\altaffilmark{7}}, \author{A. Evans\altaffilmark{8,9}}

\altaffiltext{1}{\textit{Spitzer} Science Center (SSC), California Institute of Technology, Pasadena, USA}
\altaffiltext{2}{Australia Telescope National Facility, CSIRO, POBox76, Epping NSW, 1710, AUSTRALIA}
\altaffiltext{3}{NASA \emph{Herschel} Science Center (NHSC), California Institute of Technology, Pasadena, USA}
\altaffiltext{4}{Netherlands Foundation for Research in Astronomy, Dwingeloo, The Netherlands}
\altaffiltext{5}{Kapteyn Astronomical Institute, University of Groningen, The Netherlands}
\altaffiltext{6}{Department of Physics \& Astronomy, University of Sheffield, UK}
 \altaffiltext{7}{Massachusetts Institute of  Technology, Kavli Institute for Astrophysics and Space Research, 77 Massachusetts Avenue, Cambridge, MA02139, USA}
\altaffiltext{8}{University of Virginia, Charlottesville, USA}
\altaffiltext{9}{National Radio Astronomical Observatory}


\begin{abstract}
Observations of ionized and neutral gas outflows in radio-galaxies (RG) suggest that AGN radio jet feedback has a galaxy-scale impact on the host ISM, but it is still unclear how the molecular gas is affected. Thus it is crucial to determine the physical conditions of the molecular gas in powerful RG to understand how radio sources may regulate the star formation in their host galaxies. We present deep \textit{Spitzer IRS} high-resolution spectroscopy of 8 nearby RG that show fast H{\sc i} outflows. Strikingly, all of these H{\sc i}-outflow RG have bright H$_2$ mid-IR lines that cannot be accounted for by UV or X-ray heating. This strongly suggests that the radio jet, which drives the H{\sc i} outflow, is also responsible for the shock-excitation of the warm H$_2$ gas. In addition, the warm H$_2$ gas does not share the kinematics of the ionized/neutral gas. The mid-IR ionized gas lines (with FWHM up to 1250~km~s$^{-1}$ for [Ne{\sc ii}]12.8$\mu$m) are systematically broader than the H$_2$ lines, which are resolved by the \textit{IRS} in $\approx 60$\%  of the detected lines (with FWHM up to 900~km~s$^{-1}$). In 5 sources, 3C~236, 3C~293, 3C~459, 4C~12.50 and PKS~1549-79, the [Ne{\sc ii}]$\lambda \, 12.8\mu$m line, and to a lesser extent the [Ne{\sc iii}]$\lambda \, 15.5\mu$m and  [Ne{\sc v}]$\lambda \, 14.3\mu$m lines, clearly exhibit blue-shifted wings (up to -900~km~s$^{-1}$ with respect to the systemic velocity) that match well the kinematics of the outflowing H{\sc i} or ionized gas. The H$_2$ lines do not show these broad wings, except tentative detections in 4C~12.50, 3C~459 and PKS~1549-79. This shows that, contrary to the  H{\sc i} gas, the H$_2$ gas is inefficiently coupled to the AGN jet-driven outflow of ionized gas. While the dissipation of a small fraction ($<10\%$) of the jet kinetic power can explain the turbulent heating of the molecular gas, our data show that the bulk of the warm molecular gas is not expelled from these galaxies.
\end{abstract}


%

\section{Introduction}

Active galactic nucleus (AGN) feedback is recognised to have an important effect on galaxy evolution. Widely introduced 
in numerical simulations of galaxy evolution to clear the circum-nuclear gas and halt the growth of super-massive 
black holes \citep{Silk1998, DiMatteo2005}, this mechanism would explain the correlations found between the BH mass 
and e.g. the bulge mass \citep[e.g.][]{Ferrarese2000}, and prevent the formation of too many massive galaxies in the early universe \citep[e.g.][]{Thomas2005}.

The co-evolution of massive 
black holes and their host galaxies has been well established observationally in samples at various redshifts \citep[e.g.][]{Tremaine2002, Alexander2005}. Nevertheless, detailed observations of individual active galaxies in the nearby universe, where physical processes in the central 
region can be studied in detail, are essential for answering key questions about the role 
of AGN in galaxy evolution, such as what is the magnitude of AGN feedback (e.g. gas outflow 
rates) and what is the main driving mechanism of AGN feedback (e.g. quasar wind, radio jets, circum-nuclear starbursts).


Most of the existing evidence for AGN radio jet feedback on the scales of galaxy bulges comes from observations of outflows of neutral gas \citep{Morganti2003, Morganti2005, Emonts2005, Lehnert2011} and ionized gas \citep{Nesvadba2006, Nesvadba2008, Holt2008} in radio galaxies. These observations suggest that radio jets are an efficient mechanism to convert the energy output of the AGN into an energy input into the ISM, and that radio sources may regulate gas cooling in early-type galaxies \citep{Best2005, Best2006, Donoso2009}.




%

One of major open questions about estimating mass outflow rates is whether the ionized and H{\sc i} gas traces the dominant phase of the wind. 
Therefore it is crucial to compare the masses and kinematics of each gas phase. 
It has only very recently been recognized that starburst and AGN-driven winds also include molecular components \citep[e.g.][]{Veilleux2009, Feruglio2010, Alatalo2011}.
\citet{Fischer2010} and \citet{Sturm2011} reported the discovery of molecular outflows in nearby ULIRGs (e.g. Mrk~231) through the detection of a series of OH and H$_2$O lines seen in absorption against the bright IR dust continuum with
\textit{Herschel}/PACS in the $\lambda \, 78-79\,\mu$m  and $\lambda \, 119-121\,\mu$m ranges. 
However, estimates of outflow rates from absorption lines suffer from major uncertainties regarding e.g. the geometry, and the degeneracy between covering fractions and column densities.
Moreover, most of these sources are composite: the co-existence of a starburst,
quasar, and radio source makes it difficult to infer what is
driving the wind. Starbursts \citep{Rupke2005}, radiation pressure
from the quasar \citep{Feruglio2010}, and even mechanical
interactions with the radio source \citep{Reynolds2009} have all
been proposed.

\textit{Spitzer} IRS spectroscopy opened a new perspective from which to study the impact of the 
injection of kinetic energy on the formation of molecular gas and regulation of star formation in nearby radio galaxies. \citet{Ogle2010} found that 30\% of a sample of 55 nearby 3C radio galaxies have unusually bright mid-IR line emission from warm ($10^{2-3}\,$K) H$_2$ gas, with weak tracers of star formation (SF) (e.g. PAHs). The H$_2$ to PAH luminosity ratio is more than 10 times larger than what is expected from UV and X-ray photon heating. We propose that the H$_2$ luminosity is associated with the dissipation of a small fraction of the mechanical energy of the radio jet \citep{Nesvadba2010, Ogle2010}.
These H$_2$-luminous galaxies lie off the Schmidt-Kennicutt relationship, indicating that star-formation may be suppressed relative to nearby normal star-forming galaxies \citep{Nesvadba2010}. Numerical simulations also suggest that radio jets appear to be efficient  at injecting kinetic energy into the ISM \citep[e.g.][]{Wagner2011}, possibly inhibiting star formation.

This paper reports on the mid-IR \textit{Spitzer IRS} \citep{Houck2004} spectroscopy of 8 nearby powerful radio galaxies in which fast ($\approx 1000$~km~s$^{-1}$) outflows of ionized and neutral gas have been detected \citep{Morganti2005}. These galaxies are ideal targets to study the impact of AGN feedback on the different phases of the ISM, and in particular on the molecular gas, and thus on star-formation. 
They present an environment where winds can clearly be associated with  the mechanism driving them, namely the radio jet.
How is the kinetic energy of the outflow dissipated? Is the molecular gas entrained in the ionized/neutral outflow? If yes, what are the physical processes that control the dynamical coupling between the tenuous gas outflow and the dense molecular gas?

Sect.~\ref{sec:sample} and \ref{sec:obs} present the galaxy sample and details on the data reduction and spectral analysis. Then we discuss spectroscopic diagnostics (sect.~\ref{sec:spec_diag}), focusing on the H$_2$ gas, and we compare the kinematics of the ionized, H{\sc i}, and H$_2$ gas in sect.~\ref{sec:kinematics}. Sect.~\ref{sec:discussion} discusses the excitation mechanisms and origin of the H$_2$ gas, and proposes an interpretation of the observed gas heating and kinematics of the multiphase ISM gas of the host galaxy. 

\section{The sample of H{\sc i}-outflow radio galaxies}
\label{sec:sample}
\begin{deluxetable*}{lcccccccccccc} 
\tablecolumns{13}
\tablewidth{\textwidth} 
\tablecaption{Properties of H{\sc i}-outflow radio-galaxies observed with \textit{Spitzer IRS}}
\tablehead{
\colhead{Object} & \colhead{RA\tablenotemark{a}} & \colhead{Dec\tablenotemark{a}}  & \colhead{$z\,$\tablenotemark{b}}  & \colhead{$D_L$\tablenotemark{c}}  & \colhead{K-mag} & \colhead{$M_{\star}$\tablenotemark{d}} &   \colhead{$M(H_2)$\tablenotemark{e}} & \colhead{$M(H_2)$} & \colhead{$L_{\rm 24\mu m}$\tablenotemark{f}} & \colhead{$\log L_{\rm X}$\tablenotemark{g}} & \colhead{$\log L_{\rm X}$} &  \colhead{$L(\rm H_2) / L_{\rm X}$\tablenotemark{h}} \\
	\colhead{}  & \colhead{J2000.0} & \colhead{J2000.0} & \colhead{} & \colhead{[Mpc]} & \colhead{} & \colhead{[M$_{\odot}$]} & \colhead{[M$_{\odot}$]}& \colhead{ref.}  & \colhead{[L$_{\odot}$]}  & \colhead{[erg~s$^{-1}$]}& \colhead{ref.} & }
\startdata
3C 236	 & 		   10 06 01.7 & +34 54 10.4 & 0.1004 & 449  & 12.25  & 7.4E10       & $<8$E9  & (1) & $1.94 \pm 0.15$ E10  &    43.0 & & 0.054 \\
3C 293	 & 		   13 52 17.8 & +31 26 46.5 & 0.0448 & 195  & 10.84  &  4.6E10      & 1.5E10   & (2)  &  $6.26 \pm 0.75$ E09 &    42.8 &  (4) & 0.063 \\ 
3C 305	 & 		   14 49 21.8 & +63 16 15.3 & 0.0418 & 178  & 10.64  & 4.8E10       &  2.0E9 & (1)   & $7.13 \pm 0.45$ E09  &     41.2  & (5) & 1.78 \\
3C 459	 & 		   23 16 35.2 & +04 05 18.1 & 0.2199 & 1046  & \nodata  &  \nodata  & \nodata  & \nodata  &  $5.36 \pm 0.32$ E11 &  43.2 & & 0.13   \\ 
4C +12.50 &      13 47 33.4 & +12 17 24.2 & 0.1234 & 551 &   12.21    & 1.3E11     &  1.5E10 & (3)  & $1.20 \pm 0.04$ E12  &   43.3 & (6)  & 0.16 \\ 
IC 5063 lobe &   20 52 02.1 & -57 04 06.6 & 0.0113 & 45.3 &   8.75  & 1.6E10         & \nodata  &  \nodata &  $3.74 \pm 0.08$ E10  &     42.9 & (7)  & 0.008 \\
OQ 208 &    	   14 07 00.4 & +28 27 14.7 & 0.0766 & 337 &  11.52  &  5.2E09       & 1.4E10  & (1)  & $2.31 \pm 0.03$ E11  &    42.7 & (8)  & 0.08 \\
PKS 1549-79 &  15 56 58.9 & -79 14 04.3 & 0.1521 & 699 &   \nodata  &  \nodata   & \nodata  & \nodata  & $1.31 \pm 0.04$ E12  &  44.7  & & 0.005
\enddata 
\tablerefs{(1) \citet{OcanaFlaquer2010};  (2) \citet{Evans1999a}; (3) \citet{Evans2002}; (4) \citet{Ogle2010}; (5) \citet{Massaro2009}; (6) \citet{ODea2000}; (7) \citet{Lutz2004}; (8) \citet{Guainazzi2004}}
\tablenotetext{a}{Positions of the sources used for the Spitzer IRS observations. Except for 3C~305, IC~5063 and 4C~12.50, the positions of the radio nuclei are taken from NED (see Sect.~\ref{sec:obs} for details).}
\tablenotetext{b}{Redshift derived from optical line measurements (see references for the systemic velocities in Table~\ref{table_radio_prop}).}
\tablenotetext{c}{Luminosity distance assuming $\rm H_0 = 73\,$km s$^{-1}$ Mpc$^{-1}$ and $\Omega_m = 0.27$.}
\tablenotetext{d}{Stellar mass derived from the K-band luminosity.}
\tablenotetext{e}{H$_2$ gas masses estimated from CO(1-0) measurements. For 4C~12.50, only the H$_2$ mass in the western lobe is quoted.}
\tablenotetext{f}{Rest $24\,\mu$m luminosity  measured with \textit{Spitzer} IRS.  We averaged the flux over $23-25\,\mu$m, after removal of the [Ne{\sc v}]~$24.31\,\mu$m when present.}
\tablenotetext{g}{Unabsorbed $2-10$~keV nuclear X-ray luminosity in erg~s$^{-1}$. For 3C~236 for which we extracted the X-ray spectrum from the Chandra archive (see sect.~\ref{subsec:exc_mech}). The PKS~1549-79 X-ray flux is from Chandra data (Paul O'Brien, private communication) and agrees with the Suzaku value ($4.5 \times 10^{44}\,$erg~s$^{-1}$). The 3C~459 X-ray flux is from XMM data (Martin Hardcastle, private communication).}
\tablenotetext{h}{Ratio of the H$_2$ line luminosity (summed over the S(0) to S(3) rotational transitions) to the unabsorbed $2-10$~keV nuclear X-ray luminosity.}
\label{table_sample}
\end{deluxetable*}

\begin{deluxetable*}{lcccccccccccc} 
\tablecolumns{13}
\tablewidth{\textwidth} 
\tablecaption{Radio and H{\sc i} properties of outflow radio-galaxies observed with \textit{Spitzer IRS}}
\tablehead{
\colhead{Object} & \colhead{size\tablenotemark{a}}  & \colhead{size} & \colhead{$F_{178 \,  \rm MHz}$}  & \colhead{$Q_{jet}$\tablenotemark{b}}  &   \colhead{$\tau$} & \colhead{$N$(H{\sc i})} & \colhead{$M$(H{\sc i})\tablenotemark{c}}  & \colhead{$v_{\rm sys}$\tablenotemark{d}} & \colhead{$v_{\rm sys}$}  & \colhead{$v_{\rm out}$\tablenotemark{e}} & \colhead{$\dot{M}$} & \colhead{radius}\tablenotemark{f}  \\
	\colhead{}  & \colhead{[kpc]} & \colhead{ref.} & \colhead{[Jy]} &  \colhead{[erg~s$^{-1}$]} & & \colhead{[$10^{21}\,$cm$^{-2}$]} & \colhead{[M$_{\odot}$]} & \colhead{[km s$^{-1}$]} & \colhead{ref.} & \colhead{[km s$^{-1}$]} & \colhead{[M$_{\odot}\,  \rm yr^{-1}$]} &  \colhead{[kpc]}  }
\startdata
     3C 236  &   6.0 & (1) & 15.7  & $7.6\times 10^{45}$ & 0.0033 &  5.0   &    1.2E8 &   30129 & (9)  &    750    &   47.0   &      0.5  \\
     3C 293  &  5.0	 & (2,4) & 17.1  & $5.1\times 10^{45}$ & 0.0038 &  6.0    &   1.0E8   & 13450 & (2)   &   500    &   56.0      &   1.0    \\
     3C 305  &  10.0	 & (3) &13.8  & $6.2\times 10^{45}$ & 0.0023 &  2.0  &     4.7E7  &  12550 & (7)    &  250   &    12.0      &   1.0  \\
     3C 459  &  39.0	 & (5) &  30.8 &$2.5\times 10^{46}$  & 0.0005 &  0.75  &     1.8E7  &  65990 & (10)  &    300   &     5.5         & 1.0  \\
   4C 12.50  &  0.15	 & (6) & 4.6 & $2.5\times 10^{45}$ & 0.0017 & 2.6        &    7.0E6              &  37027 & (10)   &    600    &   21.0        & $0.02- 0.2$  \\
    IC 5063   &  8.0	 & (7) & 6.3 & $2.0\times 10^{45}$ & 0.0120 & 10.0   &    2.4E8  &   3400 & (11)   &   350   &    35.0       &  0.4  \\
    OQ 208   &   0.01 & (6) & 0.12 & $5.2\times 10^{43}$ & 0.0057  & 8.3   &  2.0E4    &  22985 & (12)   &   600   &     1.2        &  0.01  \\
   PKS~1549-79 &  0.4 & (8) & 12.6 & $7.7\times 10^{45}$ & 0.02 &  40.0   &    1.0E8    &  45628 & (10)      &   250   &    30.0        &  $0.1-1.0$
\enddata 
\tablenotetext{a}{Size of the radio source. Note that for 3C~293, this is the size of the \textit{inner} radio source.}
\tablenotetext{b}{Jet cavity kinetic luminosity estimated from the 178~MHz flux (see text, Sect.$\ $\ref{sec:sample}).}
\tablenotetext{c}{Mass of H{\sc i} gas in the outflowing component (within the radius quoted in the table).}
\tablenotetext{d}{Systemic velocity of the galaxy (see text for details).}
\tablenotetext{e}{Outflow velocity, estimated as half of the full width at zero intensity of the blue-shifted part of the broad H{\sc i} component.}
\tablenotetext{f}{Radius of the H{\sc i} outflow region. Except for PKS~1549-79, the radii values are from$\ $ \citet{Morganti2005}.}
\tablerefs{(1) \citet{deKoff2000}; (2) \citet{Emonts2005}; (3) \citet{Morganti2005a}; (4) \citet{Zirbel1998};  (5) \citet{Nilsson1993}; (6) \citet{Xiang2002}; (6) \citet{Heisler1994}; (7) \citet{Morganti2005}; (8) \citet{Holt2006}; (9) \citet{Hill1996}; (10) \citet{Holt2008}; (11) \citet{Morganti1998}; (12) \citet{Marziani1993}}
\label{table_radio_prop}
\end{deluxetable*}

Except PKS~1549-79, all of the observed sources are part of the \citet{Morganti2005} sample. Most of these sources are bright compact radio sources, with sizes ranging from 0.01 to 40 kpc. 
These H{\sc i}-outflow objects appear to belong to the minority of powerful radio galaxies ($\approx 35$\%) that show evidence for recent star formation at optical wavelengths \citep[][based on spectral synthesis modelling]{Tadhunter2011} and/or mid-IR wavelengths \citep[][based on PAH detection]{Dicken2011}. Most such objects are compact radio  sources (e.g. CSS/GPS/CSO) or have unusually strong compact steep spectrum cores \citep{Tadhunter2011}. The star formation in these objects is probably linked  to the presence of an unusually rich ISM in the nuclear regions of the  host galaxies, perhaps accreted in gas-rich mergers. In such objects the  radio jets are expected to interact particularly strongly with the dense  circumnuclear gas. Therefore these are just the type of objects in which we  might expect the jets to have a direct impact on the molecular gas. 

Some properties of the sample are listed in Table~\ref{table_sample}. The stellar masses are derived from the K-band luminosities according to \citet{Marconi2003}:
\begin{equation}
\log _{10} (M_{\star}) = -2.3 + 1.21 \times \log _{10} (L_K) \ ,
\end{equation}
where $M_{\star}$ is the bulge stellar mass in $M_{\odot}$ and $L_K$ the $K$-band luminosity in $L_{\odot}$.
The unabsorbed $2-10$~keV AGN X-ray fluxes are taken from the literature, except 3C~236, 3C~459 and PKS~1549-79 which were unpublished. 
For 3C~236, we reprocessed and extracted the Chandra ACIS spectrum. We find that the spectrum is well fitted with a single power law of photon index 1.67, absorbed by a column of $N_{\rm H} = 2.3 \times 10^{23}$~cm$^{-2}$, consistent with the presence of a dust lane close to the nucleus \citep{deKoff2000}. The 3C~459 and PKS~1549-79 X-ray fluxes were provided by P. O'Brien and M. Hardcastle (private communications). The jet kinetic power is estimated from the 178~MHz flux density, according to the formula\footnote{This formula uses the 151~MHz flux density. When a direct measurement at 150~MHz was not available, we derived it from the 178~MHz measurement (taken from NED) by fitting the measured radio SED with a power-law.} derived in \citet{Punsly2005}.

These galaxies all show very broad (up to $1000\,$km~s$^{-1}$) H{\sc i} absorption profiles, with highly blue-shifted H{\sc i} gas with respect to their optical systemic velocities, which indicates a fast outflow of neutral gas \citep[e.g.][]{Morganti2003}. The associated H{\sc i} mass outflow rates are up to $\approx 60\,$M$_{\odot}$~yr$^{-1}$, up to 2 orders of magnitude higher than the mass outflow of ionized gas \citep[see e.g.][]{Emonts2005}. These outflow rates are estimated  from the gas column density, the outflow velocity, and the radius within which the flow originates \citep[see e.g. ][for a review]{Heckman2002}.  In some cases, like 3C~305 \citep{Morganti2005a}, IC~5063 \citep{Oosterloo2000, Morganti2007} and 3C~293 \citep{Morganti2003, Emonts2005}, this radius is estimated from high-resolution radio and optical observations, where the outflow is spatially resolved against the radio continuum, and co-spatial with regions where the radio emission in the propagating radio jet is enhanced. For the other sources, the radii are more uncertain. It is assumed that the outflow is coming from the brightest radio continuum region, which in principle provides only a lower limit since the H{\sc i} absorption can only be traced in regions where the continuum is bright. 
We checked that these radii are consistent with optical spectroscopy, but note that these radii estimates only provide an order of magnitude. Some of the main radio and H{\sc i} properties are summarized in Table~\ref{table_radio_prop}. 
In sect.~\ref{subsec:HI_kinematics}, we discuss in more detail the H{\sc i} kinematics, and compare them to the kinematics of the molecular and ionized gas phases.

\section{\textit{Spitzer} observations, data reduction and analysis}
\label{sec:obs}

\begin{figure*}
   \centering
    \includegraphics[width=\textwidth]{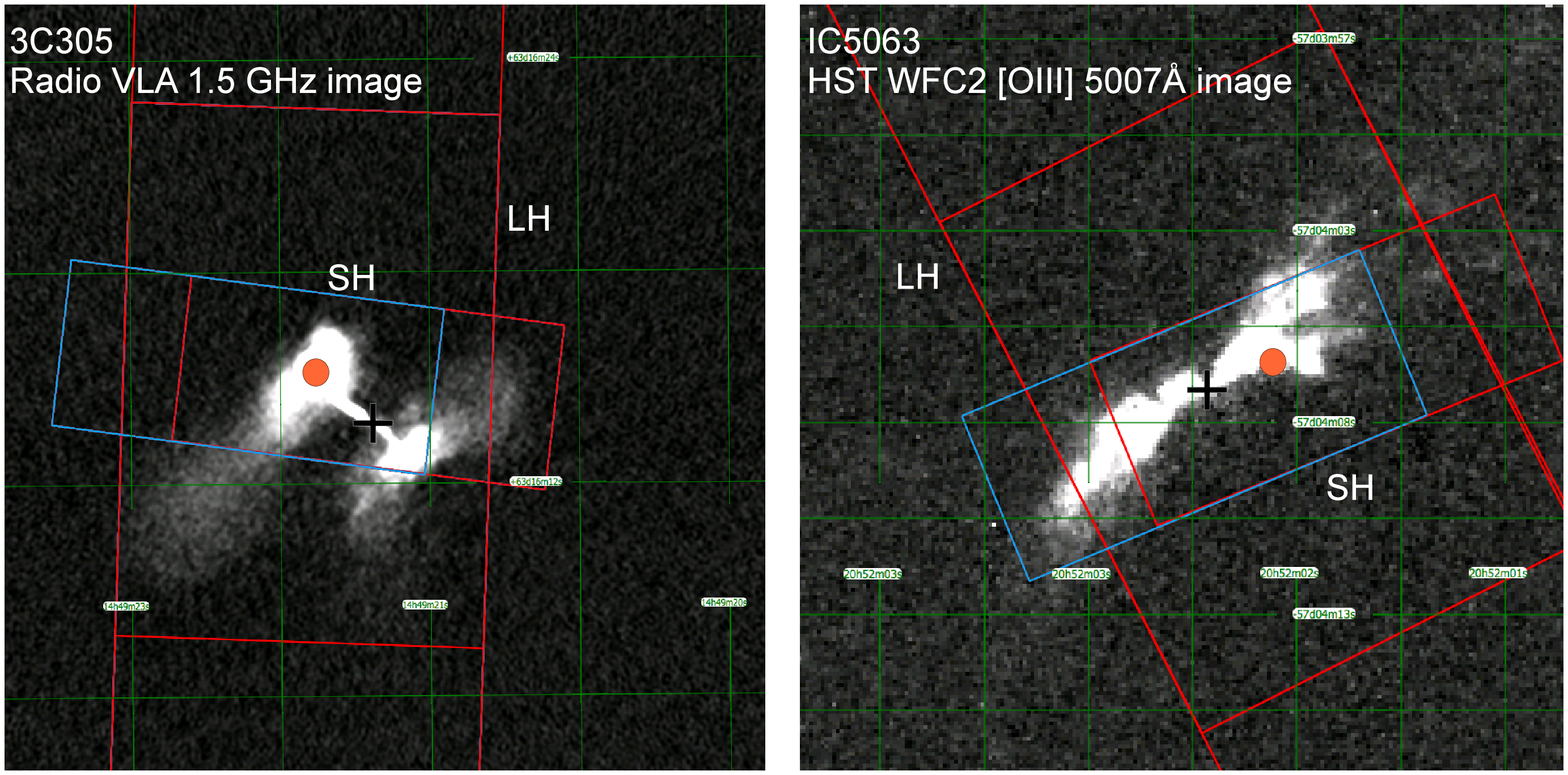}
      \caption{Footprints of the high-resolution IRS slits (SH and LH) for the 3C305 and IC5063 galaxies. For these two sources, the radio observations resolve the H{\sc i} outflow against the radio jet, at a distance of about 2 arcsec from the nucleus. Therefore, the slits were centred slightly off-nucleus, at the location of the broad, blue-shifted H{\sc i} absorption associated with the outflow (indicated by the orange circle). The position of the radio core is marked with a green cross.}
       \label{fig:AOR_IRS_hires_footprints}
   \end{figure*}

The IRS has two high-resolution modules, short high\footnote{$4.7'' \times 11.3''$ slit, observed wavelengths from 9.9 to 19.6$\,\mu$m} (SH) and long-high\footnote{$11.1'' \times 22.3''$ slit, observed wavelengths from 18.7 to 37.2$\,\mu$m} (LH), and four low-resolution modules, short-low (SL1 \& SL2) and long-low (LL1 \& LL2), spanning $5.2-38\,\mu$m.
In low resolution mode ($\lambda/\Delta \lambda \approx 57-127$), we used publicly available archival data for 7 of the 8 sources. Only PKS~1549-79 was not previously observed at low resolution, hence we observed this source as part of our \textit{Spitzer} Cycle 4 program p40453. At high resolution ($\lambda/\Delta \lambda \approx 600$), we observed 5 sources (3C~236, 3C~305, 3C~459, PKS~1549-79, and IC~5063) as part of our program p40453, and we used archival data for the remaining 3 sources (3C~293, 4C~12.50, and OQ~208) (p30877). 
The exposure times for each mode are listed in Table~\ref{table_obs_log}, and were chosen to reach a signal-to-noise ratio of at least 10 for the H$_2$ S(1) $17\,\mu$m line. 

Even if the outflow of neutral gas is occurring at a kpc-scale distance from the nucleus, this is still spatially unresolved with the IRS in high-resolution mode. However, two sources (3C~305 and IC~5063) exhibit a more extended inner radio structure, and we placed the slit slightly off-nucleus, coincident with a bright peak in the radio continuum, at a distance of 2 arcsec from the core. Note that the core is still included in all of the \textit{IRS} slits. 
The slit positions are given in Figure~\ref{fig:AOR_IRS_hires_footprints}. 4C~12.50 is a system of two interacting galaxies \citep{Axon2000}, and the SH slit was positioned only to include the western companion. For the other sources, the slits were centred on the nucleus.

\subsection{\textit{Spitzer} data reduction}

\begin{deluxetable}{lcccc} 
\tablecolumns{5}
\tablewidth{0pc} 
\tablecaption{\textit{Spitzer IRS} observation log}
\tablehead{
\colhead{Object} & \colhead{SL}  & \colhead{LL}  & \colhead{SH}  & \colhead{LH}}
\startdata
3C 236	           & $13 \times 14$ & $5 \times 14$ & $4 \times 120$ & $10 \times 60$ \\	   
3C 305	           & $13 \times 14$ & $5 \times 14$ & $4 \times 120$ & $10 \times 60$ \\	  		   
3C 293	           & $3 \times 120$ & $3 \times 120$ & $3 \times 120$ & $10 \times 60$ \\	  		   
3C 459	           & $16 \times 14$ & $2 \times 30$ & $4 \times 120$ & $10 \times 60$ \\	  		   
4C +12.50      & $3 \times 14$  & $2 \times 30$  & $6 \times 30$   & $4 \times 60$ \\	       
OQ 208          & $2 \times 60$ & $2 \times 14$ & $4 \times 120$ & $3 \times 60$  \\	     	   
IC 5063 lobe  & $2 \times 14$ & $1 \times 30$ & $8 \times 30$   & $4 \times 60$ \\	    
PKS 1549-79 & $2 \times 60$ & $3 \times 30$ & $4 \times 120$ & $5 \times 60$   
\enddata
\tablecomments{Exposure times for the \textit{IRS} modules given as number of cycles $\times$ ramp duration in seconds.}
\label{table_obs_log}
\end{deluxetable}

The galaxies were observed in staring mode, and placed at two nod positions along the slit (1/3 and 2/3 of the length of the slit). The data were processed by the \textit{Spitzer Science Center} S18.7 pipeline. We started the data reduction from the basic calibrated data products, which include corrections for flat-fielding, stray light and non-linearity of the pixels response. All but one (4C~12.50) galaxies had dedicated sky background (off) observations for all high-resolution modules. When available, the 2D \textit{off} spectra were subtracted from the \textit{on} observation for each nod position. For 4C~12.50, the LH module was used to remove the background emission.  For the LH observations, we corrected for the time- and position-dependence of the dark current using the DARK\_SETTLE routine\footnote{\url{http://irsa.ipac.caltech.edu/data/SPITZER/docs/dataanalysistools/tools/darksettle/}}. 

The cleaning of  rogue pixels was performed using the IRSCLEAN routine\footnote{\url{http://irsa.ipac.caltech.edu/data/SPITZER/docs/dataanalysistools/tools/irsclean/}}. Rogue pixel masks of the corresponding \textit{IRS} campaigns were used, and a further manual cleaning was done for each nod position and module. The \textit{Spitzer} IRS Custom Extractor (SPICE) software was used to extract the one-dimensional SH and LH spectra, with optimal extraction. 
The individual orders within the high-resolution modules were very well aligned in flux and no scaling was applied between the orders. 
Then the orders were trimmed (by removing the end of each order, where the response drops) and stitched together. Finally a median combination was used to produce the final stacked spectra. 

For all but one source (4C~12.50), the continuum levels of the SH and LH matched and no scaling was necessary. For 4C~12.50, we scaled the SH spectrum by a  factor of 1.1 to match the LH module, since it is the most accurate due to its background subtraction. 

The 3C236 and 3C305 low-resolution spectra were extracted from spectral maps with the CUBISM\footnote{\url{http://irsa.ipac.caltech.edu/data/SPITZER/docs/dataanalysistools/tools/cubism/}} software. For the other galaxies, the low-resolution observations were done in IRS staring mode, and the SMART package\footnote{\url{http://isc.astro.cornell.edu/IRS/SmartRelease}} was used to perform optimal extraction. We checked on 3C~293 that the SPICE and SMART extractions give very similar results.

\subsection{Spectral analysis}
\label{subsec:spectral_analysis}

To measure the fluxes and linewidths of emission lines, we first fitted the spectra with the IDL PAHFIT\citep{Smith2007} tool, which decomposes the spectra into starlight, thermal blackbody emission (extinguished by a uniform dust screen if specified), resolved PAH features (fitted by Drude profiles), and emission lines. Originally designed to fit low-resolution \textit{Spitzer} IRS spectra, we adapted the routine to handle high-resolution data. In particular, we allow the emission lines to be spectrally resolved and we fit them by Gaussian profiles. 

We use the PAHFIT decompositions to measure the fluxes of the PAH complexes, and to produce starlight-free and dust-free spectra. 
Except for OQ~208 and PKS~1549-79, we do not include any extinction in the fit. For OQ~208, 4C~12.50 and PKS~1549-79, we introduced the capability of fitting silicate absorption features at 9.7 and 18$\,\mu$m. 
Then we ran PAHFIT a second time to fit accurately the emission lines. We also fitted the lines with Gaussian profiles using the non-linear least-squares IDL fitting routine MPFIT \citep{Markwardt2009} on the continuum-free spectra. The two methods agree very well and give similar line fluxes (within the uncertainties). We find significant differences (up to a factor of 3 in line flux) when trying to measure the line fluxes in one go with PAHFIT, the many free parameters making the fit inaccurate.

For each of undetected emission lines, we computed the flux of  an unresolved Gaussian line with a peak flux equal to twice the rms noise in a $1\,\mu$m-wide band centred at the wavelength of the line. We took this flux value as a $2\, \sigma$ upper limit.

We also compute the $24\,\mu$m rest-frame fluxes by averaging the IRS line-free spectrum over a narrow band ($23-25\,\mu$m).  
These narrowband $24\,\mu$m fluxes are listed in Table~\ref{table_sample}.

\section{Spectroscopic diagnostics}
\label{sec:spec_diag}

\begin{figure*}
   \centering
    \includegraphics[height=0.49\textwidth, angle=90]{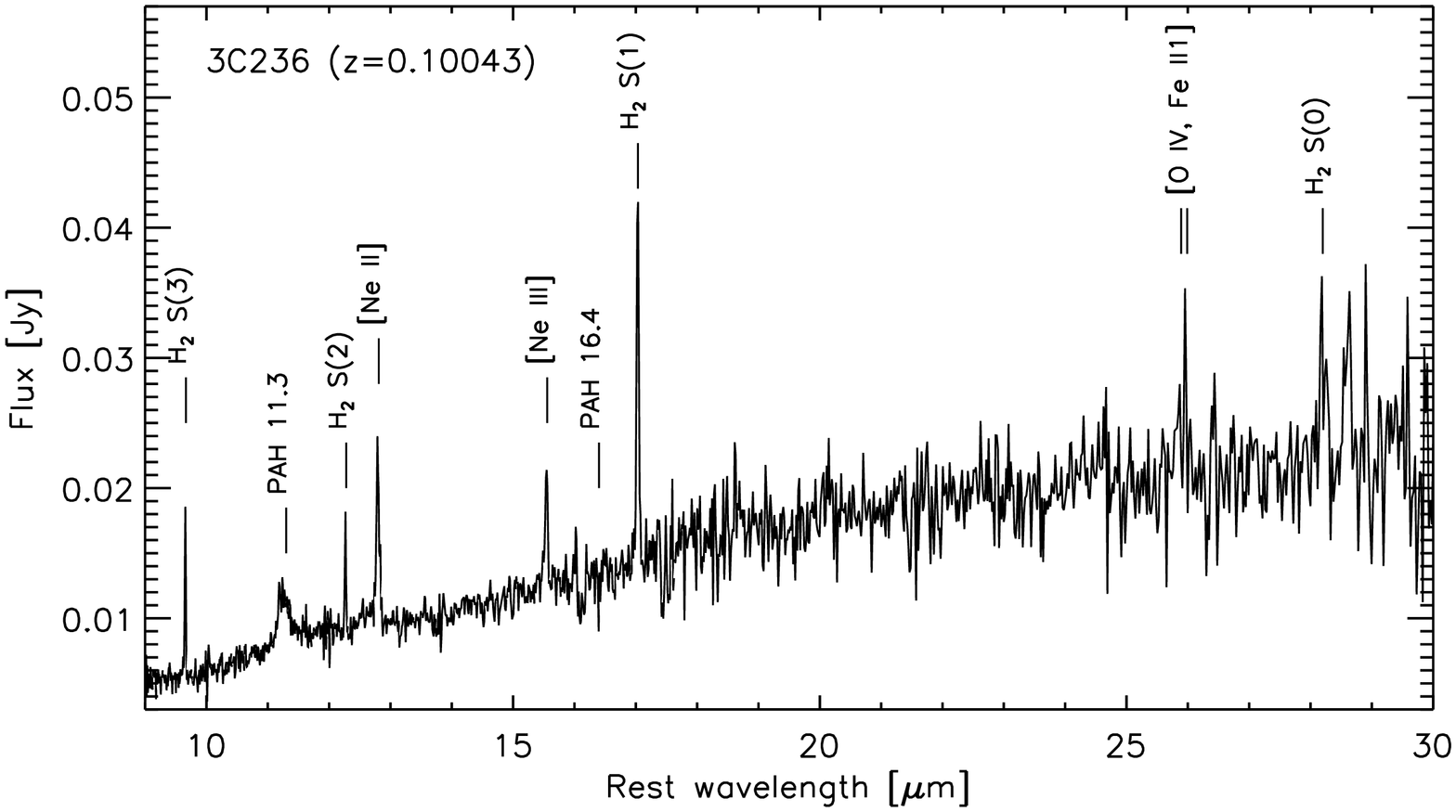}
    \includegraphics[height=0.49\textwidth, angle=90]{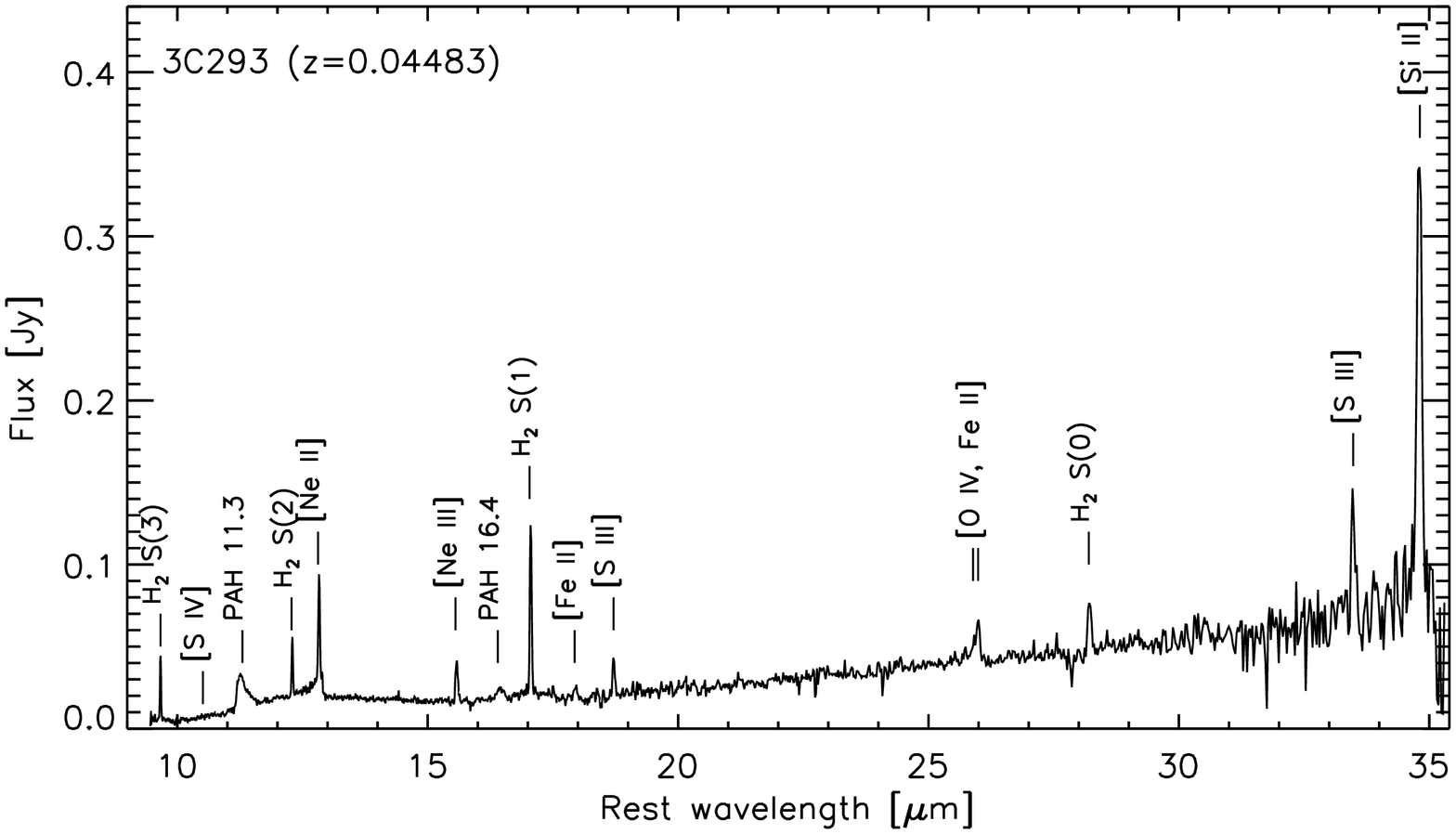}
    \includegraphics[height=0.49\textwidth, angle=90]{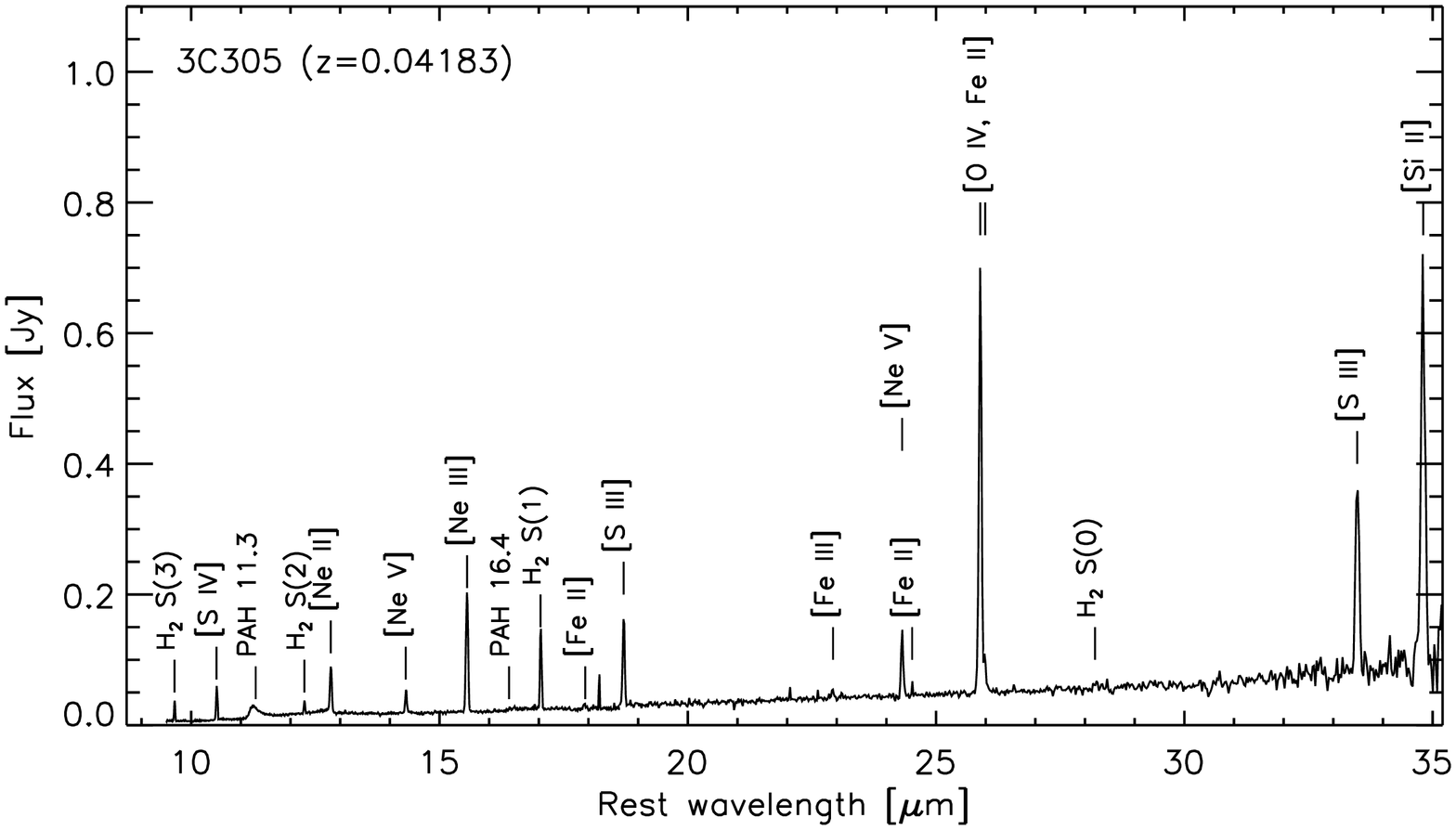}
    \includegraphics[height=0.49\textwidth, angle=90]{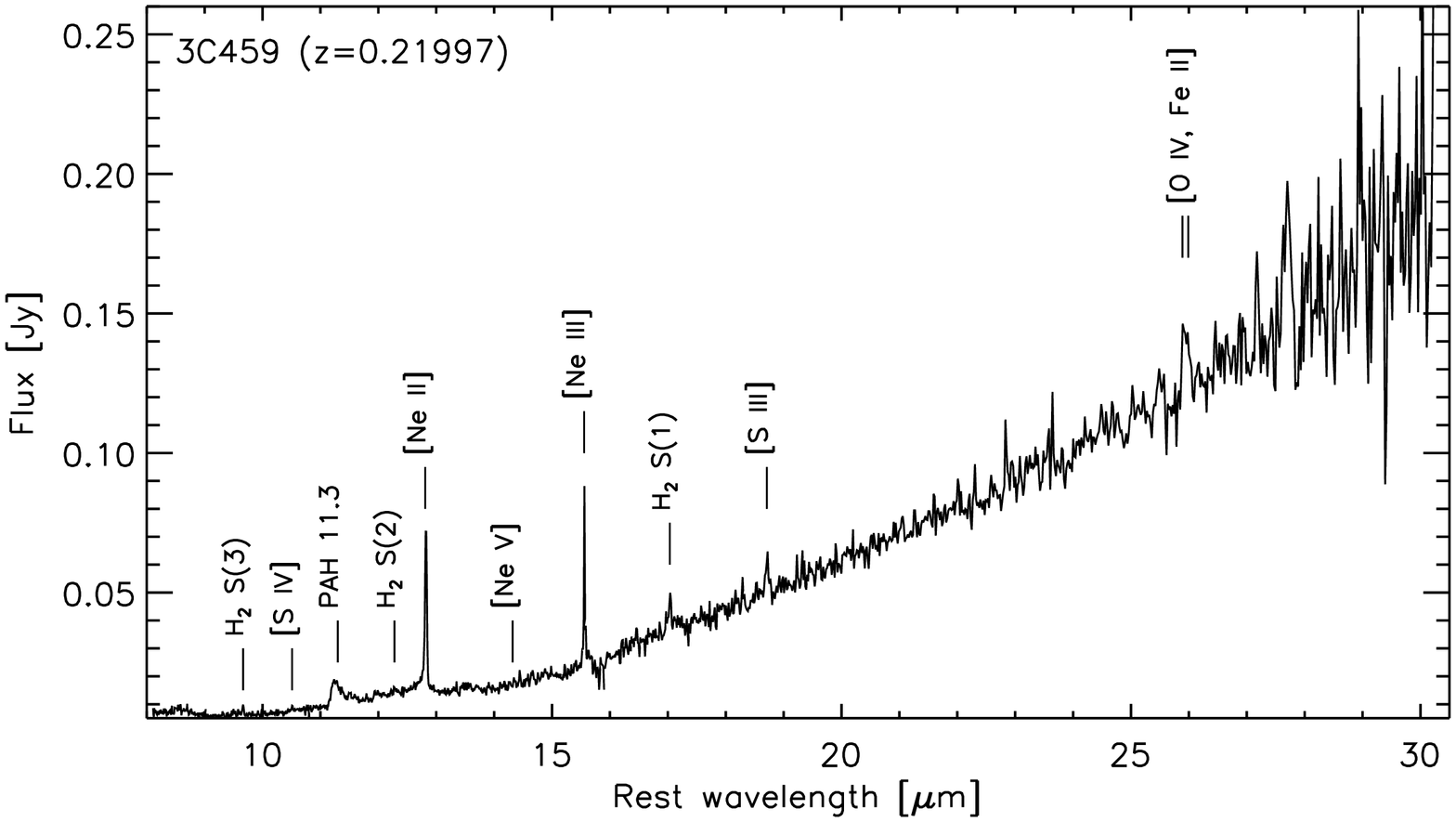}
    \includegraphics[height=0.49\textwidth, angle=90]{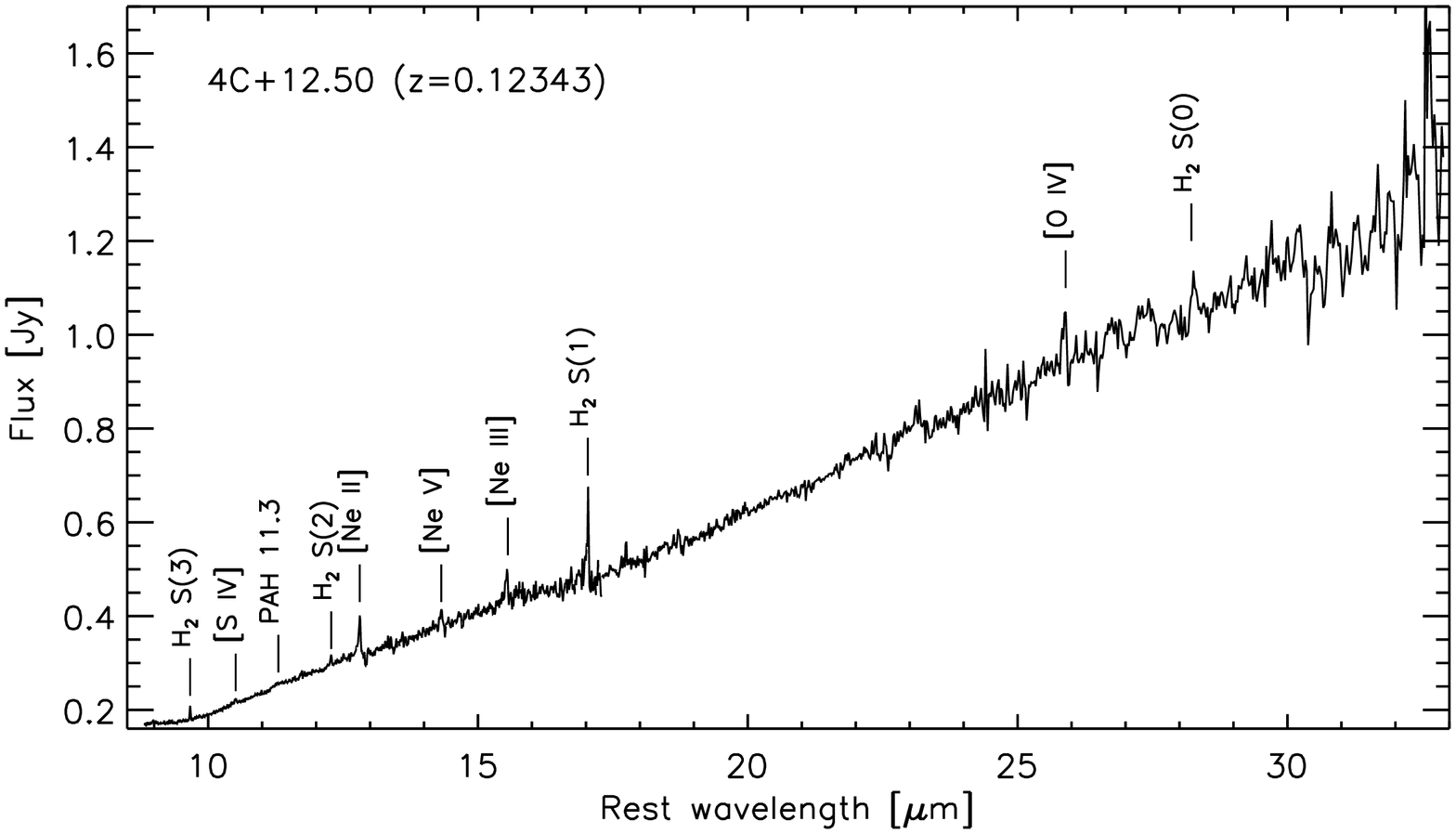}
    \includegraphics[height=0.49\textwidth, angle=90]{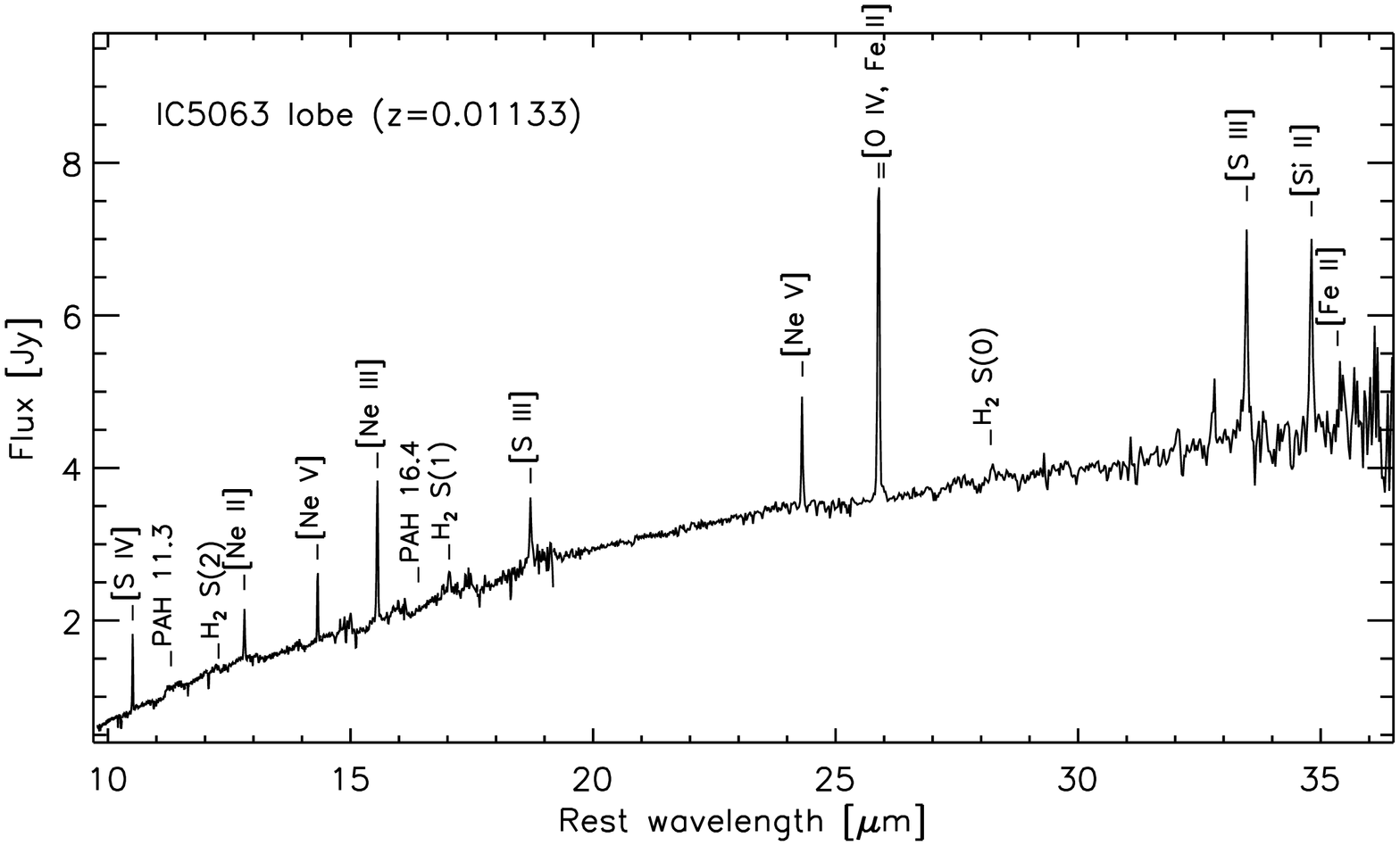}
    \includegraphics[height=0.49\textwidth, angle=90]{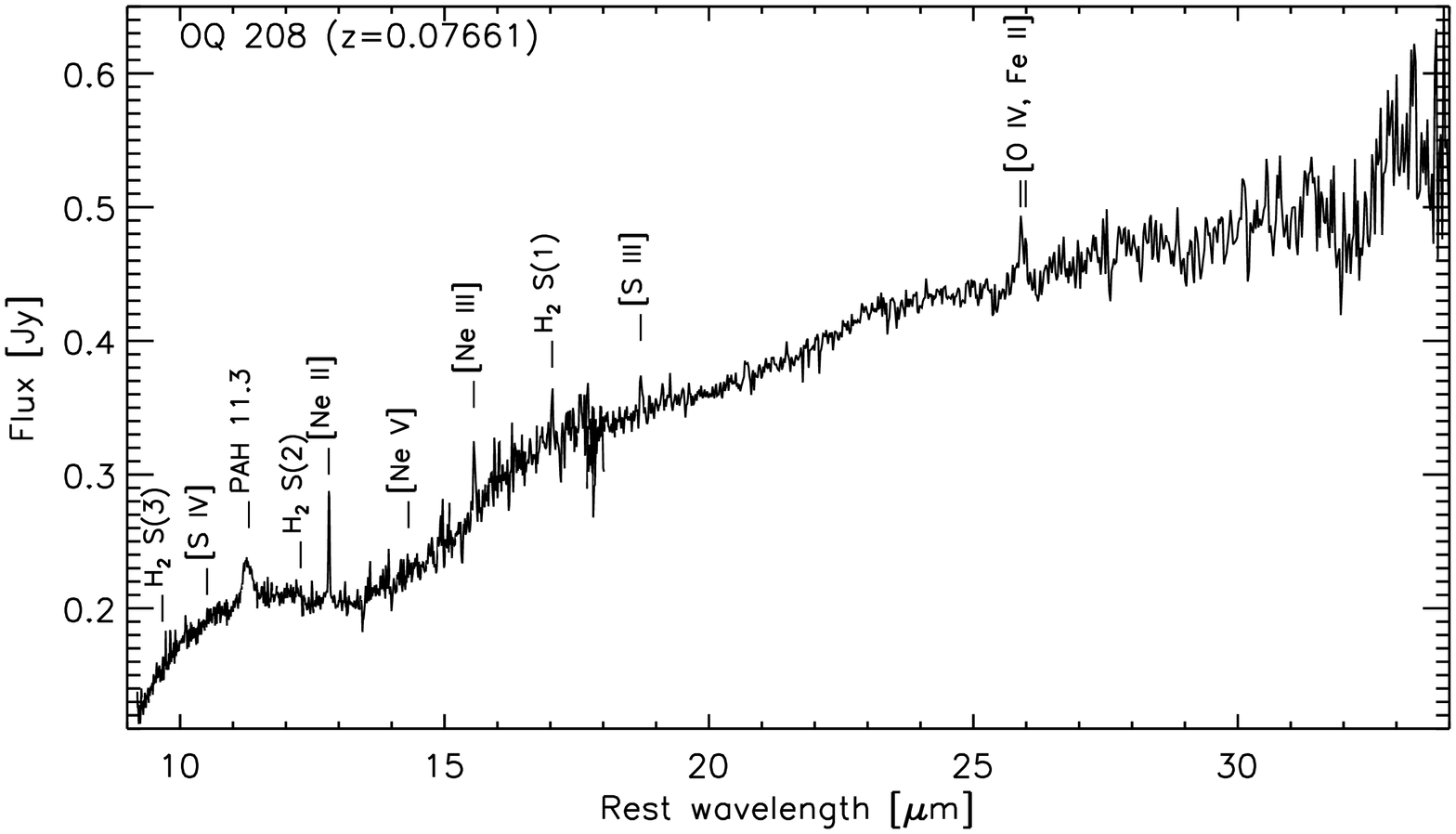}
    \includegraphics[height=0.49\textwidth, angle=90]{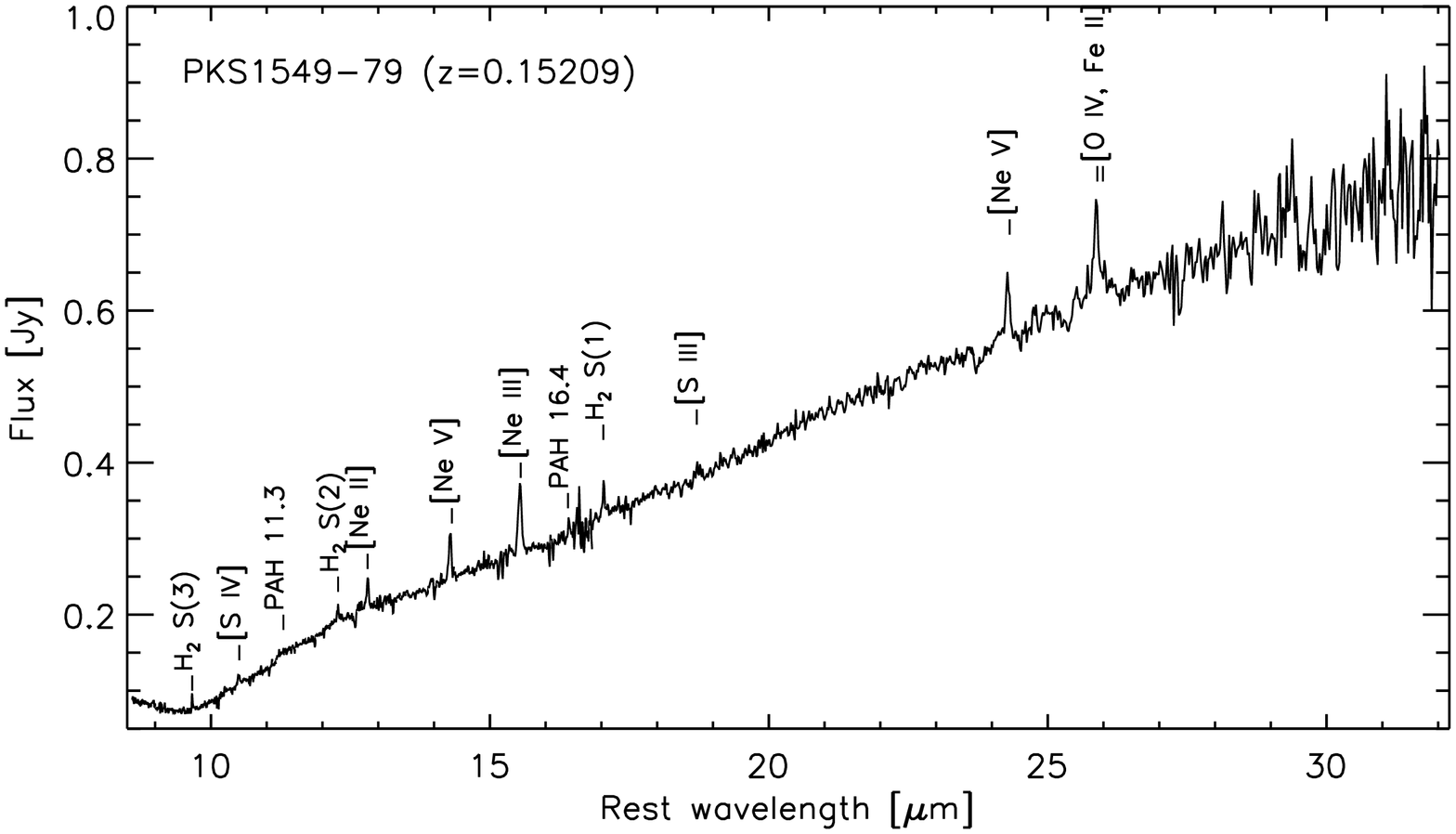}
      \caption{\textit{Spitzer} IRS high-resolution (SH+LH) spectra of the H{\sc i}-outflow radio-galaxies. The locations of PAH features and emission lines are marked for each object.}
       \label{fig:IRS_spec}
   \end{figure*}

\begin{figure*}
   \centering
    \includegraphics[height=0.49\textwidth, angle=90]{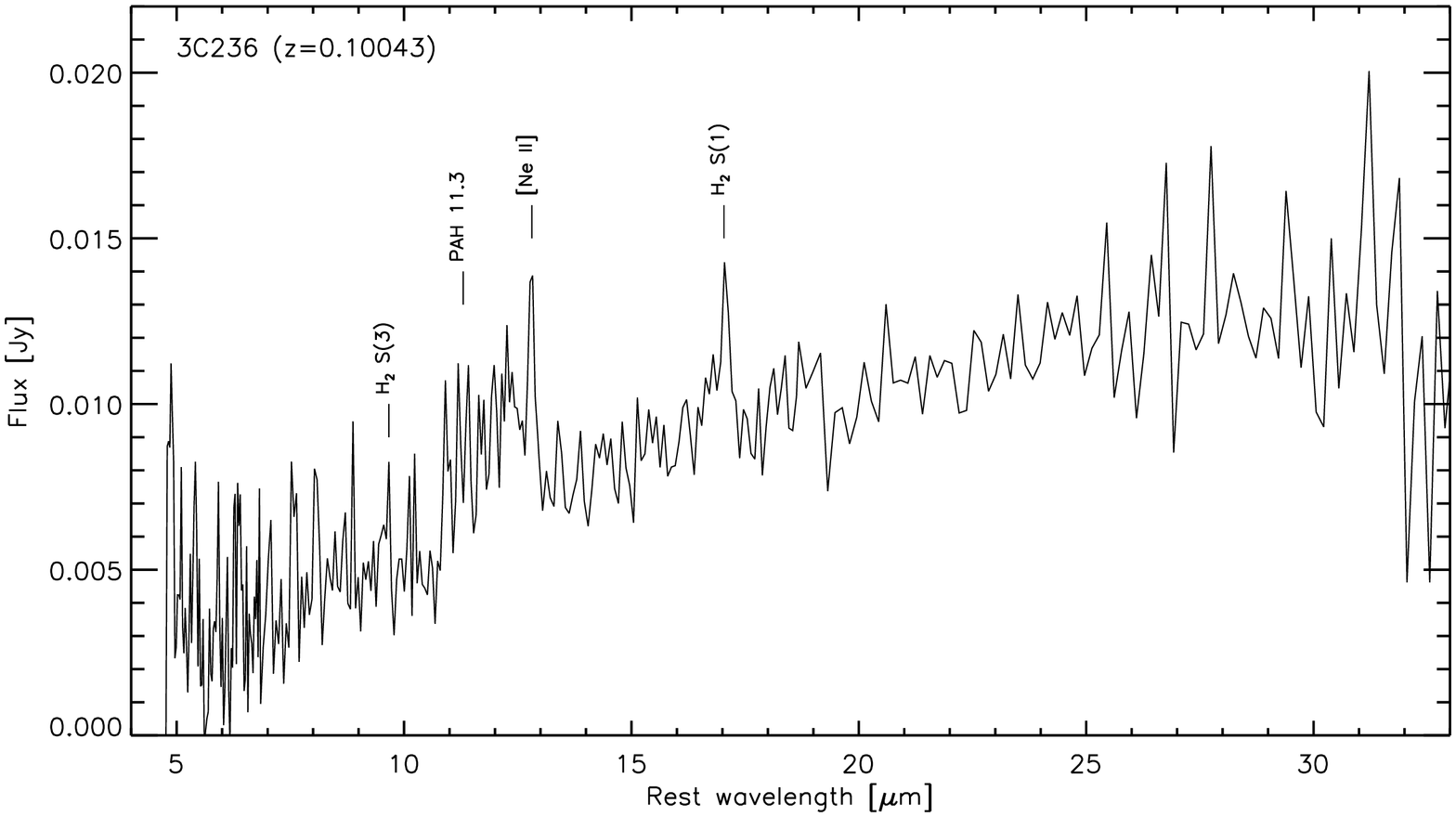}
    \includegraphics[height=0.49\textwidth, angle=90]{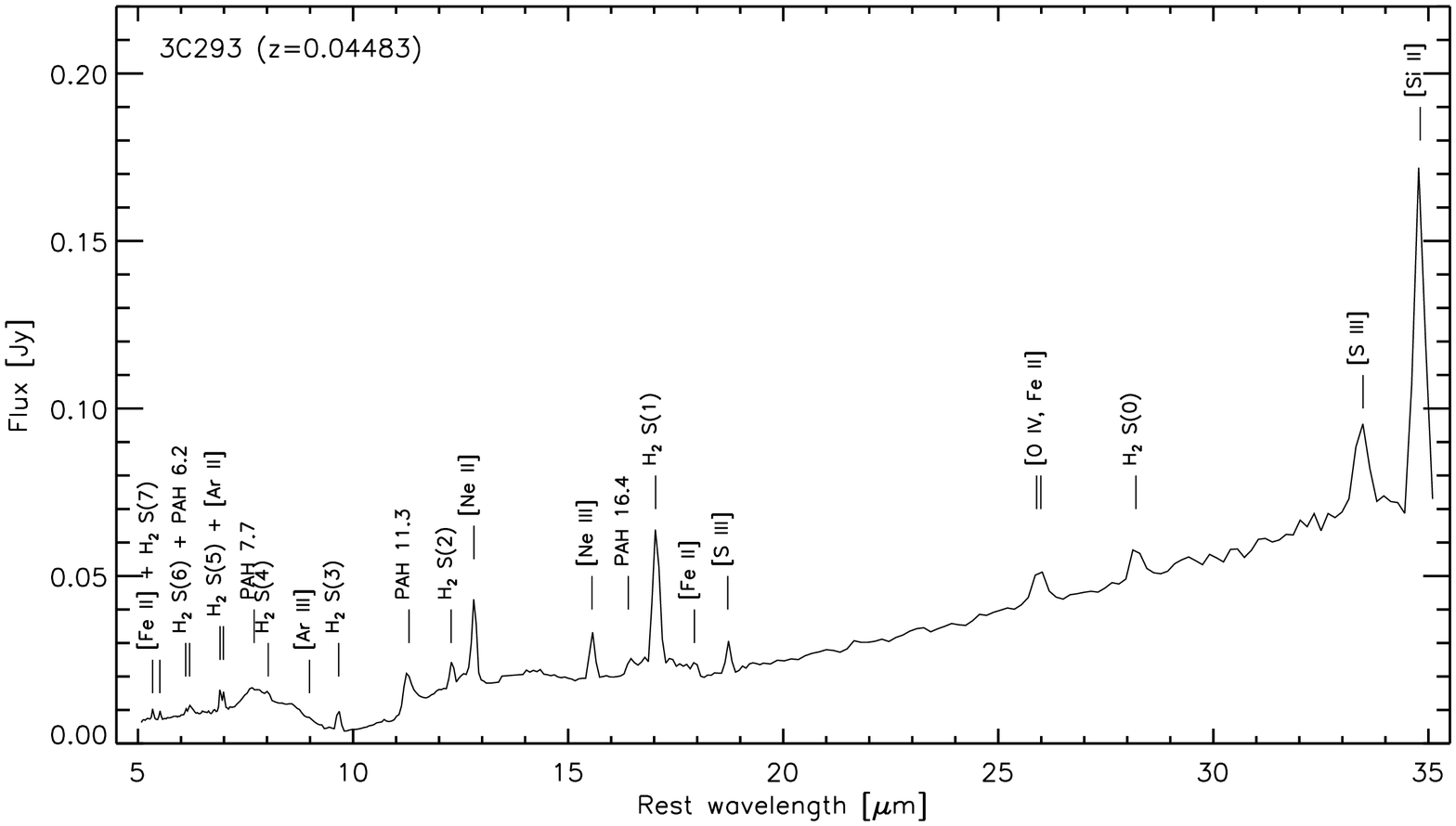}
    \includegraphics[height=0.49\textwidth, angle=90]{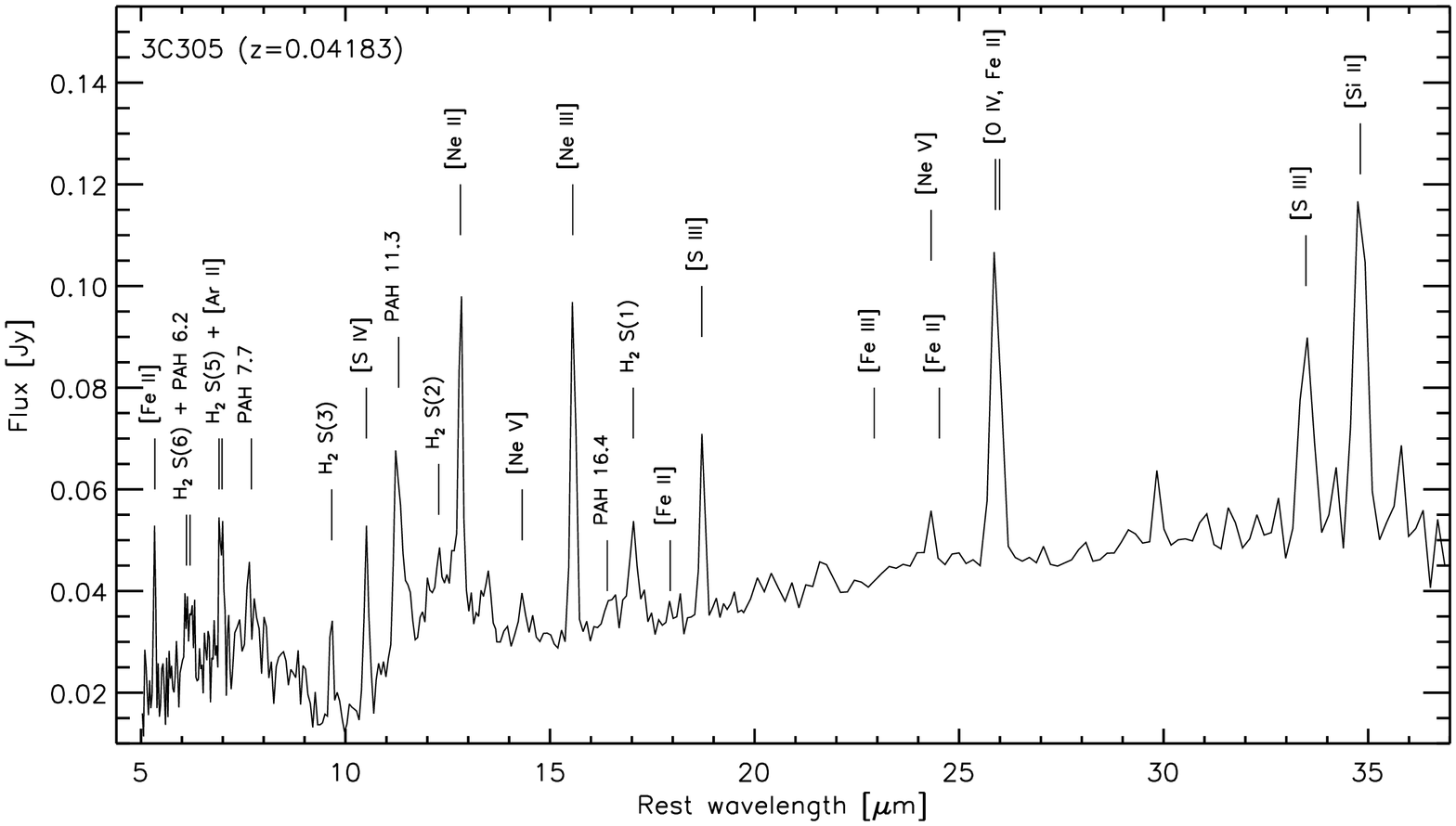}
    \includegraphics[height=0.49\textwidth, angle=90]{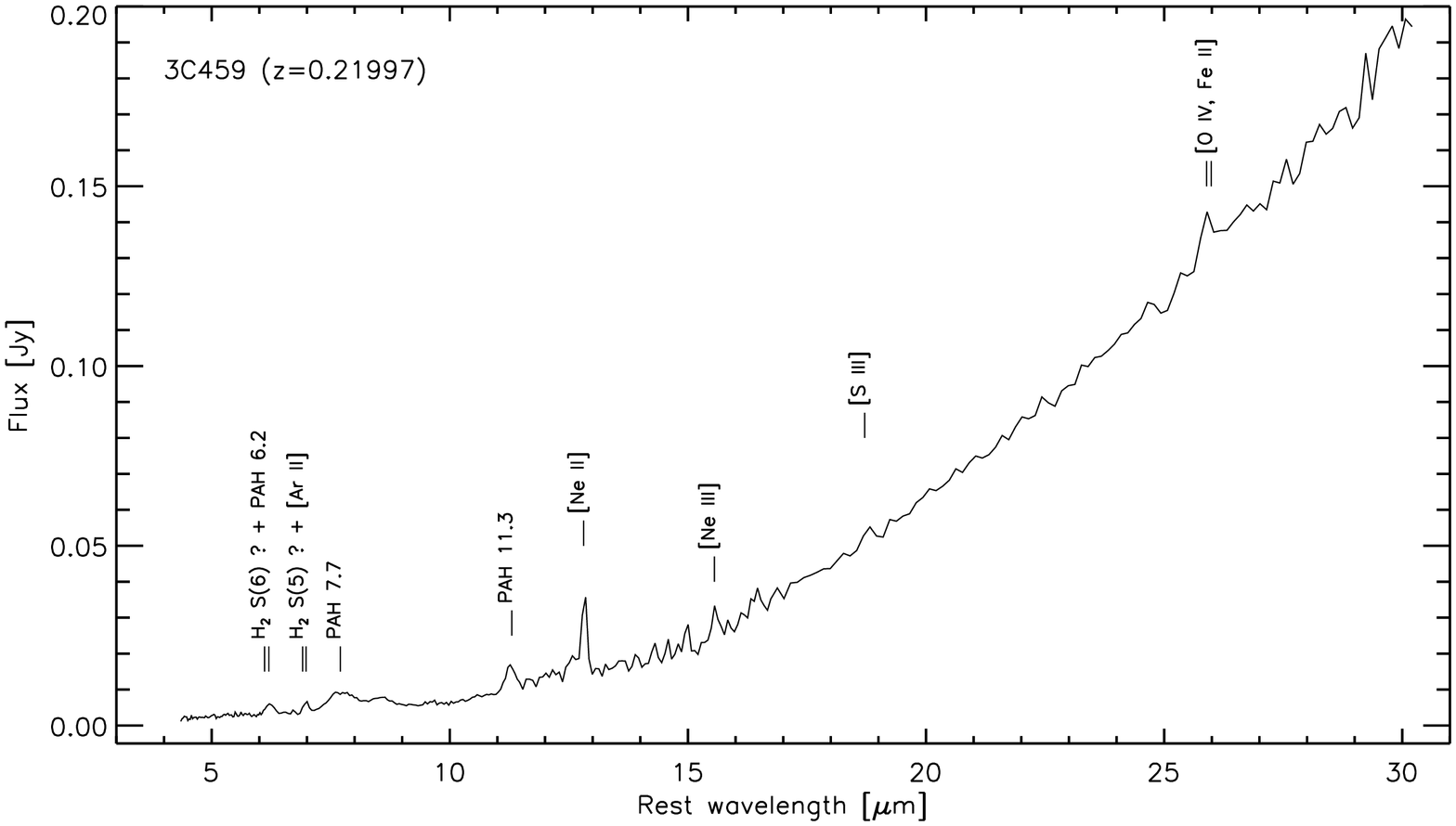}
    \includegraphics[height=0.49\textwidth, angle=90]{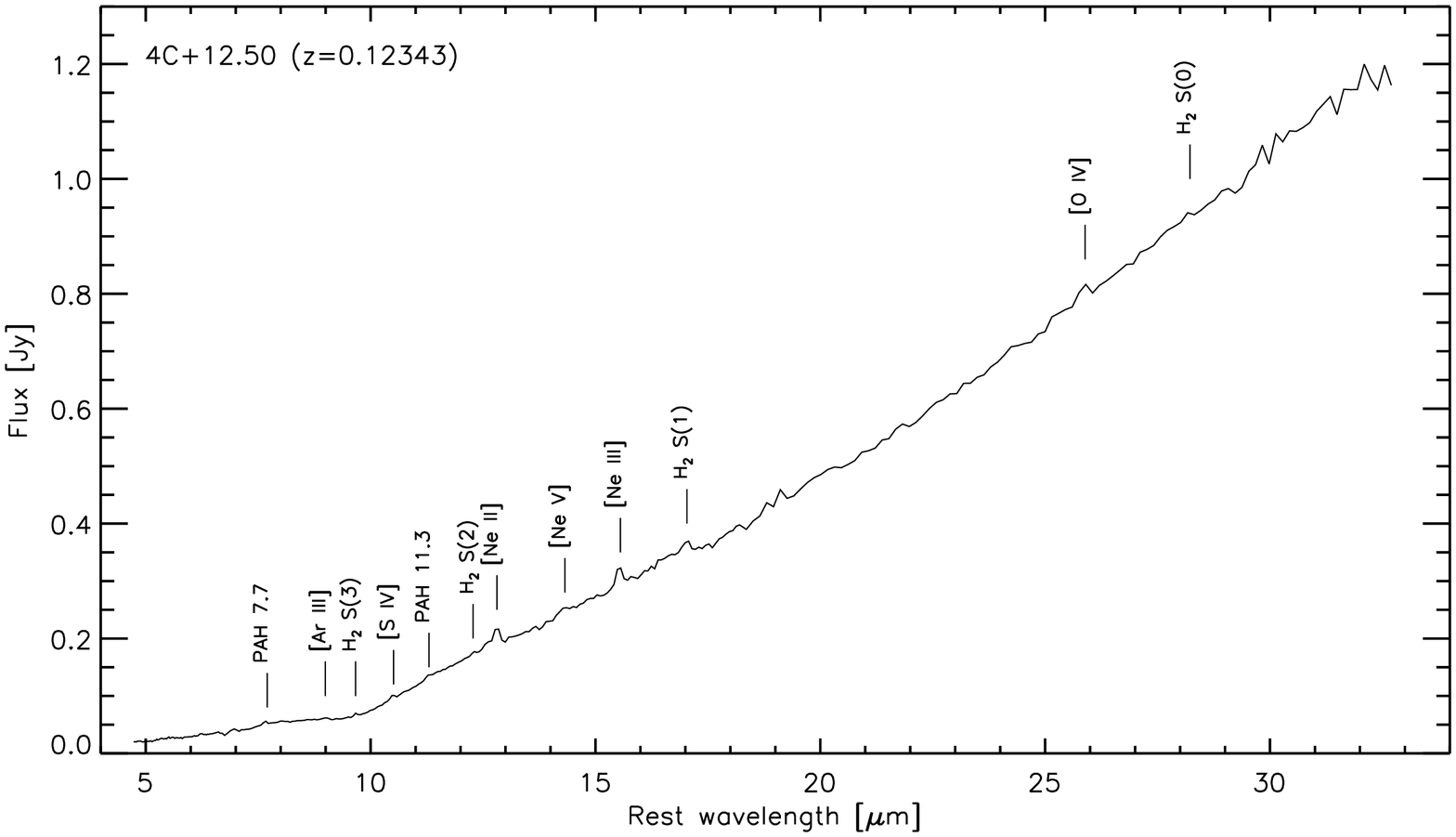}
    \includegraphics[height=0.49\textwidth, angle=90]{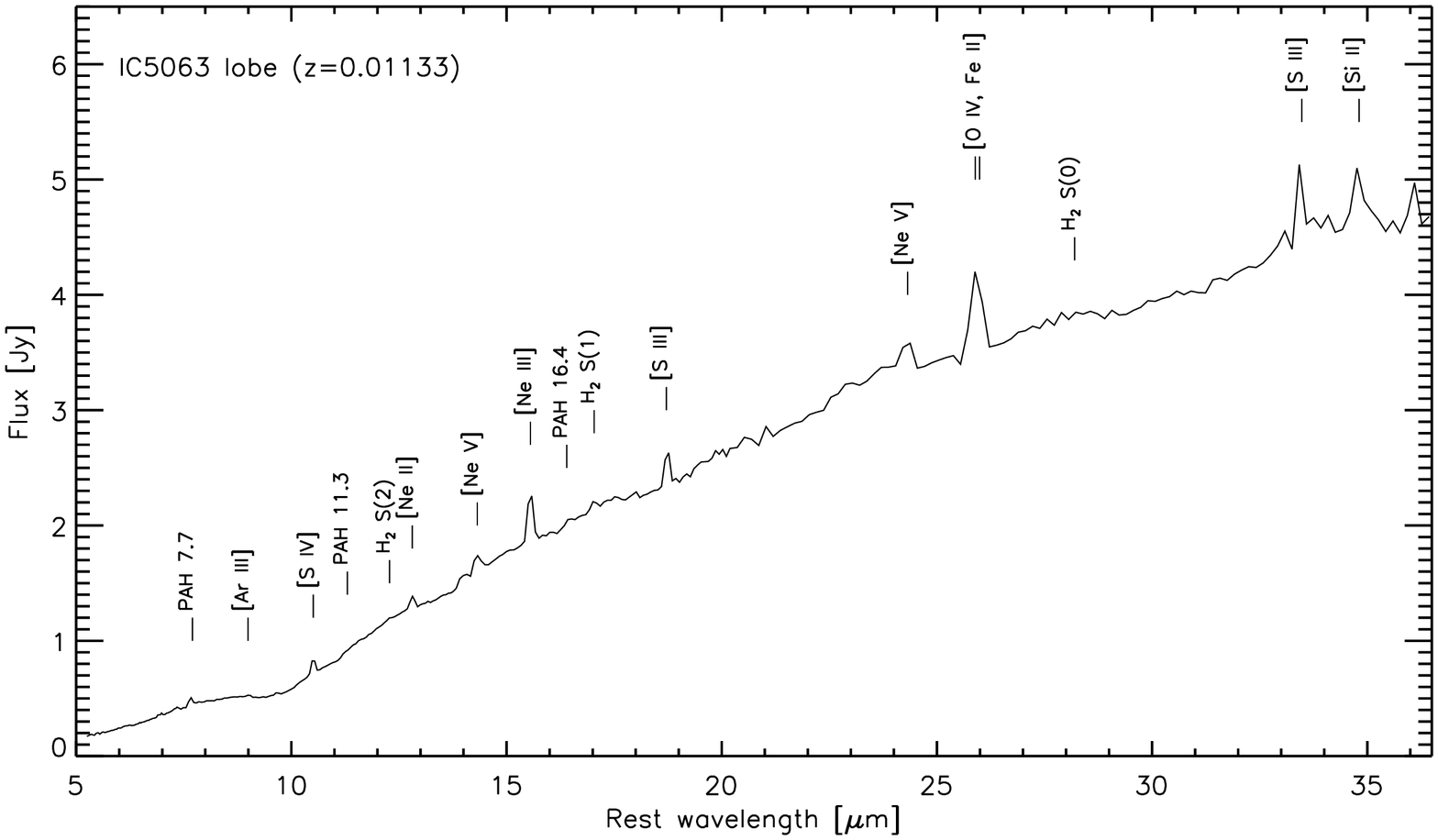}
    \includegraphics[height=0.49\textwidth, angle=90]{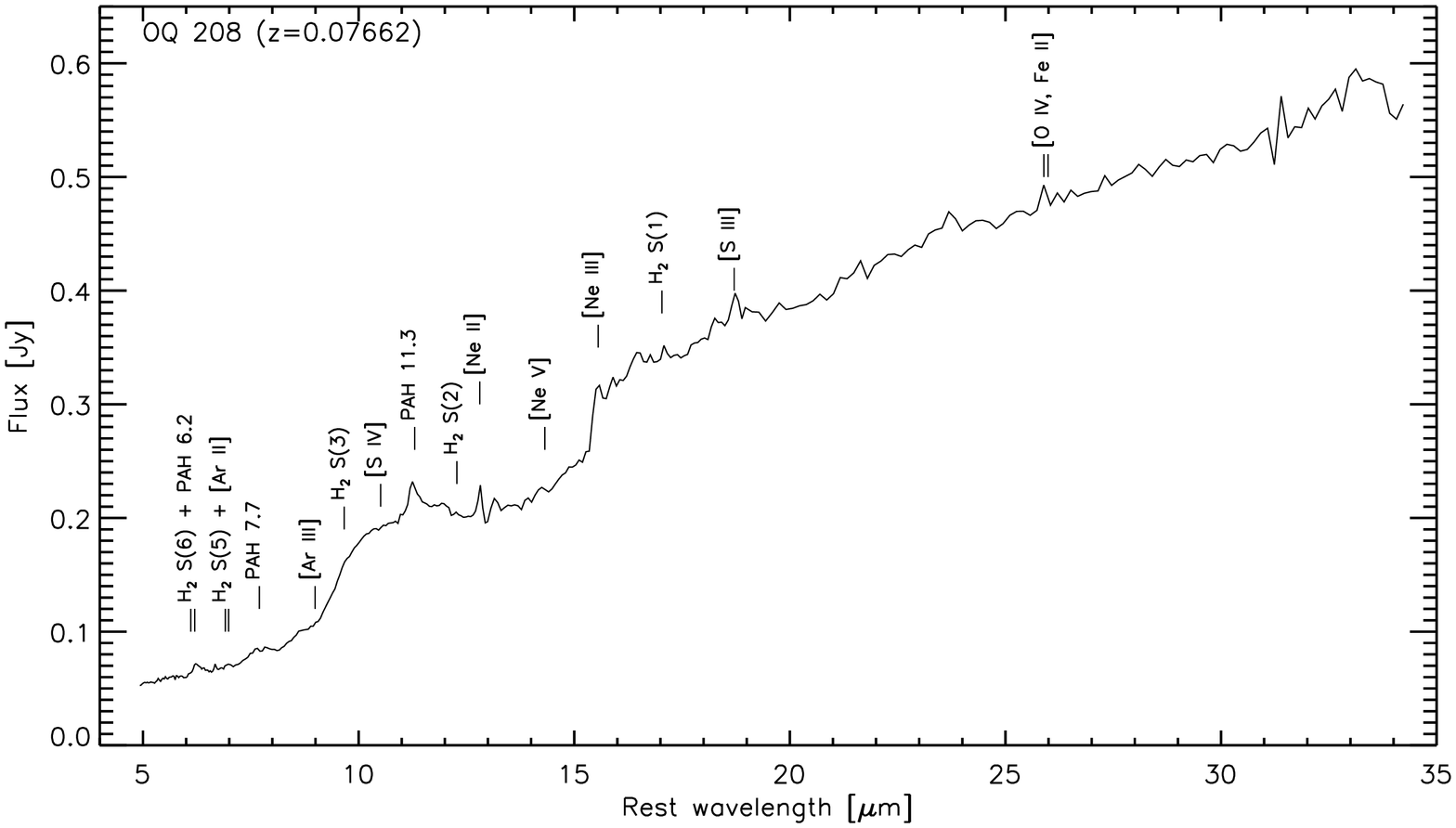}
    \includegraphics[height=0.49\textwidth, angle=90]{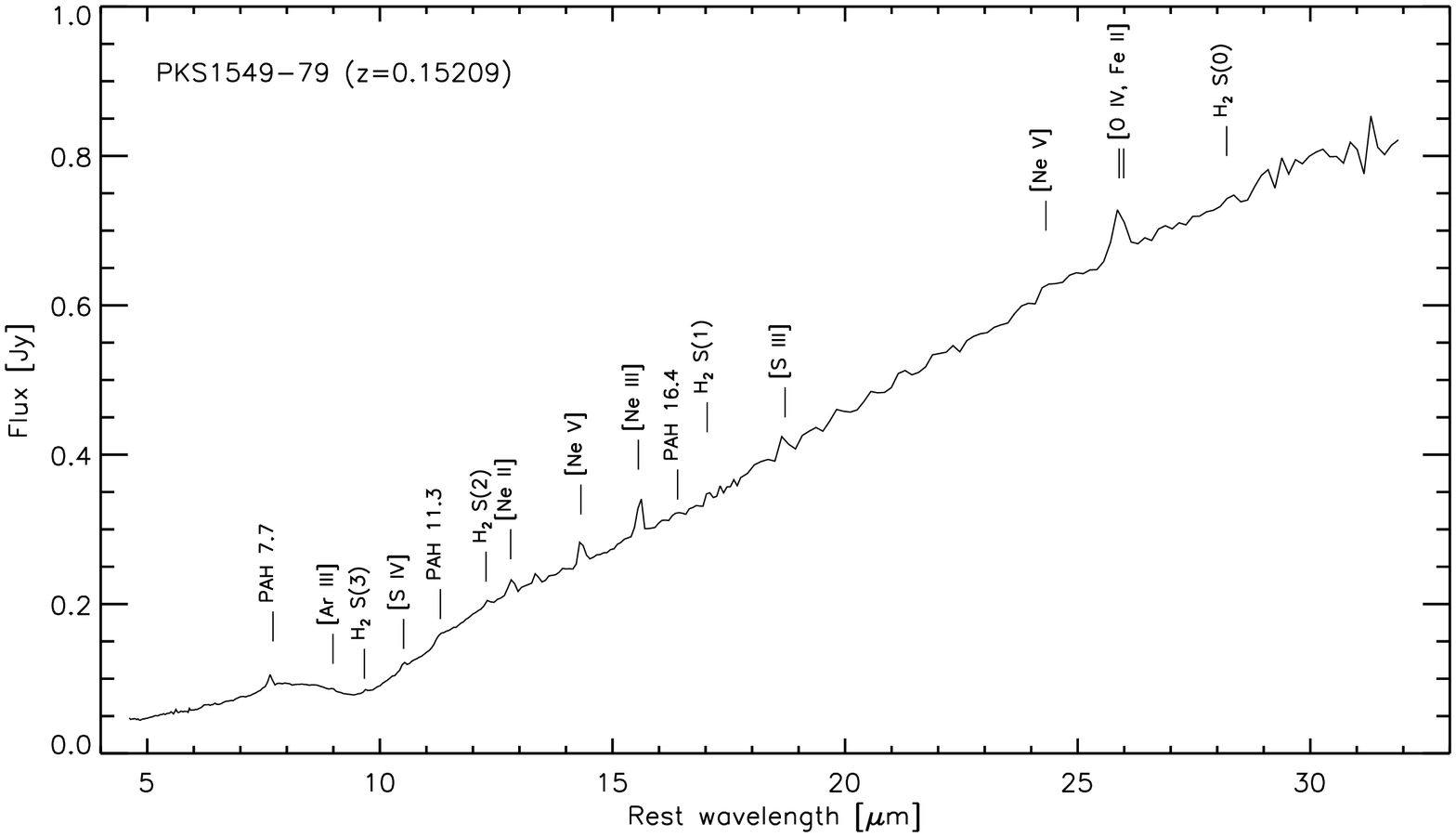}
      \caption{\textit{Spitzer} IRS low-resolution (SL+LL) spectra of the H{\sc i}-outflow radio-galaxies. The locations of PAH features and emission lines are marked for each object. The low-resolution spectra for 3C~293, OQ~208 and PKS~1549-79 were already presented
by$\ $ \citet{Ogle2010} and$\ $ \citet{Willett2010}.}
       \label{fig:IRS_spec_low-res}
   \end{figure*}

The full  high-resolution (SH+LH, i.e. observed wavelengths $9.9-37.2\,\mu$m) spectra are shown in Figure~\ref{fig:IRS_spec}, and the low-resolution spectra are presented on Figure~\ref{fig:IRS_spec_low-res}.  The location of the detected PAH complexes and emission lines are marked on each spectrum. Overall, the spectra are of very good quality, with noise levels expected for such exposure times. Some minor spurious features still remain, in particular in the LH spectra (noisier than the SH), which are due to bad pixels or instrumental artefacts that were not cleaned during the data reduction. In appendix, we provide zooms on the SH wavelengths (Figure~\ref{fig:SH_spectra}), and on all individual detected emission lines (Figure~\ref{fig:lines_1} and \ref{fig:lines_2}).

\subsection{Molecular hydrogen}
\label{subsec:H2}

Powerful emission lines from H$_2$ are detected in all of the 8 high-resolution spectra. The H$_2$ 0-0~S(1) $17\,\mu$m and 0-0~S(3) $9.7\,\mu$m lines are detected in all radio-galaxies. The pure rotational H$_2$ line fluxes (or their $2\,\sigma$ upper limits) are listed in Table~\ref{table_H2line fluxes}. 
In some cases, like 3C~236, 3C~293, and 4C~12.50, the H$_2$ S(1) line dominates the SH spectrum (Figure~\ref{fig:SH_spectra}). These H$_2$ strengths are comparable to those observed in the \citet{Ogle2010} sample of H$_2$-bright radio-galaxies, in some galaxy collisions, like Stephan's Quintet \citep{Cluver2010} and the Taffy galaxies (Peterson et al. submitted).

\begin{deluxetable*}{lcccccccc}
\tablecaption{H$_2$ line fluxes from H{\sc i}-outflow radio galaxies}
\tablecolumns{9}
\tablewidth{0pt}
\tablehead{\colhead{Source} & \colhead{H$_2$ S(0)} &  \colhead{H$_2$ S(1)} & \colhead{H$_2$ S(2)} & \colhead{H$_2$ S(3)} & \colhead{H$_2$ S(4)}  & \colhead{H$_2$ S(5)} & \colhead{H$_2$ S(6)} & \colhead{H$_2$ S(7)}  \\
 & \colhead{$28.22\,\mu$m} &  \colhead{$17.03\,\mu$m} &  \colhead{$12.28\,\mu$m} &  \colhead{$9.66\,\mu$m}  &  \colhead{$8.03\,\mu$m} &  \colhead{$6.91\,\mu$m} &  \colhead{$6.11\,\mu$m} &  \colhead{$5.51\,\mu$m}}
\startdata
3C~236   &   0.20 (0.03)  &  0.95 (0.07)  &  0.32 (0.05) & 0.64 (0.05) & $<1.04$ & $<1.10$ & $<0.23$ & $< 0.31$  \\
3C~293   &   0.79 (0.06)  &  3.46 (0.29) & 1.61 (0.10)   & 2.50 (0.16) & 1.35 (0.13) &  2.84 (0.14) & 0.65 (0.15) &  1.35 (0.19)  \\
3C~305   &   $<0.31$      & 3.82 (0.42) &  0.88 (0.09)  & 1.89 (0.12) & 1.03 (0.39) & 2.93 (0.69) & 1.83 (0.47) & $<1.40$ \\
3C~459   &   $<0.25$      & 0.55 (0.03) &  $<0.36$      & 0.37 (0.02) & $<0.29$ & $<1.04$ & $<0.15$ & $<0.49$ \\
4C~12.50   &  2.31 (0.13) & 3.44 (0.39) & 0.96 (0.09)   & 1.44 (0.09) & $<0.72$ & $<1.41$ & $<0.93$ & $<2.63$  \\
IC~5063   &   4.63 (0.24) & 10.65 (0.44) & $<1.8$       &  3.86 (0.49) & $<3.12$ & $<5.10$ & $<3.38$ & $<5.20$         \\
OQ~208  &    $<0.38$     &  1.11 (0.09) &  $<0.35$     &  0.49 (0.06) &  $<0.75$ & $<4.13$ & $<2.14$ & $<2.76$   \\
PKS~1549-79   &   $<0.46$  &  1.45 (0.12) & 0.63 (0.08)  & 0.95 (0.09)   & $<1.29$ & $<2.23$ & $<1.32$ & $<3.36$
\enddata
\tablecomments{H$_2$ mid-IR line fluxes (and $1\sigma$ error in parenthesis) in units of $10^{-17}\,$W~m$^{-2}$ measured with \textit{Spitzer} IRS. In case of non-detection, the $2\,\sigma$ upper limit is indicated. The H$_2$ S(0) to S(3) lines are measured in the high-resolution modules (except for IC~5063, where the S(3) line was measured in the SL module because it falls out of the SH wavelength coverage), and the   S(4) to S(7) lines in the low-resolution modules. Note that the S(5) and S(6) lines are severely blended with the PAH 6.2$\,\mu$m  and [Ar{\sc ii}]$\lambda \, 6.98\,\mu$m features.}
  \label{table_H2line fluxes}
\end{deluxetable*}

To compute the masses of the warm ($\gtrsim 100\,$K) H$_2$, we follow two different approaches. The first one is similar to that described by  \citet{Ogle2010} and consists of assuming a thermal distribution of the H$_2$ levels. We constructed excitation diagrams, shown on Figure~\ref{fig:H2exc_Tfit}, by plotting the  logarithm of the column densities of the upper H$_2$ levels divided by their statistical weights, $\ln (N_{vJ}/g_J)$, against their excitation energies, $E_{vJ}/k_{\rm B}$, expressed in K. For a uniform excitation temperature, the values $\ln (N_{vJ}/g_J)$ should fall on a straight line plotted versus $E_{vJ}/k_{\rm B}$, with a slope proportional to $T_{\rm exc}^{-1}$. In a situation of local thermal equilibrium (LTE), the excitation temperature $T_{\rm exc}$ equals that of the gas. We fitted the data with two or three excitation temperature components, although the gas is at a continuous range of temperatures. We constrained the temperature range to be $100 < T < 1500$~K. At each fitting iteration, the statistical weights of the ortho transitions were adjusted to match the LTE value.  

The results are given in Table~\ref{table_H2params_3T}, where we list the fitted H$_2$ excitation temperatures, ortho-to-para ratios, column densities, masses, and total luminosities for each temperature components. The bulk of the warm H$_2$ is constrained by the lowest temperature component, i.e. by the S(0) and S(1) lines. Thus, if the S(0) line is not detected, we include its upper limit in the fit as a $2\,\sigma$ detection (with $1\,\sigma$ error bar) and quote the model results as upper limits. In all galaxies where the S(0) line is detected, we measure warm H$_2$ masses, from $\approx 3\times 10^8$~M$_{\odot}$ in 3C~236, up to $\approx 3\times 10^{10}$~M$_{\odot}$ in 4C~12.50. Similar extremely large warm H$_2$ masses were detected in 3C~433 and 3C~436 \citep{Ogle2010}. The H$_2$ column densities range from $<4 \times 10^{20}$ to $\approx 2 \times 10^{22}$~cm$^{-2}$. These column densities depend on the assumed size of the emission  region.

\begin{deluxetable*}{lcccccc}
\tablecaption{H$_2$ excitation diagram fitting for a thermal distribution: results}
\tablecolumns{7}
\tablewidth{0.8\textwidth}
\tablehead{\colhead{Source} & \colhead{Size\tablenotemark{a}} & \colhead{T(K)\tablenotemark{b} }& \colhead{O/P\tablenotemark{c}}&\colhead{$N($H$_2)$\tablenotemark{d}} & \colhead{$M($H$_2)$\tablenotemark{e}}& \colhead{$L($H$_2)$\tablenotemark{f}} \\
&  [''] & [K] &  & [cm$^{-2}$] &  [M$_{\odot}$] & [L$_{\odot}$] 
}
\startdata
\multirow{3}{*}{3C~236} &   \multirow{3}{*}{3.0} &   100 ( 0) & 1.587 & 3.3E+21 & 1.6E+09 (0.5) & 1.5E+07  \\
                                          &                                       &   241 (27) & 2.947 & 1.9E+20 & 9.5E+07 (4.0) & 8.2E+07  \\
 										&    										&  1045 (80) & 3.000 & 1.5E+18 & 7.3E+05 (1.8) & 1.8E+08  \\
\hline
\multirow{3}{*}{3C~293} &   \multirow{3}{*}{4.7} &   100 ( 0) & 1.587 & 6.1E+21 & 1.7E+09 (0.2) & 1.2E+07  \\
 											&    									&   285 (14) & 2.983 & 1.9E+20 & 5.3E+07 (1.1) & 7.4E+07  \\
 											&    									&  1048 (47) & 3.000 & 1.9E+18 & 5.2E+05 (0.8) & 1.1E+08  \\
\hline
\multirow{2}{*}{3C~305} &   \multirow{2}{*}{4.7} &   226 (13) & 2.922 & $<$4.2E+20 & $<$1.0E+08 & 5.3E+07  \\
										  & 									     &  1253 (93) & 3.000 & 1.3E+18 & 3.1E+05 (0.5) & 1.2E+08  \\
\hline
\multirow{3}{*}{3C~459} &   \multirow{3}{*}{1.4} &   100 ( 0) & 1.587 & $<$2.1E+22 & $<$8.3E+09 & 1.1E+08  \\
 										&    										&   284 (32) & 2.983 & 1.5E+20 & 6.1E+07 (3.0) & 1.6E+08  \\
 											&   									 &  1500 ( 0) & 3.000 & 1.6E+18 & 6.3E+05 (1.1) & 7.9E+08  \\
\hline
\multirow{3}{*}{4C~12.50} &   \multirow{3}{*}{4.7} &   100 ( 0) & 1.587 & 2.2E+22 & 3.8E+10 (0.2) & 3.7E+08  \\
 											&    										&   275 (16) & 2.978 & 1.6E+20 & 2.8E+08 (0.9) & 4.4E+08  \\
 											&    										&  1500 ( 0) & 3.000 & 5.5E+17 & 9.2E+05 (1.5) & 8.3E+08  \\
\hline
\multirow{3}{*}{IC~5063} &   \multirow{3}{*}{4.7} &   100 ( 0) & 1.587 & 3.9E+22 & 7.3E+08 (0.9) & 4.7E+06  \\
 											&    										&   210 (34) & 2.883 & 1.4E+21 & 2.6E+07 (1.9) & 8.8E+06  \\
 											&    										&  1251 (103) & 3.000 & 3.1E+18 & 5.9E+04 (2.5) & 1.9E+07  \\
\hline
\multirow{3}{*}{OQ~208} &   \multirow{3}{*}{2.0} &   100 ( 0) & 1.587 & $<$9.2E+21 & $<$1.2E+09  & 1.0E+07  \\
 											&    										&   186 (29) & 2.784 & 2.1E+21 & 2.8E+08 (2.5) & 7.1E+07  \\
 											&    										&  1500 ( 0) & 3.000 & 2.5E+18 & 3.4E+05 (0.4) & 2.6E+08  \\
\hline
\multirow{3}{*}{PKS~1549-79} &   \multirow{3}{*}{1.9} &   100 ( 0) & 1.587 & $<$2.0E+22 & $<$7.7E+09  & 8.4E+07  \\
 												&    									&   270 (21) & 2.975 & 5.1E+20 & 2.0E+08 (0.7) & 3.3E+08  \\
 												&    									&  1500 ( 0) & 3.000 & 3.6E+18 & 1.4E+06 (0.2) & 1.4E+09 
  \enddata
\tablenotetext{a}{Assumed size (diameter) of the H$_2$ emitting source for the calculation of the column densities.}
\tablenotetext{b}{Fitted excitation temperature assuming LTE (see text for details). }
\tablenotetext{c}{H$_2$ ortho-to-para ratio (fitted self-consistently to fulfill the LTE approximation).}
\tablenotetext{d}{H$_2$ model column densities.}
\tablenotetext{e}{Warm H$_2$ masses (and $1\,\sigma$ uncertainties) derived from the fit. In case of non-detection of the S(0) line, a $2\,\sigma$ upper limit is quoted.}
\tablenotetext{f}{Total H$_2$ luminosity, summed over all rotational transitions (including those not observed with Spitzer, which includes a $\approx 40$\% correction at 200~K).}
  \label{table_H2params_3T}
\end{deluxetable*} 

The second approach to fit the H$_2$ line fluxes and derive the H$_2$ physical parameters is similar to that described in \citet{Guillard2009} and \citet{Nesvadba2010}. It assumes that the H$_2$ emission is powered by the dissipation of mechanical energy in the molecular gas (see sect.~\ref{subsec:exc_mech} for a discussion of the H$_2$ excitation mechanisms). We model this dissipation with magnetic shocks, using the MHD code described in \citet{Flower2010}. The gas is heated to a range of post-shock temperatures that depend on the shock velocity, the pre-shock density, and the intensity of the magnetic field (which is perpendicular to the shock propagation). We use a grid of shock models (varying shock speeds) similar to that described in \citet{Guillard2009}, at pre-shock densities $n_{\rm H} = 10^3$~cm$^{-3}$ and $10^4$~cm$^{-3}$. The initial ortho-to-para ratio is set to 3, and the intensity of the pre-shock magnetic field is $30\,\mu$G. The H$_2$ line fluxes are computed when the post-shock gas has cooled down to a temperature of 100~K. At a given pre-shock density, the shock velocity is the only parameter we allow to vary. 

A combination of two shock velocities is required to match the observed H$_2$ line fluxes.  We show the best-fitting shock combination in Figure~\ref{fig:H2exc_shocks_nH1e3} and \ref{fig:H2exc_shocks_nH1e4} for the two pre-shock densities.  These fits are not unique, but they provide an estimate of the range of shock velocities and pre-shock densities needed to reproduce the H$_2$ excitation. This phase-space is well constrained when six or more H$_2$ lines are detected. The lowest density and shock velocity is required to fit the low-excitation lines S(0) and S(1), which probe the bulk of the warm H$_2$ mass, whereas the high density and shock velocity is needed to fit the high-excitation lines, above S(4). In reality, the H$_2$ emission arises from a distribution of densities and shock velocities. With the present data we cannot exclude the presence of very high density gas ($n_{\rm H} > 10^5$~cm$^{-3}$). 

As discussed in \citet{Guillard2009}, the H$_2$ masses are derived by multiplying the gas cooling time (down to 100~K) by the gas mass flow (the mass of gas swept by the shock per unit time) required to match the H$_2$ line fluxes. The shock model parameters, gas cooling times, mass flows, and warm H$_2$ masses are quoted in Tables~\ref{table_H2params_shocks_nH1e3} and \ref{table_H2params_shocks_nH1e4}. 
The warm H$_2$ masses derived from the shock modeling are larger than from the LTE model. This is mainly because of the different values of the H$_2$ ortho-to-para ratio, and because at $n_{\rm H} = 10^3$~cm$^{-3}$, the S(0) and S(1) lines are not fully thermalized.

\begin{deluxetable}{lcccc}
\tablecaption{H$_2$ fluxes fitted with shock models: results for $n_{\rm H}=10^3$~cm$^{-3}$}
\tablecolumns{5}
\tablewidth{\columnwidth}
\tablehead{\colhead{Source} & \colhead{V$_{\rm s}$\tablenotemark{a}} & \colhead{Mass flow\tablenotemark{b} }& \colhead{t$_{\rm cool}$(50 K)\tablenotemark{c}} &  \colhead{$M($H$_2)$\tablenotemark{d}} \\
&  [km s$^{-1}$] & [M$_{\odot}$ yr$^{-1}$] & [yrs] &   [M$_{\odot}$] 
}
\startdata
 \multirow{2}{*}{3C~236} &     3 & 8.0E+04 & 1.5E+05 & 1.2E+10 \\
											 &    23 & 2.4E+03 & 1.8E+04 & 4.4E+07 \\
\hline
 \multirow{2}{*}{3C~293} &     5 & 4.2E+04 & 8.1E+04 & 3.4E+09 \\
  											 &    40 & 8.3E+02 & 7.8E+03 & 6.4E+06 \\
\hline
 \multirow{2}{*}{3C~305} &    11 & 7.3E+03 & 4.4E+04 & 3.2E+08 \\
   											 &    40 & 3.7E+02 & 7.8E+03 & 2.9E+06 \\
\hline
 \multirow{2}{*}{3C~459} &     4 & 2.6E+05 & 1.1E+05 & 2.9E+10 \\
   											 &    40 & 1.8E+03 & 7.8E+03 & 1.4E+07 \\
\hline
 \multirow{2}{*}{4C~12.50} &     4 & 1.1E+06 & 1.1E+05 & 1.2E+11 \\
											 &    23 & 8.6E+03 & 1.8E+04 & 1.5E+08 \\
\hline
 \multirow{2}{*}{IC~5063} &     4 & 1.9E+04 & 1.1E+05 & 2.1E+09 \\
											 &    40 & 7.7E+01 & 7.8E+03 & 6.0E+05 \\
\hline
 \multirow{2}{*}{OQ~208} &     5 & 5.4E+04 & 8.1E+04 & 4.3E+09 \\
											 &    40 & 3.6E+02 & 7.8E+03 & 2.8E+06 \\
\hline
 \multirow{2}{*}{PKS~1549-79} &    15 & 2.3E+04 & 2.9E+04 & 6.7E+08 \\
											   &    40 & 1.9E+03 & 7.8E+03 & 1.5E+07                                          
  \enddata
\tablenotetext{a}{Shock velocity.}
\tablenotetext{b}{Mass of gas that is traversed by the shock per unit time required to match the observed H$_2$ line fluxes.}
\tablenotetext{c}{Gas cooling time, from the peak post-shock temperature down to 100~K.}
\tablenotetext{d}{Warm H$_2$ masses.}
  \label{table_H2params_shocks_nH1e3}
\end{deluxetable} 

\begin{deluxetable}{lcccc}
\tablecaption{H$_2$ fluxes fitted with shock models: results for $n_{\rm H}=10^4$~cm$^{-3}$}
\tablecolumns{5}
\tablewidth{\columnwidth}
\tablehead{\colhead{Source} & \colhead{V$_{\rm s}$\tablenotemark{a}} & \colhead{Mass flow\tablenotemark{b} }& \colhead{t$_{\rm cool}$(50 K)\tablenotemark{c}} &  \colhead{$M($H$_2)$\tablenotemark{d}} \\
&  [km s$^{-1}$] & [M$_{\odot}$ yr$^{-1}$] & [yrs] &   [M$_{\odot}$] 
}
\startdata
 \multirow{2}{*}{3C~236} &     4 & 2.2E+05 & 2.5E+04 & 5.6E+09 \\
											 &    29 & 2.8E+03 & 1.8E+03 & 5.0E+06 \\
\hline
 \multirow{2}{*}{3C~293} &     8 & 4.0E+04 & 1.1E+04 & 4.4E+08 \\
										 &    29 & 1.5E+03 & 1.8E+03 & 2.7E+06 \\
\hline
 \multirow{2}{*}{3C~305} &     5 & 5.5E+04 & 1.8E+04 & 1.0E+09 \\
										 &    32 & 1.3E+03 & 1.5E+03 & 1.9E+06 \\
\hline
 \multirow{2}{*}{3C~459} &     8 & 8.3E+04 & 1.1E+04 & 9.2E+08 \\
											 &    36 & 6.2E+03 & 1.2E+03 & 7.5E+06 \\
\hline
  \multirow{2}{*}{4C~12.50} &     4 & 1.8E+06 & 2.5E+04 & 4.4E+10 \\
												 &    13 & 3.2E+04 & 5.3E+03 & 1.7E+08 \\
\hline
 \multirow{2}{*}{IC~5063} &     4 & 3.1E+04 & 2.5E+04 & 7.8E+08 \\
												&    38 & 1.5E+02 & 1.1E+03 & 1.6E+05 \\
\hline
 \multirow{2}{*}{OQ~208} &     3 & 1.6E+05 & 3.7E+04 & 5.7E+09 \\
											&    40 & 1.6E+03 & 1.0E+03 & 1.7E+06 \\
\hline
 \multirow{2}{*}{PKS~1549-79} &     8 & 1.7E+05 & 1.1E+04 & 1.8E+09 \\
												 &    40 & 1.1E+04 & 1.0E+03 & 1.2E+07 
  \enddata
\tablenotetext{a}{Shock velocity.}
\tablenotetext{b}{Mass of gas that is traversed by the shock per unit time required to match the observed H$_2$ line fluxes.}
\tablenotetext{c}{Gas cooling time, from the peak post-shock temperature down to 100~K.}
\tablenotetext{d}{Warm H$_2$ masses.}
  \label{table_H2params_shocks_nH1e4}
\end{deluxetable} 

\subsection{Thermal dust continuum and aromatic features (PAHs)}

Our sample reveal two types of spectra, those with a flat thermal dust continuum (3C~236, 3C~293, 3C~305), and those with a steep rising continuum at long wavelengths (3C~459, 4C~12.50, IC~5063, OQ~208, and PKS~1549-79). 
There is a remarkable similarity between the spectra of 3C~236, 3C~293 and 3C~305 with the 3C~326 radio-galaxy \citep{Ogle2007}, the Stephan's Quintet \citep{Cluver2010} and the Taffy galaxies (Peterson et al., submitted), where the dust continuum is very weak compared to emission lines. The narrow-band $24\,\mu$m luminosities (see sect.~\ref{subsec:spectral_analysis} and Table~\ref{table_sample}), measured on the LH spectra, span three orders of magnitude, from $5.4\times 10^9$~L$_{\odot}$ (3C~305) to ULIRG-like luminosities of $1.1\times 10^{12}$~L$_{\odot}$ (4C~12.50). For 4C~12.50 and OQ~208, our measurements are in agreement with those of \citet{Willett2010} on low-resolution IRS data.

Emission from aromatic infrared bands attributed to Polycyclic Aromatic Hydrocarbons (PAHs) excited by UV photons are detected in all galaxies. In general, in the high-resolution data, the  $11.3\,\mu$m band is the most prominent feature. The fluxes of the PAH complexes, given in Table~\ref{table_PAH},  are measured with PAHFIT on the emission line-free spectra (see sect.~\ref{subsec:spectral_analysis}), by combining the fluxes of the blended features that contribute to the main complexes (around 8, 11.3, 12.6 and 17$\,\mu$m).

\begin{deluxetable*}{lccccc}
\tablecaption{Fluxes of the main PAH complexes from H{\sc i}-outflow radio galaxies}
\tablewidth{0pt}
\tablecolumns{6}
\tablehead{\colhead{Source} & \colhead{PAH $6.2\, \mu$m} & \colhead{PAH $7.7\, \mu$m} &  \colhead{PAH $11.3\, \mu$m} & \colhead{PAH $12.6\, \mu$m} & \colhead{PAH $17\, \mu$m}}
\startdata
3C~236      & $<2.5$      & $<4.5$   &   3.7 (0.1) &  3.8 (1.2)    & 3.2 (1.3)    \\
3C~293      &  11.1 (0.6)  &  55.7 (2.5)  & 21.5 (0.6) &  10.4 (0.4)  & 25.6 (3.2)    \\
3C~305      &  15.1 (3.4) &  63.1 (8.3)  &   19.2 (0.1) &  16.1 (0.2)  &  11.2 (2.6)    \\
3C~459      &   7.2 (1.3)  & 22.5 ( 1.3)  &    9.9 (0.2) &  9.7 (0.2)  &  5.7 (0.3)         \\
4C~12.50   &   9.8 (2.3)  & 38.3 (1.6)   &   10.2 (0.2) &   18.1 (0.4) & $<9.5$    \\
IC~5063     &   $<43$    & 189 (1.9)  &  104.9 (2.6) & 230.6 (0.8) & 53.3 (6.7)                \\
OQ~208     & 40.8 (2.9)   &  51.5 (1.4)  &   55.3 (2.8) & $<61$      &  101 (8)          \\
PKS~1549-79  & 25 (3.3)  & 187 (5.3)   &    37.4 (0.3) &   31.7 (0.7) & 4.8 (0.6)
\enddata
\tablecomments{PAH fluxes (and $1\sigma$ error in parentheses) in units of $10^{-17}\,$W~m$^{-2}$ measured with \textit{Spitzer} IRS. In case of non-detection, the $2\,\sigma$ upper limit is indicated. }
  \label{table_PAH}
\end{deluxetable*}

All of the radio-galaxies have a   $11.3\,\mu$m PAH to $24\,\mu$m luminosity ratio that ranges from $L_{\rm PAH\,11.3} / L_{24\,\mu m} = 4\times 10^{-2}$ for 3C~293 to  $8 \times 10^{-4}$ for 4C~12.50, which is similar to the ratios observed in the \citet{Ogle2010} sample of 3C radio-galaxies. This ratio is significantly lower than the SINGS star-forming galaxies, where the median ratio is $\approx 0.1$. 3C459, PKS~1549-79 and 4C~12.50 are ULIRGs, but their $L_{\rm PAH\,11.3} / L_{24\,\mu m}$ ratios are at the lower end of the observed values ($\approx 10^{-3}$).  The mid-IR continuum of these radio-galaxies is therefore dominated by dust emission from the accretion disk or synchrotron emission from the jet, but not by star formation. The weakness of the PAH emission relative to the continuum emission is related to the weakness of the stellar UV radiation field \citep[see][and sect.~\ref{subsec:exc_mech} for a discussion of this claim]{Ogle2010}. Although PAH emission is expected to be diminished because of PAH destruction by hard UV or X-ray radiation close to the AGN \cite{Voit1992}, or because of gas suppression close to the nuclei, it is unlikely that their abundances are reduced significantly at galactic scales by AGN effects \citep{Ogle2010}. 

Silicate absorption at 9.7 and 18$\,\mu$m is present in the low-resolution spectra of OQ~208 and 4C~12.50 \citep{Willett2010}, as well as PKS~1549-79.  OQ~208 is the only galaxy showing silicate emission \citep{Willett2010}.

\subsection{Ionic fine-structure lines}

\begin{deluxetable*}{lcccccccccc}
\tablecaption{Fine-structure line fluxes in H{\sc i}-outflow radio-galaxies}
\tablecolumns{11}
\tablewidth{\textwidth}
\tablehead{
\colhead{Object} &  \colhead{[Ar{\sc ii}]} &  \colhead{[Ar{\sc iii}]} &  \colhead{[S{\sc iv}]} &  \colhead{[Ne{\sc ii}]}  &  \colhead{[Ne{\sc iii}]} &  \colhead{[S{\sc iii}]}  & \colhead{[O{\sc iv}]} &  \colhead{[Fe{\sc ii}]}  &  \colhead{[S{\sc iii}]} &  \colhead{[Si{\sc ii}]} \\
 & \colhead{$6.98\,\mu$m} & \colhead{$8.99\,\mu$m} & \colhead{$10.51\,\mu$m} & \colhead{$12.81\,\mu$m}  & \colhead{$15.55\,\mu$m} & \colhead{$18.71\,\mu$m}  & \colhead{$25.89\,\mu$m} & \colhead{$25.99\,\mu$m} & \colhead{$33.48\,\mu$m} & \colhead{$34.81\,\mu$m}
}
\startdata
    3C~236 & $<1.6$    & $<1.9$  &     $<0.13$    & 0.92 (0.06) &  0.44 (0.03) &    $<0.14$ &    $<0.14$ &   $<0.13$ &   $<0.09$ &  \nodata \tablenotemark{a}   \\
    3C~293 & 2.6 (0.3)  & 0.70 (0.12)  &     $<0.18$    & 4.07 (0.29) &  1.27 (0.09) &    $<0.42$  &      0.27 (0.02) &     0.57 (0.05) &  1.32 (0.12) &     5.55 (0.48)  \\
    3C~305 & 2.8 (0.4)  & $<1.5$  &     3.21 (0.24) & 4.15 (0.29) &  8.62 (0.59) &     4.11 (0.33) &      13.07 (1.15) &     0.90 (0.07) &   5.14 (0.46) &    11.24 (0.97)  \\
    3C~459 & 1.0 (0.1)  & $<0.15$  &     0.11 (0.01) & 3.11 (0.22) &  1.00 (0.07) &     0.39 (0.03) &      0.46 (0.04) &     0.58 (0.05) &    \nodata  \tablenotemark{a} &   \nodata  \tablenotemark{a} \\
  4C~12.50 & $<2.7$   & 2.4 (0.2) &      $<0.7$      & 5.45 (0.39) &  2.99 (0.21) & $<0.83$ &      2.88 (0.25) &  $<0.21$ &   \nodata    \tablenotemark{a} &   \nodata  \tablenotemark{a} \\
   IC~5063 & 4.4 (0.2)  & 9.5 (0.3)  &    37.60 (3.05) & 22.83 (1.62) & 56.85 (3.92) &    23.88 (1.93) &     69.32 (6.10) &     3.97 (0.33) &    31.89 (2.68) &  28.3 (1.1) \tablenotemark{b}  \\
    OQ~208 & 1.2 (0.2)  & $<1.8$  &     $<1.6$       & 3.33 (0.24) & 2.60 (0.18) &   $<1.12$ &      1.21 (0.11) &     0.45 (0.04) &  $<1.4$ \tablenotemark{b}  &    \nodata   \tablenotemark{a} \\
  PKS~1549-79 & 3.2 (0.3)  & 3.3 (0.2)  &   1.45 (0.12)  & 2.20 (0.17) & 5.45 (0.39) &   $<0.6$ &        3.11 (0.27) &     $<0.8$ &    \nodata  \tablenotemark{a}  &   \nodata   \tablenotemark{a}  
\enddata
\tablecomments{Ionic line fluxes (and $1\sigma$ error in parentheses) in units of $10^{-17}\,$W~m$^{-2}$ measured with \textit{Spitzer} IRS. In case of non-detections, a 2$\,\sigma$ upper limit is quoted. The [Ar{\sc ii}] and [Ar{\sc iii}] lines are measured in the low-resolution modules. All the other lines are measured in the high-resolution modules, except some of the [S{\sc iii}] and [Si{\sc ii}] lines marked with a \textit{(b)}.}
\tablenotetext{a}{Observed wavelength not visible in the IRS range.}
\tablenotetext{b}{Measured on the low-resolution LL module.}
  \label{table_ionline_int}
\end{deluxetable*}

\begin{deluxetable}{lccc}
\tablecaption{High-ionization forbidden emission lines}
\tablecolumns{4}
\tablewidth{\columnwidth}
\tablehead{
\colhead{Object} &  \colhead{[Ne{\sc vi}]$\lambda \, 7.65\,\mu$m} &    \colhead{[Ne{\sc v}]$\lambda \, 14.32\,\mu$m} &    \colhead{[Ne{\sc v}]$\lambda \, 24.32\,\mu$m}  
}
\startdata
    3C~236 & $<1.6$     &    $<0.06$ &     $<0.06$    \\
    3C~293 & $<0.53$    &    $<0.21$   &       $<0.17$   \\
    3C~305 & 0.92 (0.32)    &  1.42 (0.12) &        1.87 (0.15)  \\
    3C~459 &  $<0.29$   &  $<0.16$ &       $<0.14$   \\
  4C~12.50 & 2.66 (0.18)   &    1.68 (0.22) &     $<1.1$     \\
   IC~5063 &  3.51 (0.12)   &    22.70 (1.86) &     17.41 (1.43)   \\
    OQ~208 &  $<1.2$    &    $<0.8$  &       $<0.2$ \\
  PKS~1549-79 &  3.29 (0.21)    &     3.35 (0.27) &      2.21 (0.18) 
\enddata
\tablecomments{Line fluxes (and $1\sigma$ error in parentheses) in units of $10^{-17}\,$W~m$^{-2}$ measured with \textit{Spitzer} IRS. In case of non-detection, a 2$\,\sigma$ upper limit is quoted. The [Ne{\sc vi}]$\lambda \, 7.65\,\mu$m line is measured in the SL module. The other lines are measured in the high-resolution modules. }
  \label{table_hi_ionline_int}
\end{deluxetable}

\begin{figure*}
   \centering
    \includegraphics[width=0.49\textwidth]{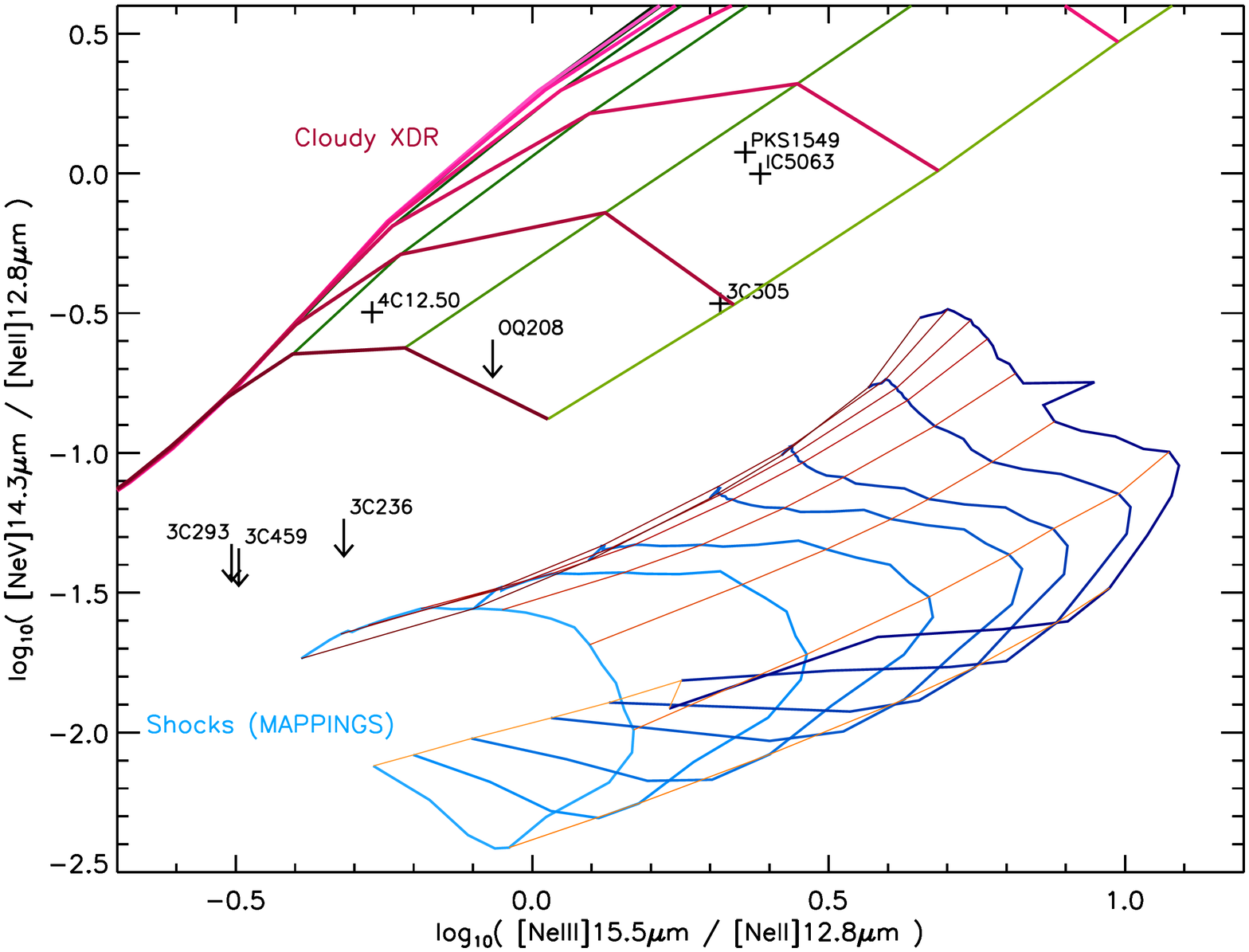}
    \includegraphics[width=0.49\textwidth]{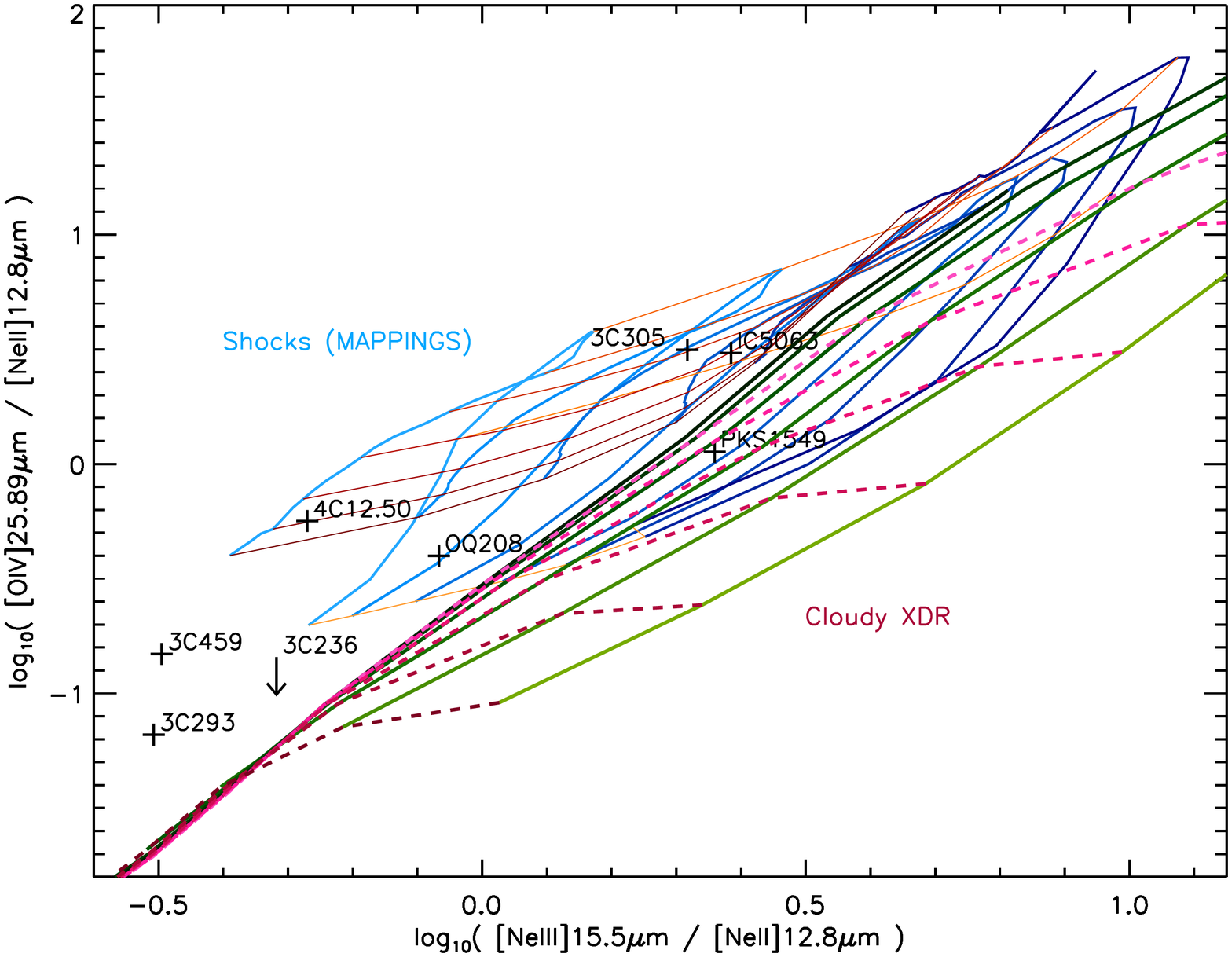}
      \caption{Mid-infrared line diagnostic diagrams for the logarithm of the model line flux ratios [Ne{\sc v}]/[Ne{\sc ii}] (left) and [O{\sc iv}]/[Ne{\sc ii}] (right) against [Ne{\sc iii}]/[Ne{\sc ii}]. The observed values for the 8 H{\sc i}-outflow radio-galaxies are compared to shock and photoionization grids of models.  The shock model grid is from the $\ $ \citet{Allen2008} MAPPINGS III shock code for combined precursor and shock models. The gas metallicity is solar and the pre-shock gas density is $n_{\rm H} = 10$~cm$^{-3}$. The orange lines indicate constant shock velocities, from 200 (light orange) to 1000 km~s$^{-1}$ (red), and the blue lines constant magnetic field strengths, from $10^{-3}$ (light blue) to 100~$\mu$G (purple). The deeper the color, the larger is the parameter. 
The grid of photoionization NLR models was computed with the CLOUDY code \citep{Ferland1998} with a standard power-law spectral index $\alpha = -1.5$ and varying ionization parameter (green lines, $U_0 = 1 - 10^{-2}$ ) and density $n_{\rm H}$ (pink lines, $n_{\rm H} = 10^3 - 10^6\,$cm$^{-3}$). On the right plot, since the shock and photoionization models partially overlap, the constant density lines are dashed for clarity.}
       \label{fig:allen_shock_models}
   \end{figure*}

The fluxes of the most commonly detected mid-IR ionized gas lines are listed in Table~\ref{table_ionline_int}.  
The [Ne{\sc ii}]$\,12.8\mu$m and [Ne{\sc iii}]$\,15.55\mu$m are detected at high signal to noise ratio  ($>6$) in all 8 radio-galaxies. The [O{\sc iv}]$\,25.89\mu$m and [Fe{\sc ii}]$\,25.99\mu$m pair is also very common, with at least one line of the doublet detected in 7/8 galaxies. The high ionization ($E_{\rm ion} = 97.12$~eV) [Ne{\sc v}]$\,14.32\mu$m line is detected in half of the sample. Although listed as non-detected by \citet{Farrah2007} and \citet{Willett2010}, we securely detect this line in 4C~12.50 at 3.2$\,\sigma$.

Three objects show detections in both [Ne{\sc v}]$\,14.32$ and $24.32\,\mu$m lines. The flux ratios of these two lines are consistent with electron densities of  $n_e \approx 3\times 10^3$~cm$^{-3}$ for IC~5063 and PKS~1549-79\footnote{\citet{Holt2006} derive $n_e \approx 430$~cm$^{-3}$ for the NLR gas using the [S{\sc ii}](6731/6717\AA) line ratio. \citet{Holt2011} derive $n_e = 3\times 10^3$~cm$^{-3}$ for 4C~12.50 using a more sophisticated technique.}, and $n_e < 10^{3}$~cm$^{-3}$ for 3C~305 \citep[see e.g. Figure~3 of ][]{Alexander1999}. These densities are below the critical densities of both lines \citep[see also][]{Dudik2009}. Using the  [S{\sc iii}]$\,18.71$ and $33.48\,\mu$m line ratios, we find $n_e \approx 2\times 10^2$~cm$^{-3}$ for 3C~305 and $n_e \approx 3 \times 10^3$~cm$^{-3}$ for IC~5063. The Neon and Sulfur line ratios give consistent estimates of electron densities, although star-forming regions may contribute to the [S{\sc iii}] line emission, whereas the [Ne{\sc v}] line is principally emitted by AGN-heated regions. 

Some more unusual lines, rarely seen in AGN, and that are not listed in Table~\ref{table_ionline_int}, are detected in the high-quality spectrum of 3C~305. The low-ionization ($E_{\rm ion} = 7.9$~eV) [Fe{\sc ii}] 17.94, 24.52, and 25.99$\,\mu$m lines are present, as well as the [Fe{\sc iii}]$\,22.93\mu$m ($E_{\rm ion} = 16.19$~eV). The [Ar{\sc v}]$\,13.10\mu$m line is also weakly detected (see the SH spectrum on Figure~\ref{fig:SH_spectra}).

In such objects, the excitation of the fine-structure lines can arise from photo-ionization by the AGN \citep[e.g. ][]{Veilleux1987, Meijerink2007} or by  shocks \citep[e.g. ][]{Dopita1996}. In practice, distinguishing between the two excitation mechanisms is difficult, since the line flux ratios can be differentially affected by the structure (homogeneous or clumpy) of the emitting gas, and since 
the model predictions for fine-structure line flux ratios are degenerate for both excitation mechanisms \citep[see also][]{Groves2006}.

In Sect.~\ref{subsec:exc_mech}, we will show that the warm H$_2$ emission is likely powered by shocks. Is it also the case for the observed ionized line emission? In Figure~\ref{fig:allen_shock_models}, we compare the observed values of three fine-structure line flux ratios to the predictions of  shock and photo-ionization models. The grid of fast ($\gtrsim 200$~km~s$^{-1}$) radiative shocks is from \citet{Allen2008}. The pre-shock gas density is $n_{\rm H} = 10$~cm$^{-3}$. Note that augmenting the density by a factor 10 reduces these line flux ratios by a factor $2-3$ on average, thus shifting the models to the bottom left of Figure~\ref{fig:allen_shock_models} by $0.3-0.5$~dex. The photo-ionization models are computed with the CLOUDY code \citep{Ferland1998}, with a setup updated from the narrow line region (NLR) models presented in \citet{Ferland1983}. We used a dusty gas slab model
 at constant pressure, a power-law ionizing continuum with $f_{\nu} \propto \nu ^{-1.5}$ and no cut-off or extinction. We varied the hydrogen density of the gas from $10^3$ to $10^6\,$cm$^{-3}$ and the ionization parameter\footnote{$U_0 = S_{\star} / (n_e \, c)$ is a direct measure of the ionization state of the gas, where $S_{\star}$ is the flux of ionizing photons, $n_e$ the electron density at the inner face of the slab of gas, and $c$ the speed of light.} from $U_0 = 10^{-2}$ to $1$. Note that these models are simplistic: they do not include geometrical and inhomogeneity effects, as well as metallicity differences (assumed to be solar here). A detailed model of the mid-IR emission of these sources is beyond the scope of this paper.

The [Ne{\sc iii}]$\,15.55\mu$m / [Ne{\sc ii}]$\,12.8\mu$m line flux ratios are in the range of $0.3-2.4$, much higher than those observed in pure starburst galaxies \citep[0.05-0.2, e.g. ][]{Bernard-Salas2009}. These ratios are consistent with shock models for $v < 250$~km~s$^{-1}$ and $n_{\rm H} < 10^2$~cm$^{-3}$, the higher values of [Ne{\sc iii}] / [Ne{\sc ii}] $\approx 2$ requiring very low gas densities ($n_{\rm H} < 1$~cm$^{-3}$).
The range of observed [Ne{\sc iii}]/[Ne{\sc ii}] ratios is also consistent with photo-ionization models. This ratio is weakly dependent on the ionization state of the gas, and sensitive to the gas density. The observed [Ne{\sc iii}]/[Ne{\sc ii}] ratios require $n_{\rm H} > 10^3\,$cm$^{-3}$.

[Ne{\sc v}]$\,14.32\,\mu$m being a high ionization potential line ($E_{\rm ion} = 97.1$~eV), the [Ne{\sc v}]$\,14.32/$[Ne{\sc ii}]$\,12.8$ line ratio is very sensitive to the ionization conditions, and therefore a strong indicator of the AGN contribution \citep[e.g.][]{Sturm2002}.
The shock radiative precursor, which ionizes the gas ahead of the shock front by UV and soft X-ray photons, adds an emission component of a highly ionized gas. However, for the sources containing a powerful AGN, the high values of the observed  [Ne{\sc v}]/[Ne{\sc ii}] ratios (left panel in Figure~\ref{fig:allen_shock_models})  rule out shocks as the dominant excitation mechanism to power the [Ne{\sc v}] luminosity. The [Ne{\sc v}] line primarily arises from photo-ionization of the gas by the AGN. According to this grid of CLOUDY models, the observed [Ne{\sc v}]$\,14.32/$[Ne{\sc ii}]$\,12.8$ ratios are consistent with $n_{\rm H} > 10^3\,$cm$^{-3}$ and $U_0 < 0.1$.

The right panel in Figure~\ref{fig:allen_shock_models} uses the lower ionization ($E_{\rm ion} = 54.93$~eV) [O{\sc iv}]$\,25.89\mu$m line. In this case, the shock and photo-ionization model grids cover similar regions of the diagram, making it difficult to distinguish between the two excitation mechanisms. The observed mid-IR line ratios are both consistent with high-velocity shocks ($v > 200$~km~s$^{-1}$) and photo-ionization, with a preference for shocks at low values of [Ne{\sc iii}]/[Ne{\sc ii}]. Some sources (3C~459, 3C~293 and 3C~236) require higher pre-shock densities ($n_{\rm H} = 10^2 - 10^3$~cm$^{-3}$) to be consistent with shock models. 
 
\section{Kinematics of the multiphase gas}
\label{sec:kinematics}

In this section we describe and compare the kinematics of the mid-IR H$_2$ and ionized gas lines (from the IRS data and optical data from the literature) to the kinematics of the outflowing H{\sc i} gas. 

\subsection{Analysis of the line profiles}
\label{subsec:line_profile_analysis}

We fitted all the mid-IR lines with Gaussians on the dust-free high-resolution spectra (sect.~\ref{subsec:spectral_analysis}). This allows us to carefully extract line profiles, especially when those are located close to underlying PAH structure, as it is the case for the [Ne{\sc ii}]$\,12.8\mu$m line for instance. We pay a special attention to the fitting of the [Ne{\sc ii}] by  checking that the PAH 12.6$\,\mu$m has been correctly removed by the PAHFIT modelling, since the PAH features at 12.62 and 12.69$\,\mu$m \citep{Hony2001} may contribute to an apparent blue-shifted wing of the line. We validated our subtraction method by simulating the presence of an artificial Lorentzian complex in the $12.6-12.9\,\mu$m range of intensity compatible with the observed PAH $6.2$ and $7.7\,\mu$m features. We find that the strength of the simulated 12.6$\,\mu$m feature measured with PAHFIT is at least a factor of 5 smaller than the strength of the [Ne{\sc ii}] blue wing. In addition, the strength of the 12.7$\,\mu$m PAH complex correlates with the 6.2$\,\mu$m feature, which is another indication that our measurement of the 12.7$\,\mu$m PAH complex is correct.

Figures~\ref{fig:lines_1} and \ref{fig:lines_2} show a detailed view of the individual lines. The blended lines, like [O{\sc iv}]$\lambda \, 25.89$ and [Fe{\sc ii}]$\lambda \, 25.99$, and the lines exhibiting wings (see sect.~\ref{subsec:HI_kinematics}) were fitted with a sum of two Gaussian components. 

We assume a Gaussian decomposition of the instrumental profile (shown on each of the lines in Figures~\ref{fig:lines_1} and \ref{fig:lines_2}), and we derive the intrinsic linewidth (FWHM $w_i$ in km~s$^{-1}$) from $w_i = (w_m ^2 - w_{\rm IRS} ^2)^{1/2}$, where $w_m$ is the FWHM in km~s$^{-1}$ directly measured on the observed spectrum, and $w_{\rm IRS}  = c / \mathcal{R}$ is the velocity resolution of the IRS. $c$  is the speed of light, and $\mathcal{R} = 600 \pm 72$ is the spectral resolution of the high-resolution module, corresponding to $w_{\rm IRS}  = 500 \pm 60$~km~s$^{-1}$. Measurements of calibration targets indicate that  the resolution of the high-resolution module of the IRS is constant over its wavelength coverage  \citep{Dasyra2008}. 
The error on $w_m$ is computed as $\epsilon _m = (\epsilon _f ^2 + \epsilon _{\rm IRS} ^2)^{1/2}$, where $\epsilon _{\rm IRS} = 60$~km~s$^{-1}$ is the error on the instrumental resolution and $\epsilon _f$ the error of the measurement, estimated from two Gaussian fits of the line profile, one with the upper flux values (adding flux uncertainties), one with the lower values (subtracting uncertainties).
A line is considered resolved when $w_m - \epsilon _m > w_{\rm IRS} + \epsilon _{\rm IRS} = 560$~km~s$^{-1}$.
Unresolved lines are assigned a conservative upper limit of 500~km~s$^{-1}$ on their intrinsic linewidths.

\subsection{Warm H$_2$ gas}

 \begin{deluxetable}{lcccc}
\tablecaption{Intrinsic H$_2$ linewidths in H{\sc i}-outflow radio-galaxies}
\tablecolumns{5}
\tablewidth{\columnwidth}
\tablehead{
\colhead{Object} &  \colhead{H$_2$ S(0)} &  \colhead{H$_2$ S(1)}  &  \colhead{H$_2$S(2)} &  \colhead{H$_2$ S(3)} \\
& \colhead{$28.22\,\mu$m} & \colhead{$17.03\,\mu$m} & \colhead{$12.28\,\mu$m} & \colhead{$9.66\,\mu$m} 
}
\startdata  
    3C~236 & $<500$ &   548 (59) & $<500$ & $<500$  \\
    3C~293 &   743 (79) &   485 (54) &   524 (63) &   586 (68)  \\
    3C~305 & \nodata &   482 (54) &   525 (63) &   566 (66)  \\
    3C~459 & \nodata &   432 (50)\tablenotemark{a}  & \nodata &   751 (81)  \\
  4C~12.50 &   906 (91) & $<500$\tablenotemark{a} &   529 (63) & $<500$  \\
   IC~5063 &   709 (76) &   721 (71) & $<500$ & \nodata  \\
    OQ~208 & \nodata & $<500$ & \nodata & \nodata  \\
  PKS~1549-79 & \nodata &   585 (62)\tablenotemark{a} & $<500$ & $<500$  
\enddata
\tablecomments{H$_2$ FWHM in units of km~s$^{-1}$ measured by single Gaussian fitting.  A Gaussian decomposition from the instrument profile is assumed to derive the intrinsic linewidths ($w_i$, see sect.~\ref{subsec:line_profile_analysis} for details).}
\tablenotetext{a}{3C~459, 4C~12.50 and PKS~1549-79 possibly exhibit a blue wing in the H$_2$ S(1) line, which is not included in the quoted linewidth here because this detection is marginal.}
  \label{table_H2line_fwhm_i}
\end{deluxetable} 

The intrinsic linewidths ($w_i$) of the H$_2$ rotational lines are listed in Table~\ref{table_H2line_fwhm_i}. 
According to the criterion defined above, more than 60\% of the detected H$_2$ lines are resolved by the \textit{Spitzer} IRS, with FWHM up to $\approx 900$~km~s$^{-1}$.  All the sources, except OQ~208, have at least one broad H$_2$ line with intrinsic FWHM$> 430$~km~s$^{-1}$. 3C~293, 3C~305, 3C~459, 4C~12.50, and IC~5063 show two or more resolved H$_2$ lines. These results show that the warm H$_2$ gas is very turbulent in most of the sources. As we will show in sect.~\ref{subsec:nrj_transfer}, a fraction of the warm H$_2$ emission is at velocities larger than the escape velocity.

Three of the sources, 3C~459,  4C~12.50 and PKS~1549-79 exhibit asymmetric H$_2$ line profiles, with blue-shifted wings (Figures~\ref{fig:NeII_HI}, \ref{fig:lines_1} and \ref{fig:lines_2}). However, the detection of these wings is tentative, at a $2.5-2.8\,\sigma$ significance for the S(1) line, and even weaker for the S(2) and S(3) lines. Therefore we do not attempt to fit these wings, and we do not include them in the FWHM quoted in Table~\ref{table_H2line_fwhm_i}. Deeper observations at higher spectral resolution are needed to confirm the blue-shifted H$_2$ emission and measure what fraction of the the warm H$_2$ gas is participating to the outflow, and to derive the outflow parameters (warm H$_2$ mass in the outflowing component, mass outflow rate, etc.).


\subsection{Ionized gas}
\label{subsec:kinematics_ionized_gas}

\begin{deluxetable*}{lcccccccccc}
\tablecaption{Intrinsic linewidths of ionic fine-structure lines in H{\sc i}-outflow radio-galaxies}
\tablecolumns{11}
\tablehead{
\colhead{Object} &  \colhead{[S{\sc iv}]} &  \colhead{[Ne{\sc ii}]} &  \colhead{[Ne{\sc v}]} &  \colhead{[Ne{\sc iii}]} &  \colhead{[S{\sc iii}]} &  \colhead{[Ne{\sc v}]} & \colhead{[O{\sc iv}]} &  \colhead{[Fe{\sc ii}]}  &  \colhead{[S{\sc iii}]} &  \colhead{[Si{\sc ii}]} \\
 & \colhead{$10.51\,\mu$m} & \colhead{$12.81\,\mu$m} & \colhead{$14.32\,\mu$m} & \colhead{$15.55\,\mu$m} & \colhead{$18.71\,\mu$m} & \colhead{$24.32\,\mu$m} & \colhead{$25.89\,\mu$m} & \colhead{$25.99\,\mu$m} & \colhead{$33.48\,\mu$m} & \colhead{$34.81\,\mu$m}
}
\startdata  
     3C~236 & \nodata &  1239 (109) & \nodata &   875 (86) & \nodata & \nodata & \nodata & \nodata & \nodata & \nodata  \\
    3C~293 & \nodata &   861 (84) & \nodata &   877 (86) & \nodata & \nodata &   723 (83) &   595 (70) & \nodata &   779 (85)  \\
    3C~305 &   709 (79) &   864 (84) &   611 (72) &   817 (82) &   543 (64) & $<500$ &   488 (62) & $<500$ &   565 (67) &   746 (82)  \\
    3C~459 &   573 (68) &   775 (77) & \nodata & $<500$ &   794 (125) & \nodata & $<500$ &   789 (86) & \nodata & \nodata  \\
  4C~12.50 & \nodata &  1033 (95) &   852 (92) &   838 (83) & \nodata & \nodata &   767 (87) & $<500$ & \nodata & \nodata  \\
   IC~5063 & $<500$ &   408 (49) & $<500$ & $<500$ &   632 (71) & $<500$ & $<500$ &   661 (75) & $<500$ & \nodata  \\
    OQ~208 & \nodata & $<500$ & \nodata &   895 (87) & \nodata & \nodata &   781 (89) & $<500$ & \nodata & \nodata  \\
  PKS~1549-79 &  1355 (132) &   848 (83) &   967 (89) &  1233 (105) & \nodata &   834 (89) &   857 (95) &   \nodata & \nodata & \nodata 
\enddata
\tablecomments{FWHM in units of km~s$^{-1}$ measured by single Gaussian fitting.  A Gaussian decomposition from the instrument profile is assumed to derive the intrinsic linewidths   ($w_i$, see sect.~\ref{subsec:line_profile_analysis} for details).}
  \label{table_ionline_fwhm_i}
\end{deluxetable*}

Table~\ref{table_ionline_fwhm_i} lists the intrinsic linewidths of the fine-structure lines. Remarkably, the [Ne{\sc ii}]$12.8\mu$m and [Ne{\sc iii}]$15.5\mu$m lines, the brightest among detected fine-structure lines, are spectrally resolved in 6/8 objects, with intrinsic FWHM up to $\approx 1250$~km~s$^{-1}$ (velocity dispersions up to $\approx 540$~km~s$^{-1}$). 

In five sources (3C~236, 3C~293, 3C~459, 4C~12.50 and PKS~1549) the [Ne{\sc ii}], and to a lesser extent [Ne{\sc iii}] and [Ne{\sc v}], line profiles are \textit{asymmetric}. We securely detect blue-shifted (and in some cases red-shifted) wings, up to 3000~km~s$^{-1}$ wide, underlying a stronger and narrower peak component centred very close to the optical systemic velocity. As explained in sect.~\ref{subsec:line_profile_analysis}, we carefully checked that the  [Ne{\sc ii}]$12.8\mu$m line wings cannot be ascribed  to an underlying PAH spectral feature. This result is confirmed by the fact that in most cases the [Ne{\sc iii}]$15.5\mu$m line profile, though noisier but located in a PAH-free part of the spectrum, is consistent with the [Ne{\sc ii}]$12.8\mu$m line profile. 
The kinematic properties of the broad [Ne{\sc ii}]$12.8\mu$m features are listed in Table~\ref{table:kin_wings}. 
Such wings in the Neon lines have been reported in 4C~12.50 (and in other ULIRGs) by \citet{Spoon2009}. 

\begin{deluxetable*}{lcccccc} 
\tablecolumns{7}
\tablecaption{Ionized gas kinematics: fit results for mid-IR lines with broad wings}
\tablehead{
\colhead{Line} &  \colhead{Comp.} &  \colhead{$\lambda _c$}  &  \colhead{$v_c$}  &  \colhead{obs. FWHM $w_m$}  &  \colhead{int. FWHM $w_i$}   &  \colhead{Flux} \\
                             &                                &  \colhead{[$\mu$m]}        &  \colhead{[km~s$^{-1}$]}   &  \colhead{[km~s$^{-1}$]} &  \colhead{[km~s$^{-1}$]} & \colhead{[$10^{-17}$~W~m$^{-2}$]}
}
\startdata
\multicolumn{7}{c}{\underline{3C~236}} \\
 $\mbox{[Ne{\sc ii}]}$   & n &     12.792   & $ -507 \pm  33 $   &  $  981  \pm  96 $   &    $ 844   \pm    82 $   &  $   0.55 \pm  0.04 $\\
 $\mbox{[Ne{\sc ii}]}$   & b &     12.796  &  $ -424 \pm 121 $  &   $ 2172 \pm  201 $   & $   2114   \pm   195 $   & $    0.38 \pm  0.03$ \\  
 \hline
\multicolumn{7}{c}{\underline{3C~293}} \\
 $\mbox{[Ne{\sc ii}]}$  & n &    12.818   & $  98 \pm 26 $  &   $  779  \pm  85 $    &  $  597    \pm   65  $   & $   2.66 \pm  0.19$ \\
 $\mbox{[Ne{\sc ii}]}$  & b &   12.835   & $  494 \pm  51 $  &   $ 2338  \pm 215 $  &  $   2284    \pm  209 $  &  $    1.91 \pm  0.16$ \\
  \hline
\multicolumn{7}{c}{\underline{3C~459}} \\
 $\mbox{[Ne{\sc ii}]}$  & n &    12.824   & $  227  \pm  21 $  &    $ 804  \pm  86 $   &    $ 630   \pm    67  $   &  $  2.50  \pm 0.18$ \\
 $\mbox{[Ne{\sc ii}]}$  & b &    12.797   & $ -372 \pm 89 $  &   $ 1721  \pm 163 $  &   $  1647   \pm   155 $    &  $  0.80 \pm  0.06$ \\
  \hline
\multicolumn{7}{c}{\underline{4C~12.50}} \\
 $\mbox{[Ne{\sc ii}]}$   & n &     12.810   & $  -88  \pm  15 $   &  $  786  \pm  85  $  &   $  606  \pm     65   $  &   $ 3.35 \pm  0.24$ \\
 $\mbox{[Ne{\sc ii}]}$   & b &     12.779  &  $ -812 \pm  99 $  &  $  2097 \pm  194  $  & $   2037   \pm   189 $   &   $  2.46 \pm  0.20$ \\
  \hline
\multicolumn{7}{c}{\underline{PKS~1549-79}} \\
 $\mbox{[Ne{\sc ii}]}$  & n &    12.814   &  $ -5  \pm 12  $ &   $  793  \pm  83  $ &  $    616   \pm    66  $   &  $  1.86  \pm  0.13$ \\
 $\mbox{[Ne{\sc ii}]}$  & b &    12.775   & $ -906  \pm 131  $  &  $ 885 \pm  142  $  & $   730   \pm   82  $  &   $  0.48 \pm  0.04$ 
\enddata
\tablecomments{Results from a two-component Gaussian decomposition of the [Ne{\sc ii}]$12.8\mu$m 
emission line profiles into narrow (n) and broad (b) components. $\lambda _c$ is the central wavelength of 
the component and corresponds to a velocity shift $v_{c}$ of the component with respect to systemic velocity. 
The statistical error on the velocity shift does not include corrections for the uncertainty in the wavelength 
calibration, which is $1/5$ of a resolution element (100~km~s$^{-1}$). $w_m$ (respectively $w_i$) is the observed (resp. intrinsic) FWHM before (resp. after) removal of the instrumental profile in quadrature (see sect.~\ref{subsec:line_profile_analysis} for details).}
  \label{table:kin_wings}
\end{deluxetable*}

To characterize the asymmetry and study the kinematics of the outflowing ionized gas, we decomposed the asymmetric line profiles into two Gaussian components,  which provides a satisfactory fit. The resulting fit and the broad component are shown on Figures~\ref{fig:lines_1} and \ref{fig:lines_2}. The detailed results of the fits are listed in Table~\ref{table:kin_wings}. The intrinsic FWHM of the broad [Ne{\sc ii}]$12.8\mu$m  components range from $\approx 730$~km~s$^{-1}$ (PKS~1549-79) to $\approx 2300$~km~s$^{-1}$ (3C~293). In three sources (3C~459, 4C~12.50, and PKS~1549-79), the broad component is clearly blue-shifted with respect to the systemic velocity, with velocity shifts up to $\approx 800$~km~s$^{-1}$. For 3C~236, the entire [Ne{\sc ii}]$12.8\mu$m line and  the broad H{\sc i} absorption profile are redshifted by $\approx 500$~km~s$^{-1}$ with respect to systemic, whereas the H$_2$ line is centred on the systemic velocity. This is a clear case where a large fraction of the ionized and H{\sc i} gas is in an outflow, but the H$_2$ gas is not.

\subsection{H{\sc i} gas}
\label{subsec:HI_kinematics}

\begin{deluxetable*}{lccccccccc} 
\tablecolumns{10}
\tablecaption{H{\sc i} gas kinematics and energetics: broad H{\sc i} component associated with the neutral outflow}
\tablehead{
\colhead{Object} & \colhead{FWHM\tablenotemark{a}}  & \colhead{Velocity\tablenotemark{b}} & \colhead{timescale\tablenotemark{c}} &  \colhead{$\tau$\tablenotemark{d}} & \colhead{$E_{\rm kin}^{(turb)}$ \tablenotemark{e}} & \colhead{$E_{\rm kin}^{(bulk)}$ \tablenotemark{f}}  &  \colhead{$L_{\rm kin}^{(turb)}$ \tablenotemark{g}} & \colhead{$L_{\rm kin}^{(bulk)}$ \tablenotemark{h}} & \colhead{$\dot{E}({\rm HI})$ \tablenotemark{i}} \\
	\colhead{}  & \colhead{[km s$^{-1}$]} &\colhead{[km s$^{-1}$]} & \colhead{[yrs]} & \colhead{} & \colhead{[Joules]}  &\colhead{[Joules]} & \colhead{[L$_{\odot}$]} &  \colhead{[L$_{\odot}$]}  &  \colhead{[L$_{\odot}$]} }
\startdata
3C 236     &  932         & 29367 & 6.5E05 & 0.0033  &  5.8E49    & 2.5E50 & 7.2E09  &  1.6E10  &  2.0E10    \\

3C 293     & 852  & 13156 & 2.0E06  & 0.0038  & 3.9E49 & 3.9E50 & 1.6E09   & 1.1E10  &  1.4E10      \\

3C 305     & 689          & 12402 & 3.9E06 & 0.0023  & 1.2E49  & 3.9E50 & 2.5E08   & 1.2E10     & 5.9E08      \\

3C 459     & 411          & 65934 & 3.3E06 & 0.0005  & 1.6E48 & 1.2E50 &  4.1E07  & 4.0E09   &  3.8E08      \\
4C 12.50  & 970         & 36617 & 3.3E05 & 0.0017  & 5.1E48 & 2.7E49 & 1.3E09   & 5.7E09   &  2.1E09   \\
IC 5063   & 133        & 2785 & 1.1E06 & 0.0120  & 2.3E48 & 1.4E50 & 1.7E08   &  8.2E09   &  3.2E09     \\
OQ 208   & 482        & 22568 & 1.6E04 & 0.0057  & 5.0E48 & 2.0E47 & 2.5E10   &  6.5E08  &  3.4E08     \\
PKS 1549-79 & 329      & 45553 & 3.9E06 & 0.0200  & 5.9E48 & 3.3E51 &1.2E08   & 1.0E11   & 1.2E10     
\enddata 
\tablenotetext{a}{Full width at half maximum of the broad H{\sc i} component (fitted with a Gaussian).}
\tablenotetext{b}{Central velocity of the broad H{\sc i} component.}
\tablenotetext{c}{Outflow timescale derived from the radius $r$ and the outflow velocity.}
\tablenotetext{d}{Optical depth of the broad H{\sc i} component for each velocity components.}
\tablenotetext{e}{Turbulent kinetic energy computed from $3/2 M({\rm HI}) \sigma ^2$.}
\tablenotetext{f}{Bulk kinetic energy computed by integrating the H{\sc i} profile (see sect.$\ $\ref{subsec:HI_kinematics} for details).}
\tablenotetext{g}{Turbulent kinetic luminosity.}
\tablenotetext{h}{Bulk H{\sc i} kinetic luminosity computed from Eq.$\ $\ref{eq:bulk_kin_lum}}
\tablenotetext{i}{H{\sc i} kinetic energy loss rate computed from Eq.$\ $\ref{eq:nrj_loss_rate}}
\label{table:HI_kinematics}
\end{deluxetable*}

In this section we use the WSRT H{\sc i} absorption profiles \citep{Morganti2005} to quantify the kinematics and energetics of the H{\sc i} gas. We recall that this H{\sc i} absorption, detected against the radio continuum, is spatially unresolved, except for IC~5063 and 3C~305. The profiles all exhibit a deep and narrow (FWHM$<$200~km~s$^{-1}$) component that traces quiescent gas (likely located in a large scale disk), plus  shallow and broad components (with FWHM up to 970~km~s$^{-1}$), mostly blueshifted, that indicate outflowing gas \citep[e.g.][]{Morganti2003}. Two or three Gaussians were necessary to accurately fit the profiles (see Figure~\ref{fig:HIgauss_fit}). The Gaussian fitting parameters and energetics of the H{\sc i} gas are summarized in Table~\ref{table:HI_kinematics}.

At each given velocity, the optical depth ($\tau$) is calculated from $e^{-\tau} = 1- S_{\rm abs} / S_{\rm cont}$ , 
where $S_{\rm abs}$ is the absorption flux and $S_{\rm cont}$ is the underlying radio continuum flux. Then, the H{\sc i} column density is derived from:
\begin{equation}
\frac{N_{HI}}{10^{21} \, \rm cm^{-2}} = 1.822 \times  \frac{T_{\rm spin}}{10^3 \, {\rm K}} \int \tau (v) dv \ ,
\end{equation}
where $v$ is the velocity and $T_{\rm spin}$ is the spin temperature of the H{\sc i} gas, assumed to be 1000~K, given that the H{\sc i} gas is heated by X-ray photons and shocks \citep[see e.g.][]{Bahcall1969, Maloney1996}. Note that this value of $T_{\rm spin}$ is an order of magnitude, and could well be within a $10^2 - 10^4$~K range.
The mass of H{\sc i} gas is derived from $N_{HI}$ and the size of the H{\sc i} absorption region (see sect.~\ref{sec:sample} and Table~\ref{table_radio_prop}).

The turbulent kinetic energy of the H{\sc i} gas (associated with the velocity dispersion of the gas) is estimated from $E_{\rm kin}^{(turb)} = 3/2 M_{\rm HI} \sigma _{\rm HI}^{2}$, where  $\sigma _{\rm HI}= \rm FWHM / 2.36$ is the velocity dispersion of the H{\sc i} gas. The factor of 3 in $E_{\rm kin}^{(turb)}$ takes into account the three dimensions.

%

The bulk (radial) mechanical energy of the entrained gas in the outflow is calculated directly from the broad H{\sc i} profile (after subtraction of the narrow component) by integrating the following quantity relative to the systemic velocity ($v_{\rm sys}$):
\begin{equation}
\label{eq:bulk_kin_lum}
\frac{L_{\rm kin}^{bulk}}{{\rm L}_{\odot}} = \mbox{1.23E4} \frac{r}{1\,{\rm kpc}}   \frac{T_{\rm spin}}{10^3 \, {\rm K}} \int \ln  \frac{S_{\rm cont}}{S_{\rm cont} - S_{\rm abs}}  \left(\frac{v - v_{\rm sys}}{10^3\, \mbox{km/s}} \right)^2 \frac{dv}{1\,\mbox{km/s}} \ ,
\end{equation}
where $r$ is the H{\sc i} outflow radius \citep[see Table~\ref{table_radio_prop} and ][]{Morganti2005}. 

For comparison, one can assume an outflow with a constant velocity and mass flow $\dot{M}$, and derive the energy loss rate as \citep{Heckman2002}:
\begin{equation}
\label{eq:nrj_loss_rate}
\dot{E} = 8.1 \times 10^8 C_f  \frac{\Omega}{4 \pi} \frac{r}{\mbox{1 kpc}} \frac{N_{HI}}{10^{21}\, {\rm cm^{-2}}} \left(\frac{v}{350\, {\rm km s^{-1}}} \right)^3 \ ,
\end{equation}
where $C_f$ is the covering fraction and $\Omega$ the opening solid angle in which the gas the outflowing from a radius $r$.

The energetic quantities defined above are listed in Table~\ref{table:HI_kinematics}. The conversion from the kinetic energies to the kinetic luminosities are done via a timescale defined with the radius $r$ and the outflow velocity (or velocity dispersion). The bulk (radial) H{\sc i} kinetic luminosities and energy loss rates are larger than the turbulent (velocity dispersion only) kinetic luminosities, except for OQ~208. The size of the OQ~208 radio source is small and the H{\sc i} absorption spectrum exhibits three broad components, which makes the identification of the outflow component difficult. The bulk H{\sc i} kinetic luminosities represent about a few percent of the Eddington luminosities. This is two orders of magnitude larger than the ionized outflow kinetic power \citep{Holt2006, Morganti2007, Morganti2010}. We will further discuss these results in sect.~\ref{subsec:nrj_transfer} when examining the transfer of kinetic energy from the jet to the H{\sc i} and H$_2$ gas.

\subsection{Comparison of the multiphase gas kinematics}
\label{subsec:kin_comp}
\begin{figure}
   \centering
    \includegraphics[width=0.49\textwidth]{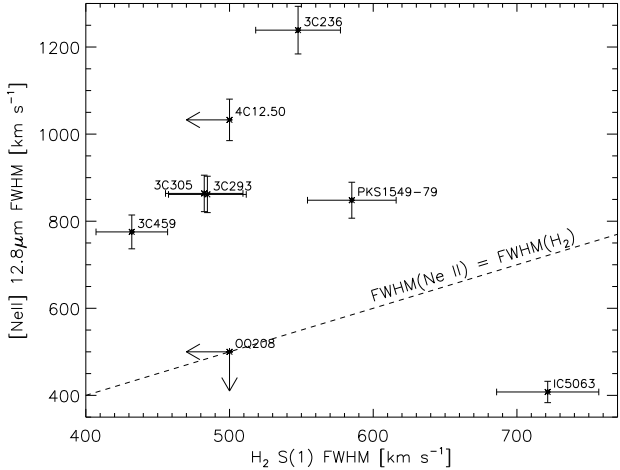}
      \caption{[Ne{\sc ii}]$\, \lambda \, 12.8\,\mu$m versus H$_2$ S(1) 17$\,\mu$m linewidths (FWHM in km~s$^{-1}$) measured from 
      the \textit{Spitzer} IRS spectra with the high-resolution module ($w_m$ in sect.~\ref{subsec:line_profile_analysis}). The linewidths are intrinsic, i.e. 
      they have been corrected for the instrumental Gaussian profile.}
       \label{fig:NeII_vs_H2S1_fwhm}
   \end{figure}


\begin{figure*}
     \includegraphics[width=0.85\columnwidth]{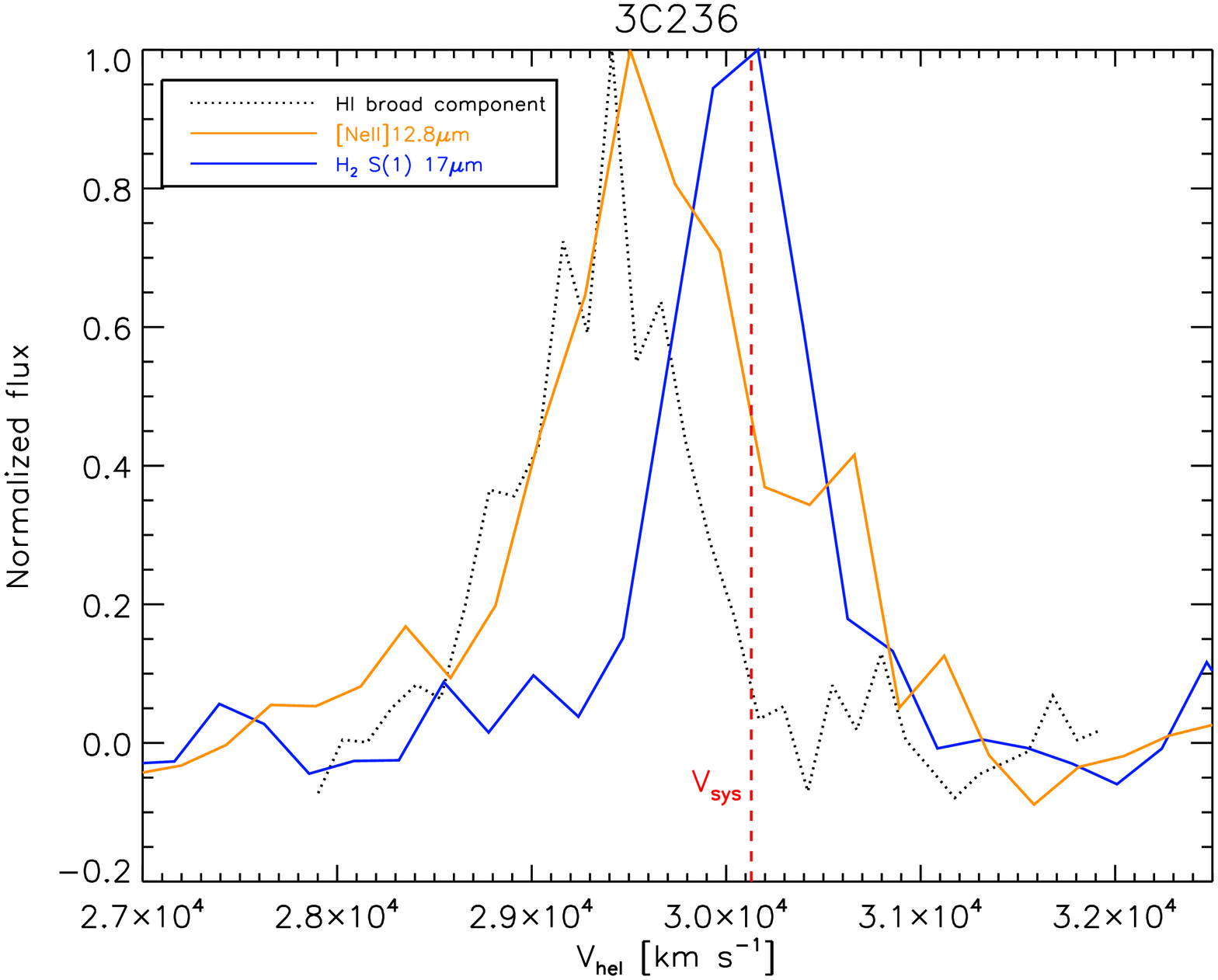}
     \includegraphics[width=0.85\columnwidth]{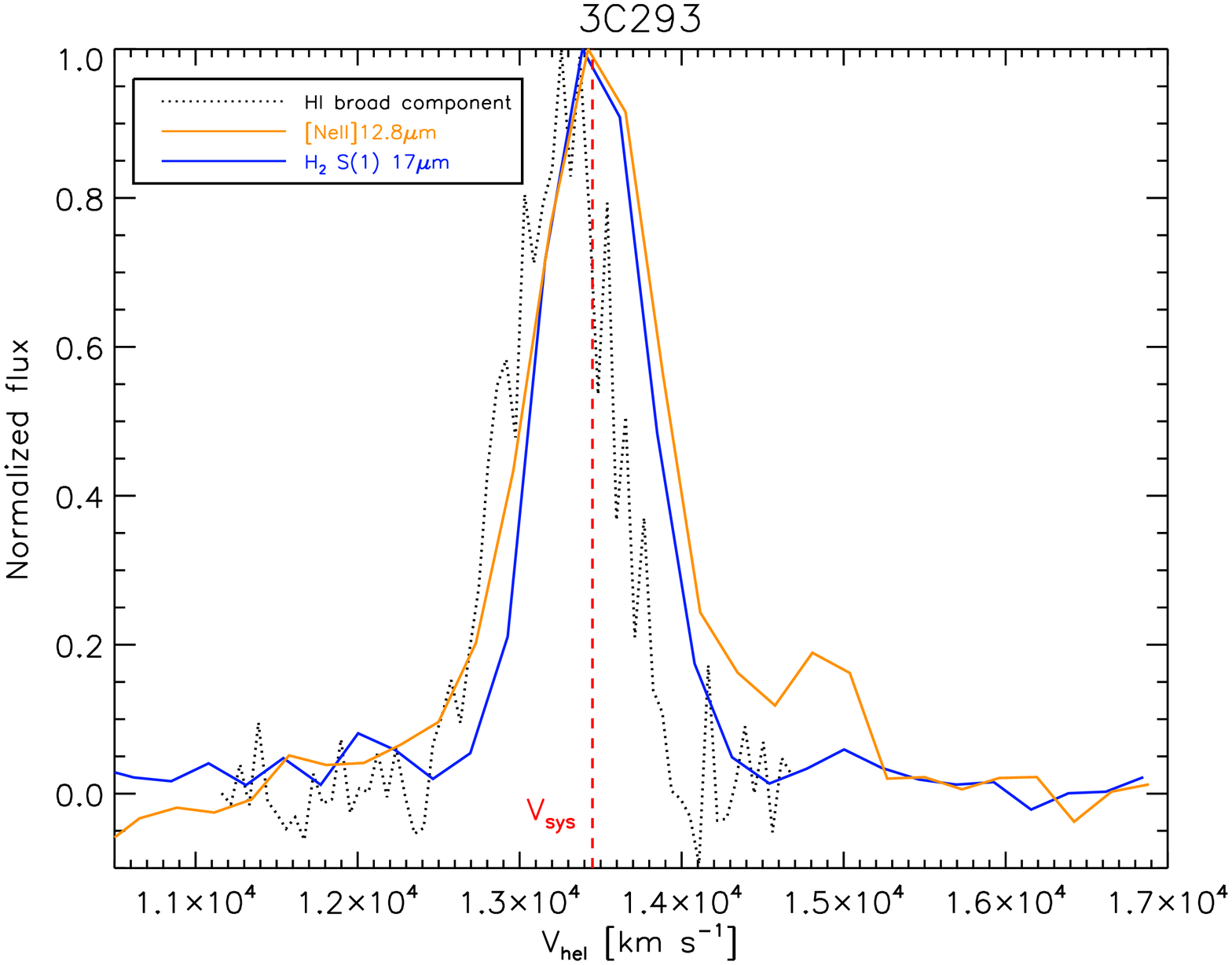}
     \includegraphics[width=0.85\columnwidth]{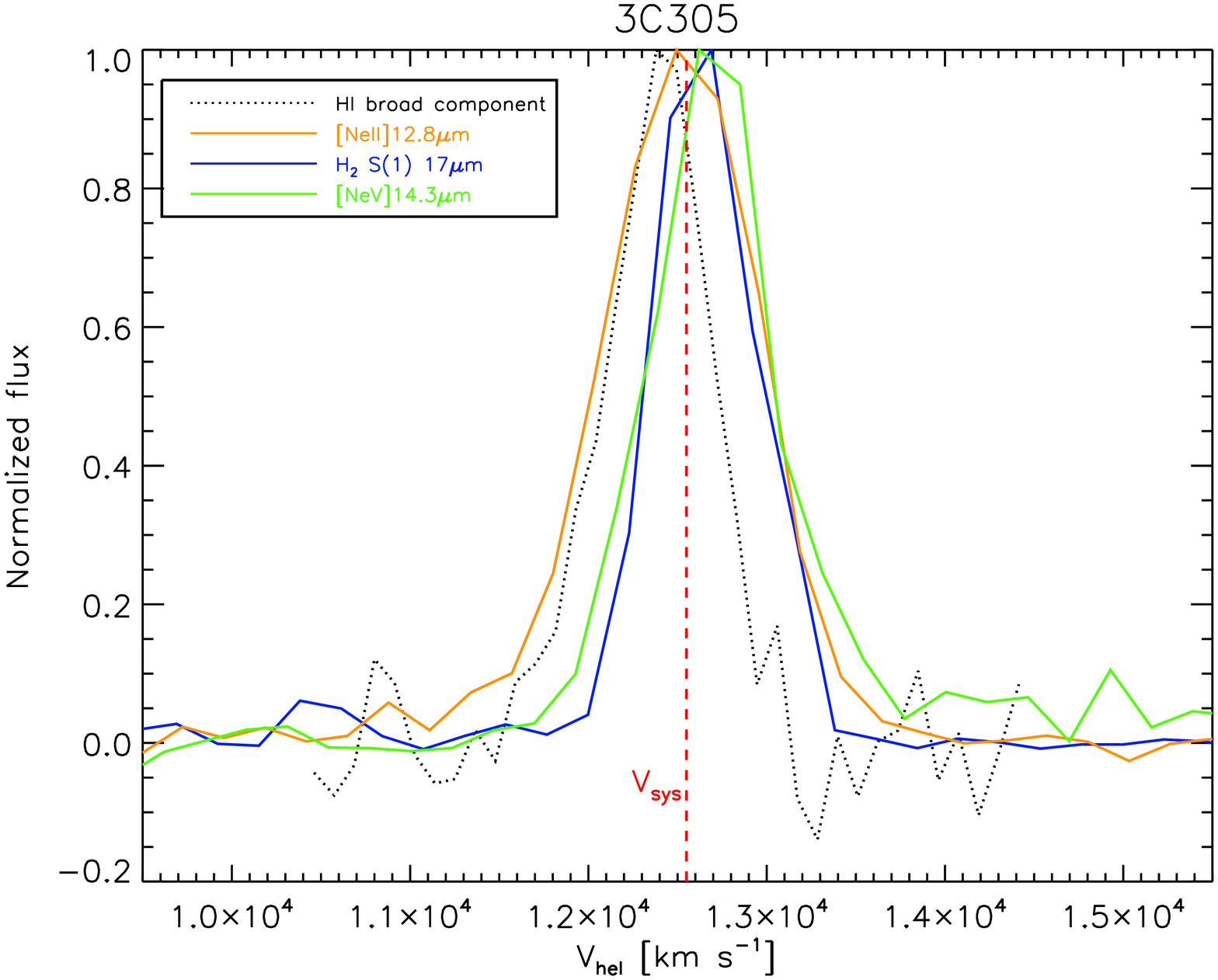}
     \includegraphics[width=0.85\columnwidth]{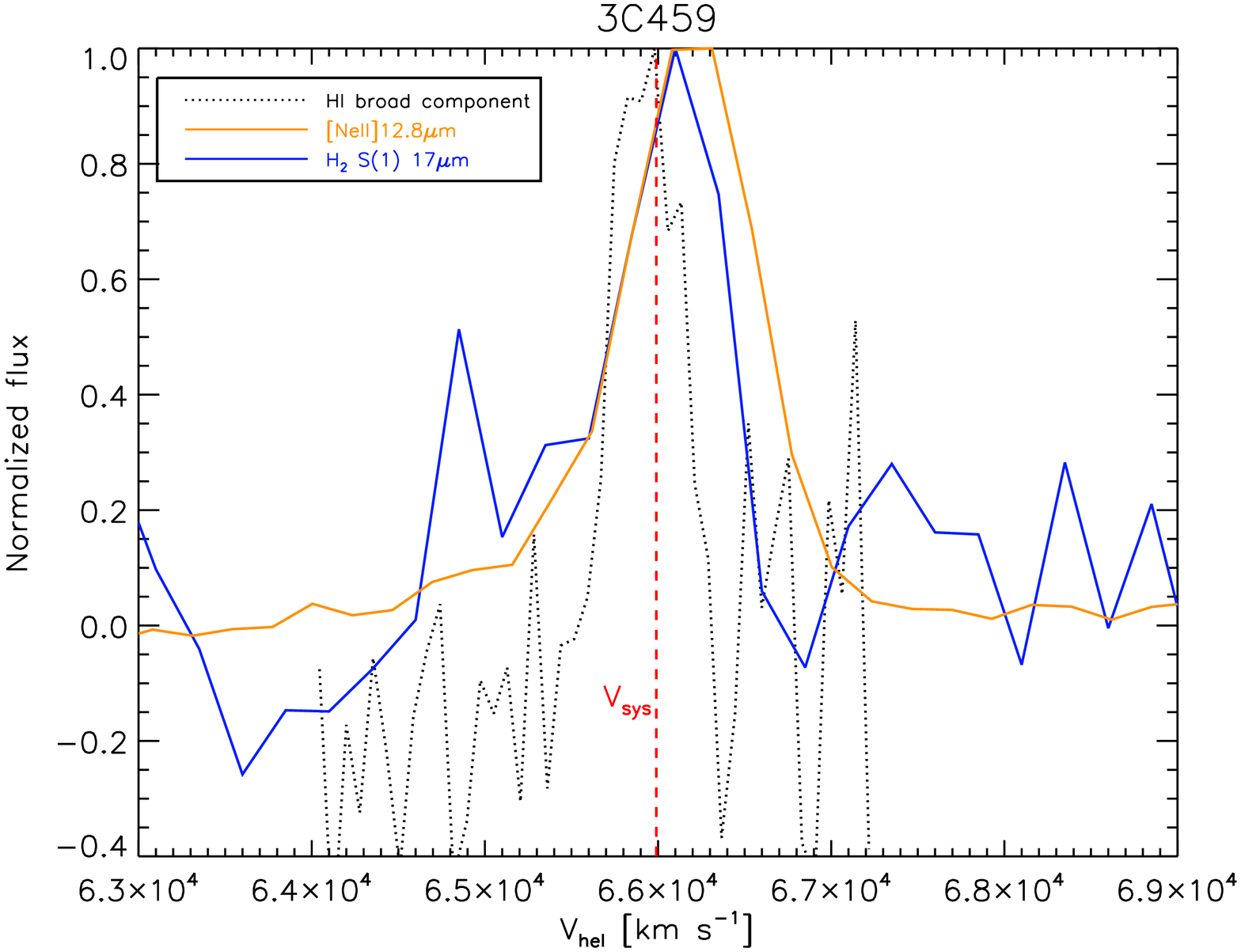}
      \includegraphics[width=0.85\columnwidth]{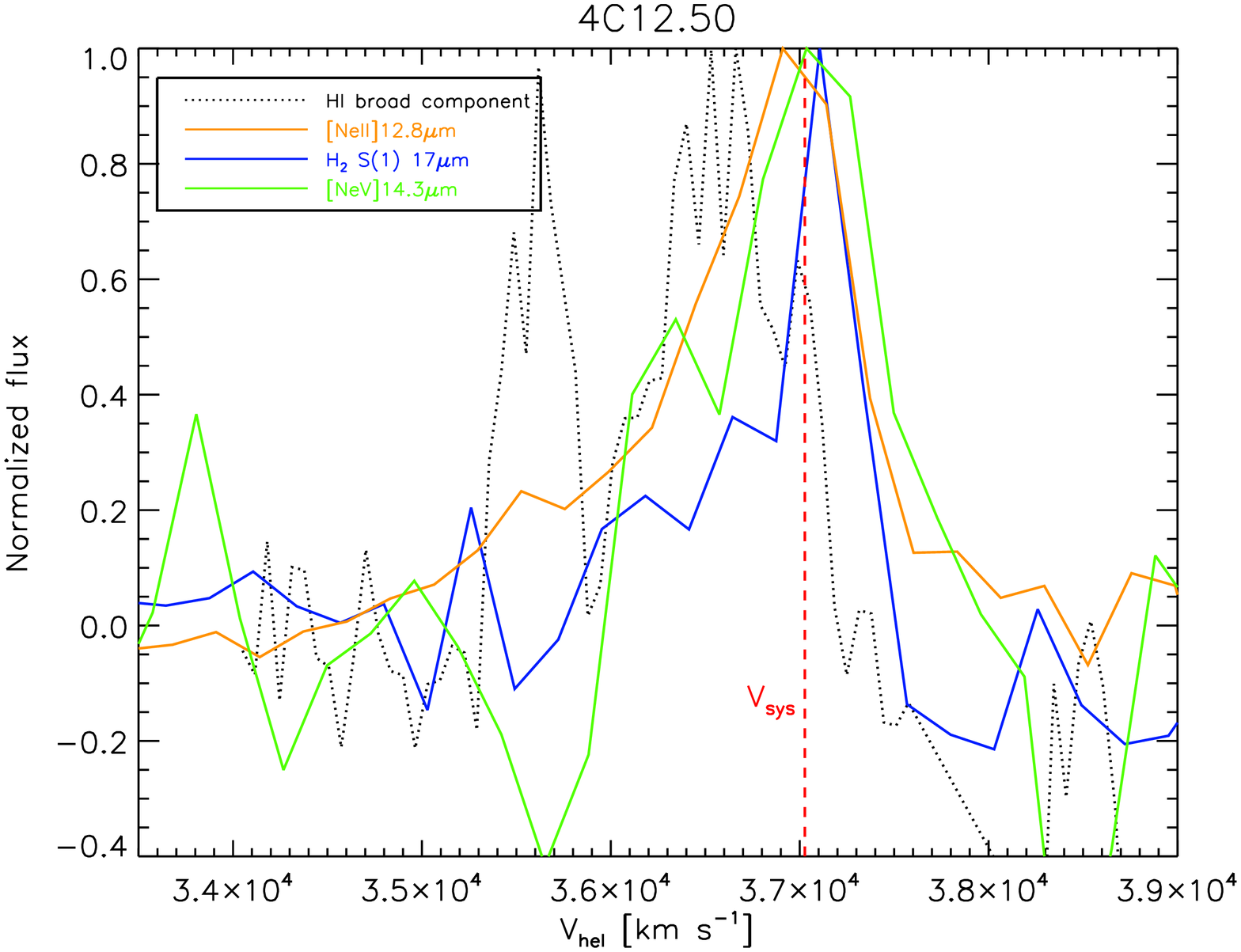}
      \includegraphics[width=0.85\columnwidth]{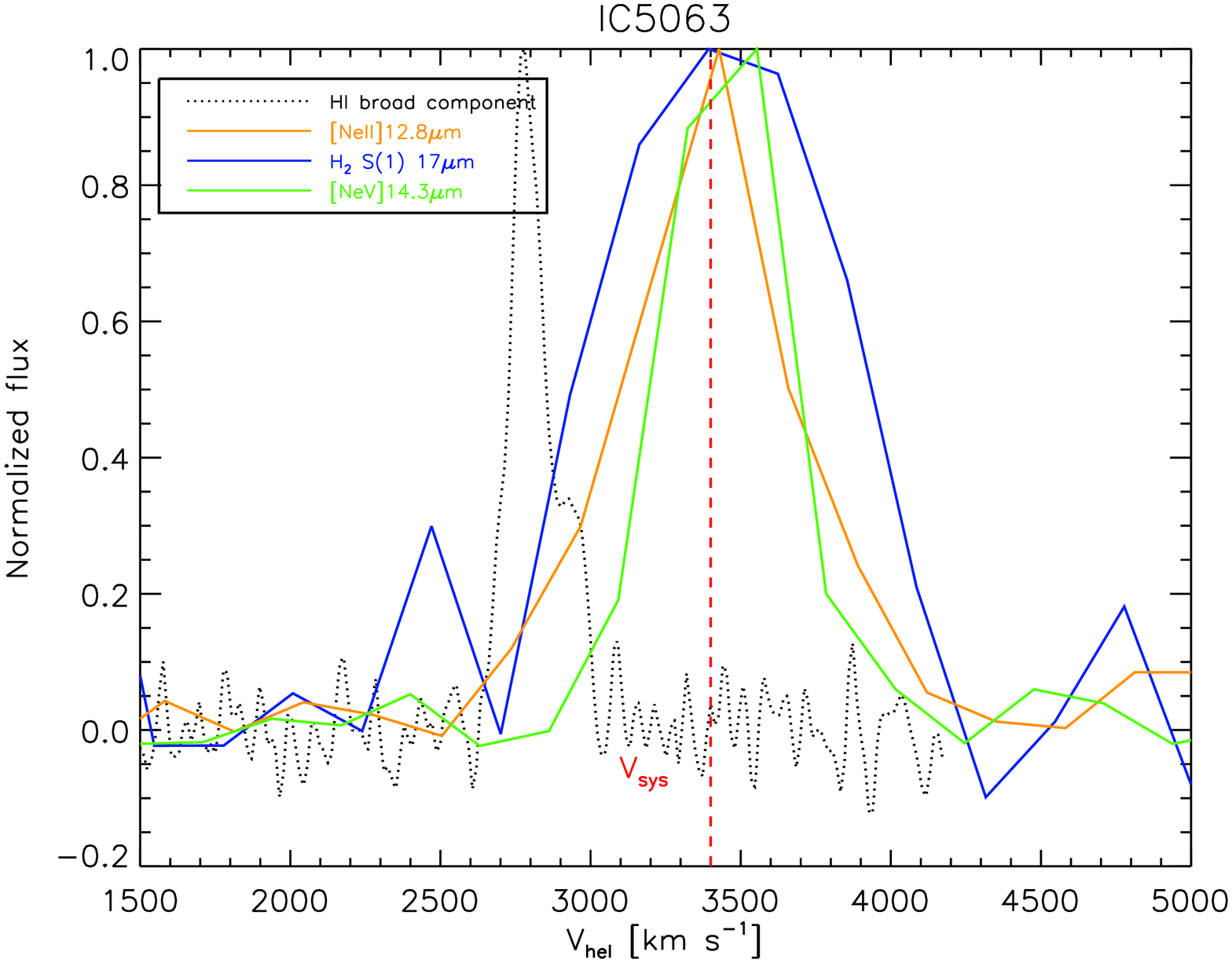}
      \includegraphics[width=0.85\columnwidth]{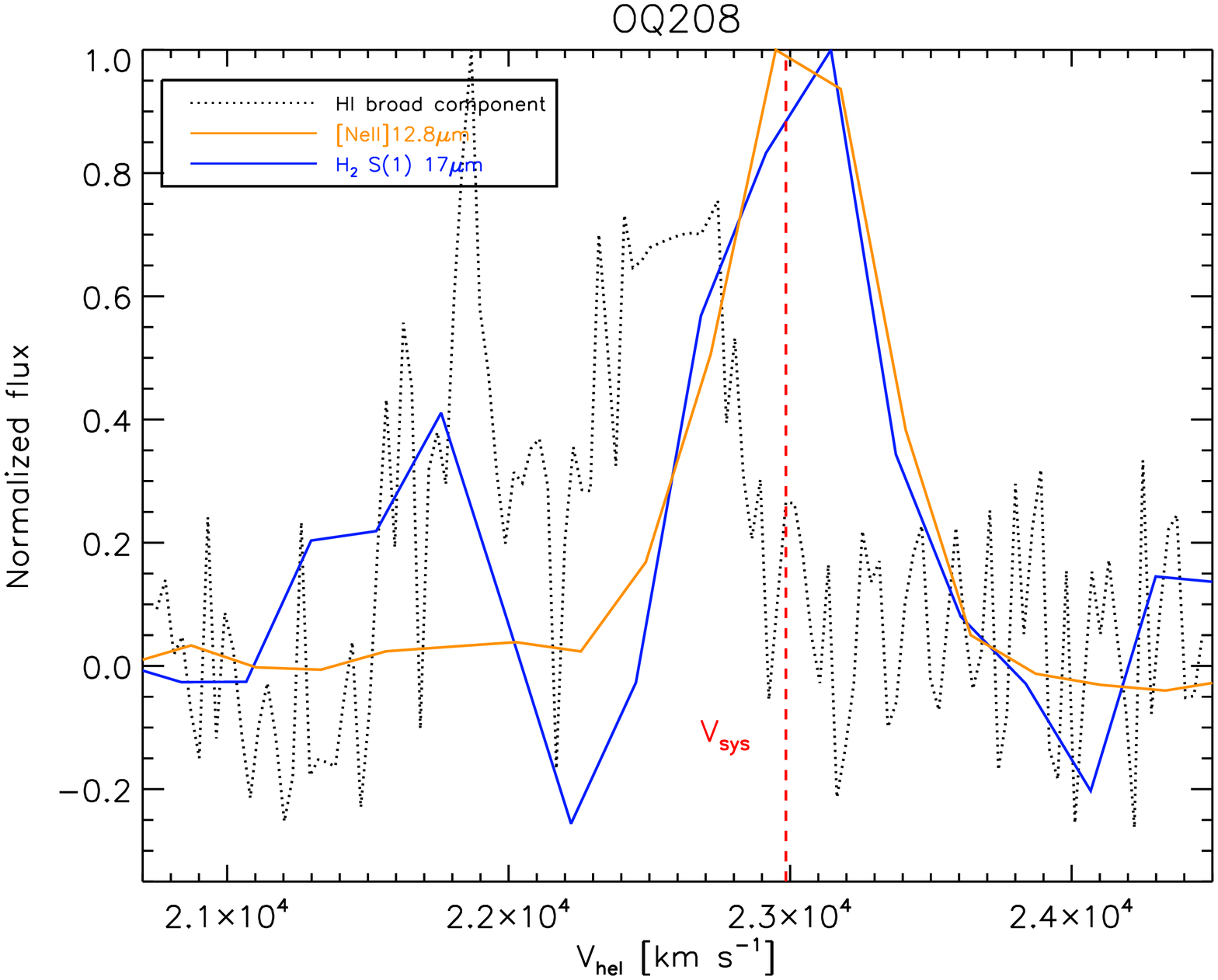}
      \qquad \qquad \qquad \qquad  \qquad \qquad 
      \includegraphics[width=0.85\columnwidth]{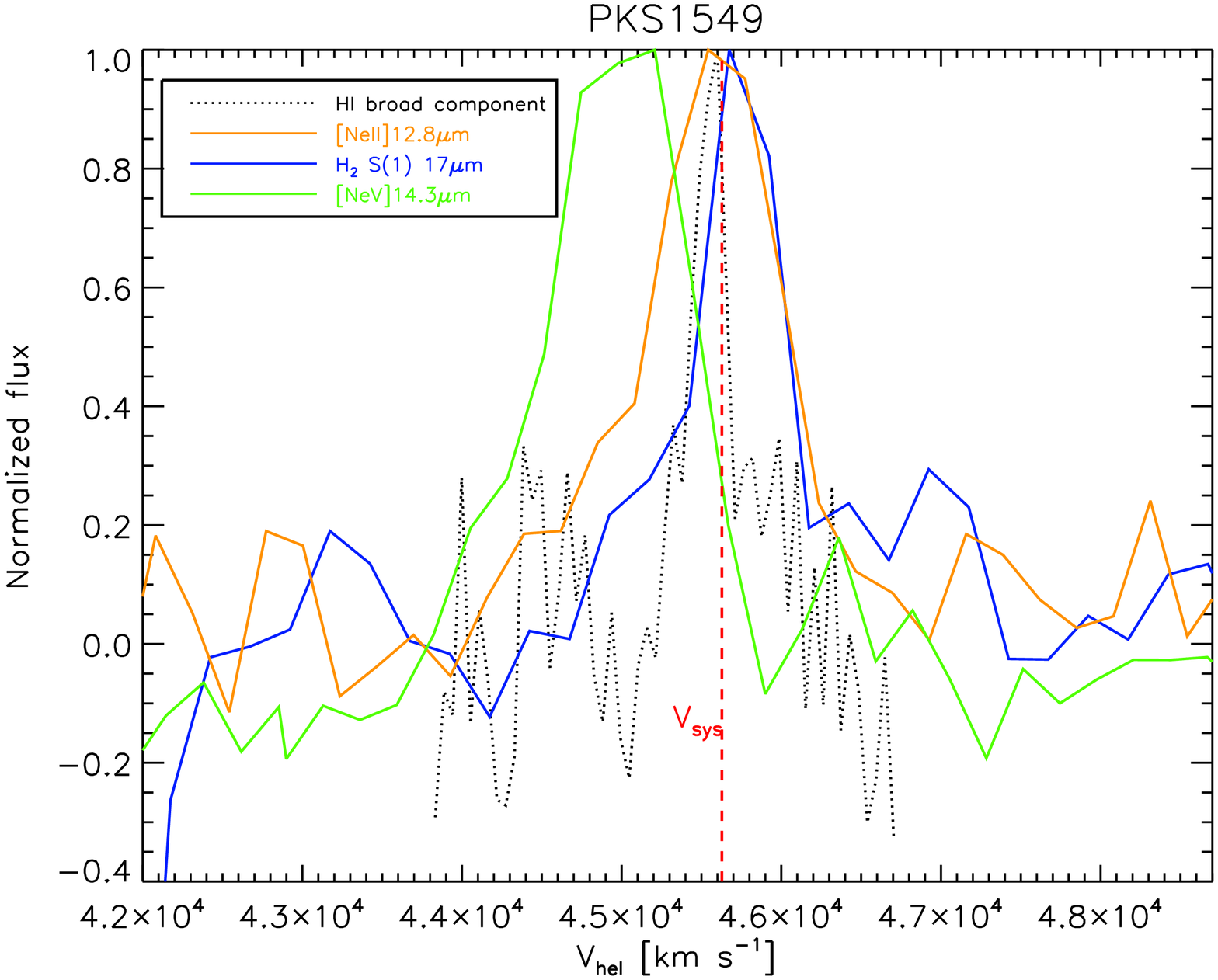}
      \caption{Comparison of the normalized H{\sc i} absorption \citep[data from][]{Morganti2005} (shown on a positive scale) and the \textit{Spitzer} IRS H$_2$ S(1),  [Ne{\sc ii}]$\lambda \, 12.8\,\mu$m and [Ne{\sc v}]$\lambda \, 14.3\,\mu$m (when detected) emission profiles. The dotted black line is the shallower broad H{\sc i} component associated with the outflowing gas, after removal of the deep, narrower absorption peak. The optical systemic velocity is indicated by a vertical red dashed line. For clarity, the H{\sc i} spectra of  3C~236, 3C~293, 3C~305, 3C~459 and 4C~12.50 (respectively IC~5063, OQ~208 and PKS~1549-79) have been smoothed to a resolution of $\approx 125\,$km~s$^{-1}$, (resp. $\approx 30\,$km~s$^{-1}$). Note that for the \textit{Spitzer} profiles shown here, no correction for instrumental broadening has been applied. }
       \label{fig:NeII_HI}
   \end{figure*}

In Figure~\ref{fig:NeII_vs_H2S1_fwhm}, we compare the measured FWHM of the H$_2$ S(1) and  [Ne{\sc ii}]$\lambda \, 12.8\mu$m  lines. Except for IC~5063 and OQ~208, the [Ne{\sc ii}] is systematically broader, by a factor of 40\%, than the 17$\,\mu$m H$_2$ S(1) line. This is also the case for the other ionic fine-structure lines: \textit{the velocity dispersion of the ionized gas is generally larger than that of the warm H$_2$ gas.}

Figure~\ref{fig:NeII_HI} compares the kinematics of the ionized, atomic and warm H$_2$ gas. In 3C~236, 3C~293, 3C~305, 3C~459, 4C12.50 and PKS~1549-79, the blue-shifted, broad H{\sc i} absorption profile matches the blue wing of the [Ne{\sc ii}]$\lambda \, 12.8\mu$m line. The ionized and atomic gas are well coupled dynamically and outflowing at comparable velocities. Most of these galaxies are also known to exhibit outflows from optical spectroscopy: blue-shifted 5007 \AA [O{\sc iii}] line emission have been reported in 3C~293 \citep{Emonts2005}, 3C~305 \citep{Morganti2005}, IC~5063 \citep{Morganti2007}, PKS~1549-79 \citep{Tadhunter2001, Holt2008}, 4C~12.50 \citep{Holt2003, Holt2008} and 3C~459 \citep{Holt2008}. For these three targets, the [O{\sc iii}] velocity widths and shifts with respect to the systemic velocity are very close to our measurements of the [Ne{\sc ii}]$\lambda \, 12.8\mu$m line. 
Interestingly, in PKS~1549-79, the [Ne{\sc v}]$\lambda \, 14.3\mu$m line is blueshifted by $500-700$~km~s$^{-1}$, at the position of the [Ne{\sc ii}] and [Ne{\sc iii}] blue wings. This increase of blueshift  with increasing ionization has been observed in ULIRGs \citep{Spoon2009}, and suggests that the outflow speed decreases with distance to the central ionizing source.

On the other hand, in general, the kinematics of the warm H$_2$ gas does not follow that of the ionized or atomic gas.  In the cases where a blue wing is tentatively detected on the H$_2$ S(1) line (for 4C~12.50, 3C~459 and PKS~1549-79), the velocity extent of this H$_2$ wing is smaller than that of the [Ne{\sc ii}] line. If some of the H$_2$ gas is entrained in the flow, its velocity is at least a factor of 2 to 3 smaller than that of the ionized and atomic gas, based on the comparison between the velocity extent of the blue-shifted H$_2$ S(1) signal, and the one observed for the optical and mid-IR ionized gas lines. 
Note that the broad molecular line -- e.g. CO(1-0)-- components detected in composite sources like Mrk~231 are at a $1-5$\% level of the peak line flux. The limited sensitivity and spectral resolution of the \textit{Spitzer} IRS do not allow us to securely detect such broad wings in the H$_2$ lines at comparable levels. In addition, if the molecular gas in these galaxies lies in a very turbulent rotating disk \citep[as confirmed by our VLT/SINFONI observations of 3C~326,][]{Nesvadba2011a}, H$_2$ line wings in spectra integrated over the whole galaxy could be difficult to to identify on top of the fast rotation-velocity field.

\section{Discussion}
\label{sec:discussion}

\subsection{Which dominant mechanism powers the H$_2$ emission?}
\label{subsec:exc_mech}

\begin{figure*}
\centering
\includegraphics[width=\textwidth, clip]{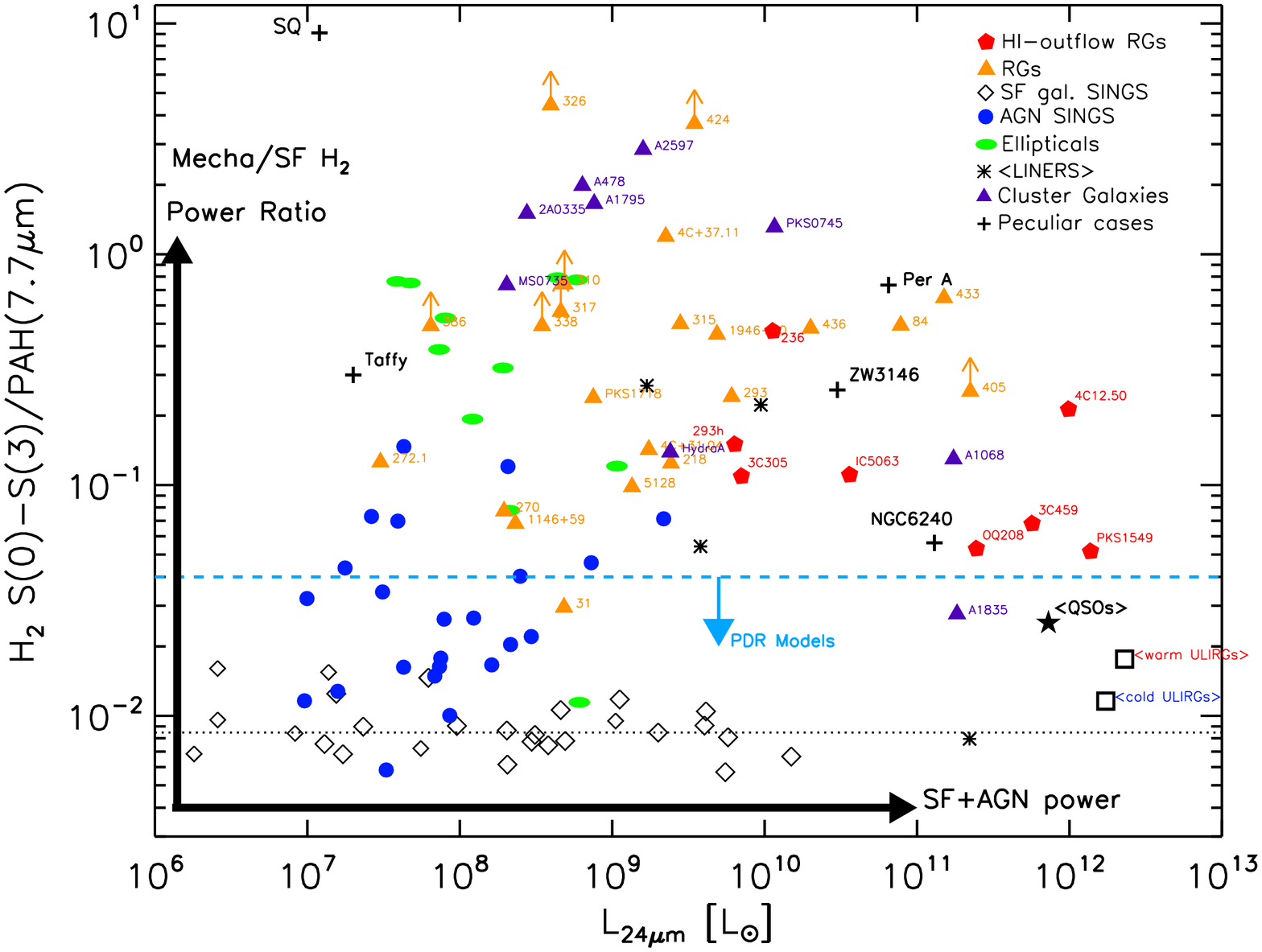}
 \caption{Ratio of the mid-IR H$_2$ line luminosities (summed over S(0) to S(3)) to the PAH 7.7$\mu$m emission vs. narrow-band 24$\mu$m continuum luminosity.
This ratio indicates the relative contribution of mechanical heating (shocks) and star-formation (SF) power. The red pentagons are the nearby radio galaxies with fast ($> \! 1000\,$km/s) H{\sc i} outflows. The orange triangles are the H$_2$-bright radio-galaxies presented by \citet{Ogle2010}, augmented by the compact symmetric objects observed by \citet{Willett2010}. The green ellipses are elliptical galaxies from the $\ $ \citet{Kaneda2008} sample. The purple triangles are the H$_2$-bright cool-core galaxy clusters from \citet{Donahue2011}. These H$_2$-luminous galaxies stand out above SF and AGN galaxies from the SINGS survey \citep[open diamonds and blue circles; data from ][]{Roussel2007}. 
The black dotted line shows the median value of the H$_2$-to-PAH luminosity ratio for the SINGs SF galaxies.
The H$_2$ emission in these sources cannot be accounted for by UV or X-ray photon heating. The blue dashed line shows the upper limit given by the Meudon PDR models \citep{LePetit2006} obtained for $n_{\rm H} = 10^4\,$cm$^{-3}$ and $G_{\rm UV}=10$. For comparison, a few other types of H$_2$-luminous galaxies are shown (black crosses): the Stephan's Quintet (SQ) and Taffy galaxy collisions \citep[data from ][and Peterson et al., sub.]{Cluver2010}, ZW~3146 \citep{Egami2006} and Perseus A \citep{Johnstone2007} clusters, and the NGC~6240 merger \citep{Armus2006}. The black squares indicate averaged values for the \citet{Higdon2006} sample of ULIRGs. The black star is an averaged value of the \citet{Schweitzer2006}sample of  QSOs. 
 }
       \label{fig:H2_PAH_ratio}
   \end{figure*}

Following \citet{Ogle2010}, we use the H$_2$ to PAH  luminosity ratio, plotted on Figure~\ref{fig:H2_PAH_ratio}, to investigate the contribution of UV photons to the total heating of the H$_2$ gas.
In ``normal'' star-forming galaxies and dwarfs, the tight correlation between the H$_2$ and PAH luminosities, and the correspondence of the H$_2$-to-PAH luminosity ratio with PDR models, suggest that most of the H$_2$ line emission is powered by stellar UV photons \citep[e.g. ][]{Rigopoulou2002, Higdon2006, Roussel2007}. Strikingly, all of the H{\sc i}-outflow radio-galaxies have a $L({\rm H_2}) / L({\rm PAH7.7})$ ratio larger (by at least a factor of 4)  than the median value observed for star-forming galaxies \citep[0.0086 for the SINGS galaxies, ][]{Roussel2007}. These H$_2$-bright galaxies fall in the MOHEGs (molecular hydrogen emission galaxies) category \citep{Ogle2010}, defined as $L({\rm H_2 \, S(0)-S(3)}) / L({\rm PAH7.7}) > 0.04$. 
Using the  Meudon PDR code \citep{LePetit2006}, we computed the H$_2$ S(0)-S(3) to PAH~7.7$\,\mu$m flux ratio as a function of the $G_{\rm UV} / n_{\rm H}$, where $G_{\rm UV}$ is the intensity of the UV radiation field (in Habing units) illuminating a slab of gas of hydrogen density $n_{\rm H}$. The H$_2$-to-PAH ratio is high for small values of $G_{\rm UV} / n_{\rm H}$. Exploring densities from $n_{\rm H} = 10^{2}$ to $10^{4}$~cm$^{-3}$ and $G_{\rm UV} = 1 - 10^4$, we find a maximum H$_2$-to-PAH flux ratio of $4\times 10^{-2}$, which is in agreement with the \citet{Kaufman2006} models. This value is precisely the limit chosen empirically by \citet{Ogle2010} to define MOHEGs. All of the H{\sc i}-outflow radio-galaxies have $L({\rm H_2}) / L({\rm PAH7.7}) > 0.04$ (see Figure~\ref{fig:H2_PAH_ratio}), showing that their H$_2$ emission cannot be accounted only by UV heating.

Interestingly enough, Figure~\ref{fig:H2_PAH_ratio} shows that the sources having the largest H$_2$ to PAH  luminosity ratio have moderate IR luminosities ($L_{24\mu m} \approx 10^9 \,$L$_{\odot}$) and are jet-dominated, suggesting that the jet plays an important role in powering the H$_2$ emission. At larger IR luminosities, the sources become dominated by star formation, hence the more moderate H$_2$ to PAH ratios.  We do not find any extreme  H$_2$ to PAH luminosity ratios ($>1$) at large IR luminosities  ($L_{24\mu m} \gtrsim 10^{11} \,$L$_{\odot}$). This may be due to the fact that all of these sources are at low ($\lesssim 0.3$) redshifts, so the most luminous quasars are not included in this sample. We note that the predominately radio-quiet \citet{Higdon2006} sample of ULIRGs ($0.02 < z < 0.93$) and the \citet{Schweitzer2006}  sample of QSOs ($z < 0.3$) have H$_2$ to PAH luminosity ratios significantly higher than the SINGs star-forming galaxies. Although their H$_2$ to PAH luminosity ratios are compatible with the range of ratios predicted by PDR models, other studies (based on extinction or H$_2$-to-CO line ratios), suggest that additional sources of H$_2$ excitation (other than UV photon heating) are required to explain the H$_2$ emission \citep[e.g. shocks, see][]{Zakamska2010, Fiolet2010}.

AGN X-ray heating is potentially a source of H$_2$ excitation \citep[e.g.][]{Maloney1996}. Assuming a characteristic H$_2$ gas temperature of 200~K and that all the X-ray flux from the AGN is absorbed by the molecular gas, \citet{Ogle2010} estimated that the maximum H$_2$-to-X-ray luminosity ratio is of the order of  $L({\rm H_2 \ 0-0 S(0)-S(3)}) / L_X({\rm 2-10\,keV}) < 0.01$. We use the c08.01 version of the CLOUDY  code, last described by \citet{Ferland1998}, to compute this ratio for a cloud of density $n_{\rm H} = 10^{4}$~cm$^{-3}$ and a total ionizing photon luminosity corresponding to $L_X({\rm 2-10\,keV}) = 10^{43}$~erg~s$^{-1}$, which is the average X-ray luminosity measured for our sample. The covering and filling factors are set to unity. For this set of parameters we find $L({\rm H_2 \ 0-0 S(0)-S(3)}) / L_X({\rm 2-10\,keV}) = 7 \times 10^{-3}$, which is in reasonable agreement with the upper limit of 0.01 given above. All H{\sc i}-outflow radio-galaxies that have X-ray measurements have an observed H$_2$-to-X-ray flux ratio above that limit (see last column of Table~\ref{table_sample}), which shows that the X-ray heating cannot be the dominant powering source of the H$_2$ emission in H{\sc i}-outflow radio MOHEGs, except perhaps in IC~5063, where the H$_2$-to-X-ray luminosity ratio is close to 0.01.

\begin{figure}
   \centering
    \includegraphics[width=\columnwidth]{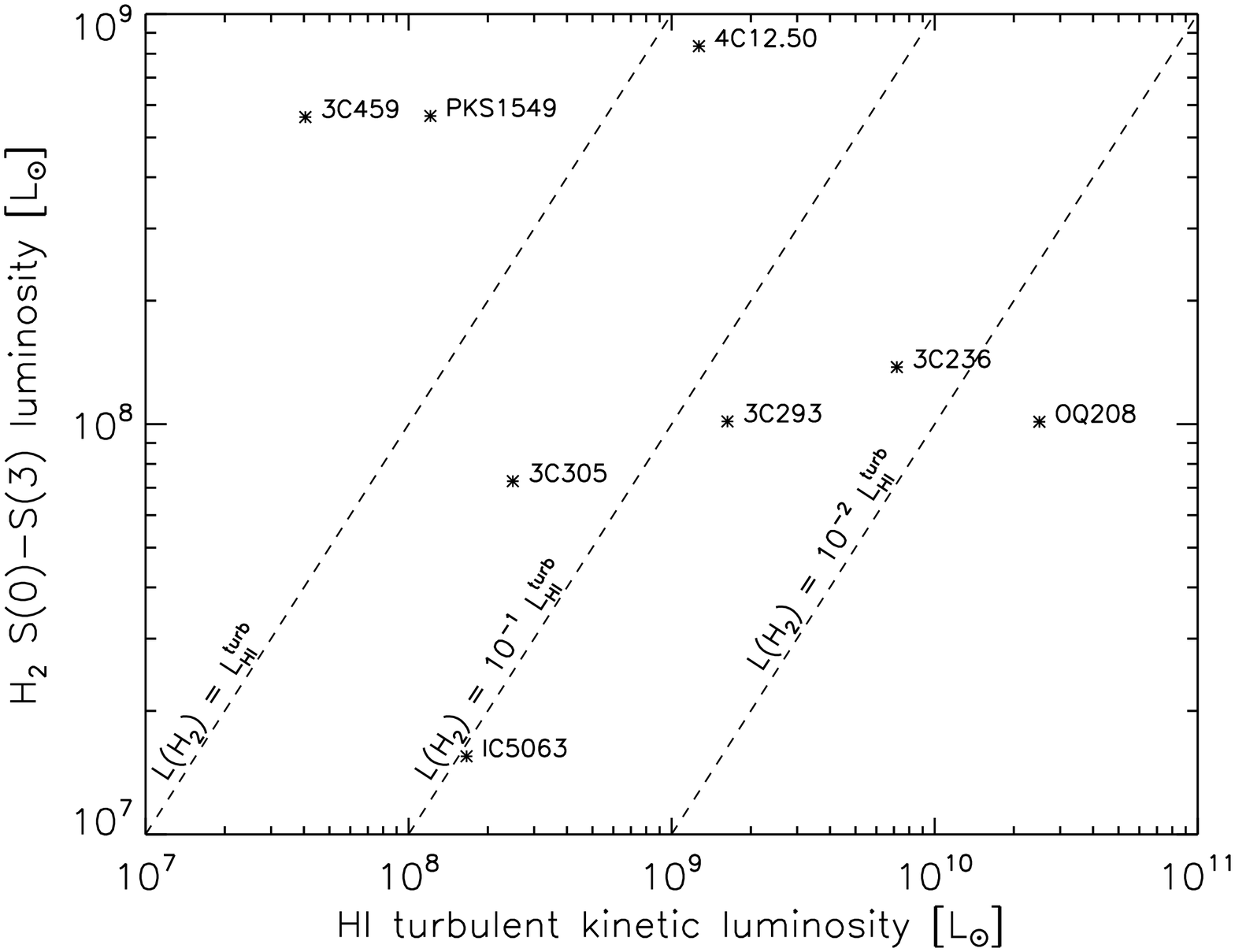}
    \includegraphics[width=\columnwidth]{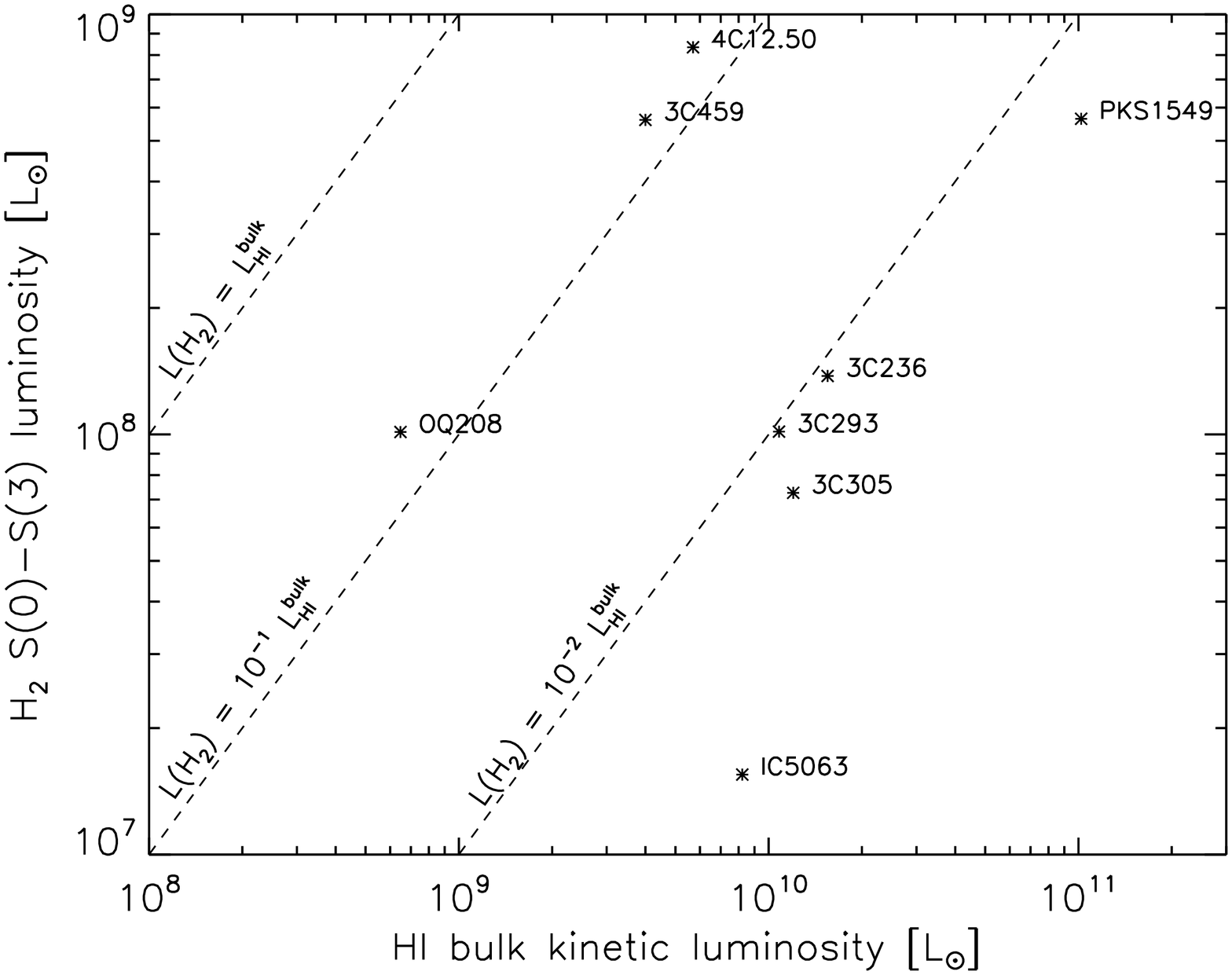}
      \caption{\textit{Top:} H$_2$ luminosity (summed over the S(0)-S(3) lines) against the H{\sc i} turbulent kinetic luminosity, derived from the linewidth of the broad H{\sc i} absorption component, associated with the outflowing gas. \textit{Bottom:} H$_2$ luminosity as a function of the bulk  kinetic luminosity of the H{\sc i} gas entrained in the flow.}
       \label{fig:LH2_HIkin}
   \end{figure}

Another source of heating for the warm H$_2$ gas is cosmic rays, as proposed by \citet{Ferland2008} to explain the mid-IR H$_2$ emission observed from the H$\alpha$-emitting filament in the Perseus~A cool-core cluster \citep{Johnstone2007}. Following, the discussion by \citet{Ogle2010}, we find the observed H$_2$ S(0)-S(3) line cooling  (i.e. the H$_2$ luminosity per H$_2$ molecule, using the modelled H$_2$ masses given in Table~\ref{table_H2params_shocks_nH1e3}) could be balanced by cosmic ray heating if the cosmic ray ionization rate (per H) is $\zeta_{\rm H} = 2 \times 10^{-14} - 3 \times 10^{-13}$~s$^{-1}$, which is $10^3 - 10^4$ times higher than the standard Galactic rate \citep[$\zeta_{\rm H} = 2.5 \times 10^{-17}$~s$^{-1}$,][]{Williams1998}. This would require a cosmic ray pressure $1-10$ times higher than the thermal pressure of the warm H$_2$ gas ($\approx 10^7$~K~cm$^{-3}$). Although we cannot exclude it for some of our sources, in the most extreme cases (3C~305, 3C~459, PKS~1549-79) that require the highest ionization rates, it is unlikely that such a departure from the equipartition of energy can be supported on galactic scales.  

\citet{Ogle2010} and \citet{Nesvadba2010} argued that the dissipation of the kinetic energy of the radio jet is the most probable source to power the H$_2$ emission in radio MOHEGs. The fact that \textit{all} the radio sources studied here have bright H$_2$ emission reinforces the idea that the radio jet, which drives the H{\sc i} outflow, is also responsible for the shock excitation of the warm H$_2$ gas. As in 3C~326 \citep{Ogle2007, Nesvadba2010}, all H{\sc i}-outflow radio MOHEGs in our sample have jet kinetic powers that exceed the observed H$_2$ luminosity (see Table~\ref{table_radio_prop}) and the H{\sc i} turbulent and bulk kinetic luminosities (see Table~\ref{table:HI_kinematics}). The jet provides  a huge reservoir of mechanical energy that can be dissipated over timescales of the order of $10^8$~yrs, perhaps exceeding the jet lifetime \citep{Nesvadba2010}. 
Since the cooling time of the warm H$_2$ gas ($\approx 10^4$~yr) is much shorter than the dynamical timescales  of the injection of mechanical 
energy ($\approx 10^7$~yr), the large masses of warm H$_2$ gas requires that the gas is repeatedly heated. \textit{Spitzer} H$_2$ observations thus imply an energy cascade converting bulk kinetic energy of the multi-phase ISM to turbulent motions 
of much smaller amplitude within molecular gas \citep{Guillard2009}. This is what motivated us to model the observed H$_2$ line fluxes with shock models (sect.~\ref{subsec:H2}). The turbulent heating of the molecular gas in powerful AGN is also suggested by observations of highly excited CO lines \citep[e.g.][]{Papadopoulos2010}.


\subsection{Efficiency of the transfer from the jet kinetic energy to the H{\sc i} and H$_2$ gas}
\label{subsec:nrj_transfer}

Two main mechanisms may be responsible for driving the observed outflow of H{\sc i} and ionized gas: the radiation pressure from the AGN (or the starburst if present), or the mechanical impact of the radio source. For the sources studied here, based on the spatial correspondence of the H{\sc i} broad absorption, optical outflowing component, and an off-nucleus bright spot in the radio jet for some of the sources, we favour the second mechanism. Outflows are most likely driven by the interaction between the expanding radio jets and the ionized/atomic gaseous medium enshrouding the central regions \citep[e.g.][]{Emonts2005}. This may also be the case in even more powerful, high-redshift quasars \citep{Nesvadba2011, Polletta2011} and high-redshift radio galaxies \citep{Humphrey2006}.

The details of the physics that govern the energy transfer from the radio jet to the molecular gas are not yet understood.  Being collimated, the jet itself cannot affect a significant fraction of the volume of the ISM of the host galaxy. Instead, the expanding cocoon of hot and tenuous gas inflated by the jet \citep[e.g.][]{Begelman1989} could transfer part of its kinetic energy to the molecular gas by driving shocks into the dense clouds or/and by turbulent mixing between hot and cold gas, similarly to the process proposed by \citet{Guillard2009} to explain the powerful H$_2$ emission from the Stephan's Quintet galaxy collision. 
It also possible that part of the bulk kinetic energy of the outflowing H{\sc i} gas is transferred to the molecular phase through shocks. The top (respectively bottom) panel of Figure~\ref{fig:LH2_HIkin} compares the H$_2$ luminosity to the turbulent (resp. bulk) H{\sc i} kinetic luminosity. In all sources, the bulk H{\sc i} kinetic luminosity exceeds by more than one order of magnitude the observed H$_2$ luminosities. The dissipation of a small fraction ($<10$\%) of this kinetic energy in the molecular gas could power the H$_2$ emission.

Numerical simulations of the impact of jets on galaxy evolution stress the importance of considering the multi-phase nature of the host galaxy ISM \citep[e.g.][]{Sutherland2007}. \citet{Wagner2011} showed that the efficiency of transfer of kinetic energy and momentum from the jet to the dense ($n_{\rm H} > 10^2$~cm$^{-3}$) gas can be high (10-70\%), with the momentum flux of the clouds exceeding that of the jet.
However, the mid-IR \textit{IRS} spectroscopy of the H$_2$ rotational lines presented in this paper indicates that most of the molecular gas does not share the kinematics of the outflowing atomic and ionized gas. 
If most of the H$_2$ gas was in the outflow, we would have observed blue-shifted H$_2$ lines, similar to the ionized gas lines. This suggests that the dynamical coupling between the molecular gas and the more tenuous outflowing gas is weak, probably because of the high density contrast between these phases. For density contrast of  $10^4$, it takes $\approx 4 \times 10^8$~yrs to accelerate a 10~pc cloud up to a velocity of 500~km~s$^{-1}$ \citep[e.g.][]{Klein1994}, which is likely to be longer than the jet lifetime. 
If some H$_2$ gas is effectively outflowing, the wind would remove this gas gradually over an AGN lifetime. It is very unlikely that the entire mass of the molecular disk would be entrained in the wind.
Thus the fraction of the molecular gas expelled from these galaxies could be small, which does not fit with the current assumptions made in the cosmological galaxy evolution models that assumes that AGN feedback sweep up most of the gas during the epochs of strong star-formation \citep[e.g.][]{Hopkins2006, Narayanan2008}.

The current numerical simulations show a very efficient transfer of momentum essentially because the dense ($n_{\rm H} > 1\,$cm$ ^{-3}$, $T=10^4\,$K) gas cools very fast. These calculations do not include the turbulent cascade and dissipation of the kinetic energy due to the supersonic turbulence within the dense gas phase, which would make the gas cooling longer, and therefore the dynamical coupling less efficient. 
The dynamical coupling would be much more efficient if a significant fraction of the warm H$_2$ gas is formed \textit{in situ}, i.e. in the outflow, from shocked H{\sc i} gas. On the contrary, our data suggest that the bulk of the warm H$_2$ mass is difficult to entrain in the outflow, and is rather perhaps settled in the galactic disk. A similar conclusion is reached based on near-IR IFU observations of nearby Seyfert galaxies \citep{Storchi-Bergmann2009, Riffel2011}, where the H$_2$ gas is concentrated in the galactic plane (with a velocity structure consistent with a rotating disk), whereas the ionized gas is mostly distributed along the outflowing cone. 

On the other hand, in some of the sources studied here, the H$_2$ lines are broad, showing that the H$_2$ gas is very turbulent. 
This suggests that the molecular gas is heated by the dissipation of supersonic turbulence, which  can prevent it from being gravitationally bound (and thus from forming stars), rather than expelled from the galaxy. 
In some cases, these turbulent motions are unlikely to be rotation-supported. Assuming the galaxies are pressure-supported with an isothermal mass profile, we derived the escape velocities, ${\rm v} _{\rm esc}$, from the stellar masses: ${\rm v} _{\rm esc} = \sqrt{2 \, G \ M_{\star} / (5\, r_e)}$. We assume an effective radius of $r_e = 2\,$kpc. We find escape velocities that range from 70~km~s$ ^{-1} $ (OQ~208) to 330~km~s$ ^{-1} $ (4C~12.50).
In all of the sources, the observed H$_2$ velocity dispersions (up to $\approx 300$~km~s$^{-1}$, broader than that observed in nearby Seyferts) are comparable to their escape velocities, except for IC~5063, where $\sigma (\rm H_2) \approx 3 \times v _{\rm esc}$. Thus, a fraction of the warm H$_2$ gas could become unbound and escape from these galaxies because of its high velocity dispersion. However, the fraction of the H$_2$ S(1) line emission having velocities larger than the escape velocity is small ($\lesssim10\,$\%), so it is unlikely that this process would strongly affect and empty the reservoir of molecular gas in these galaxies. 

We are complementing the observations presented here with higher resolution spectroscopy of near-IR H$_2$ lines, interferometric CO line observations, and \textit{Herschel} spectroscopy and photometry, to complete our census of the physical and dynamical state of the molecular gas in these sources. For extended objects, the use of near-IR IFUs is necessary. Hopefully, JWST/MIRI  will allow us to probe mid-IR H$_2$ lines at a much better sensitivity and $5\times$ better spectral resolution than the \textit{IRS}.

\section{Conclusions and final remarks}
\label{conclusion}

We present mid-IR \textit{Spitzer} spectroscopy of 8 nearby radio-galaxies where fast outflows of ionized and atomic gas were detected by previous optical and radio observations. Our main results and conclusions are the following:
\begin{itemize}
\item In all of the sources, we detect high equivalent width H$_2$ line emission from warm (100-5000~K) molecular gas, implying warm H$_2$ masses ranging from $10^8$ to $10^{10}$~M$_{\odot}$. The observed H$_2$-to-PAH luminosity ratios are all above the values predicted from photo-ionization models, and we suggest that the H$_2$ emission is associated with the dissipation of a fraction of the kinetic energy provided by the radio jet.

\item In 5 sources  (3C 236, 3C 293, 3C 459, 4C 12.50 and PKS~1549-79), we securely detect blue-shifted wings (up to 3000~km~s$^{-1}$) on the [Ne{\sc ii}] (and [Ne{\sc iii}] for the highest SNR spectra) that matches remarkably well with the blue-shifted, broad H{\sc i} absorption associated with the outflow. 

\item 
All but one of these sources have resolved and very broad mid-IR rotational H$_2$ lines with FWHM$ \gtrsim 500$~km~s$^{-1}$ (and up to 900~km~s$^{-1}$), compared to only 2\% overall amongst the 298 AGN with such spectra in the \textit{Spitzer} archive \citep{Dasyra2011}. This suggests that the efficiency of kinetic energy deposition into the molecular ISM is higher in radio jet sources than in other types of AGN. 

\item
The kinematics of the warm H$_2$ gas do not follow that of the ionized or H{\sc i} gas. The rotational  H$_2$ lines are systemically narrower than the mid-IR ionized gas lines, and do not exhibit asymmetric profiles with blue-shifted wings (except perhaps tentative detections in three targets, 4C~12.50, 3C~459, and PKS~1549-79). We conclude that, although very turbulent, the bulk of the  warm H$_2$ mass 
is not entrained in the wind.

\item We show that UV, X-ray and cosmic ray heating are unlikely to be the dominant source of H$_2$ excitation. We argue that the dissipation (via supersonic turbulence) of a small fraction ($< 10\%$) of the mechanical energy provided by the radio jet can power the observed emission. 

\end{itemize}

\acknowledgements

This work is based primarily on observations made with the \textit{Spitzer Space Telescope}, which is operated
by the Jet Propulsion Laboratory, California Institute of Technology under a contract
with NASA. 

This research has made use of the NASA/IPAC Extragalactic Database (NED)
which is operated by the Jet Propulsion Laboratory, California Institute of Technology, under
contract with the National Aeronautics and Space Administration.

\bibliographystyle{apj}
\bibliography{Guillard_HIout_RGs.bbl}

\appendix

\begin{figure*}
   \centering
    \includegraphics[height=0.49\textwidth, angle=90]{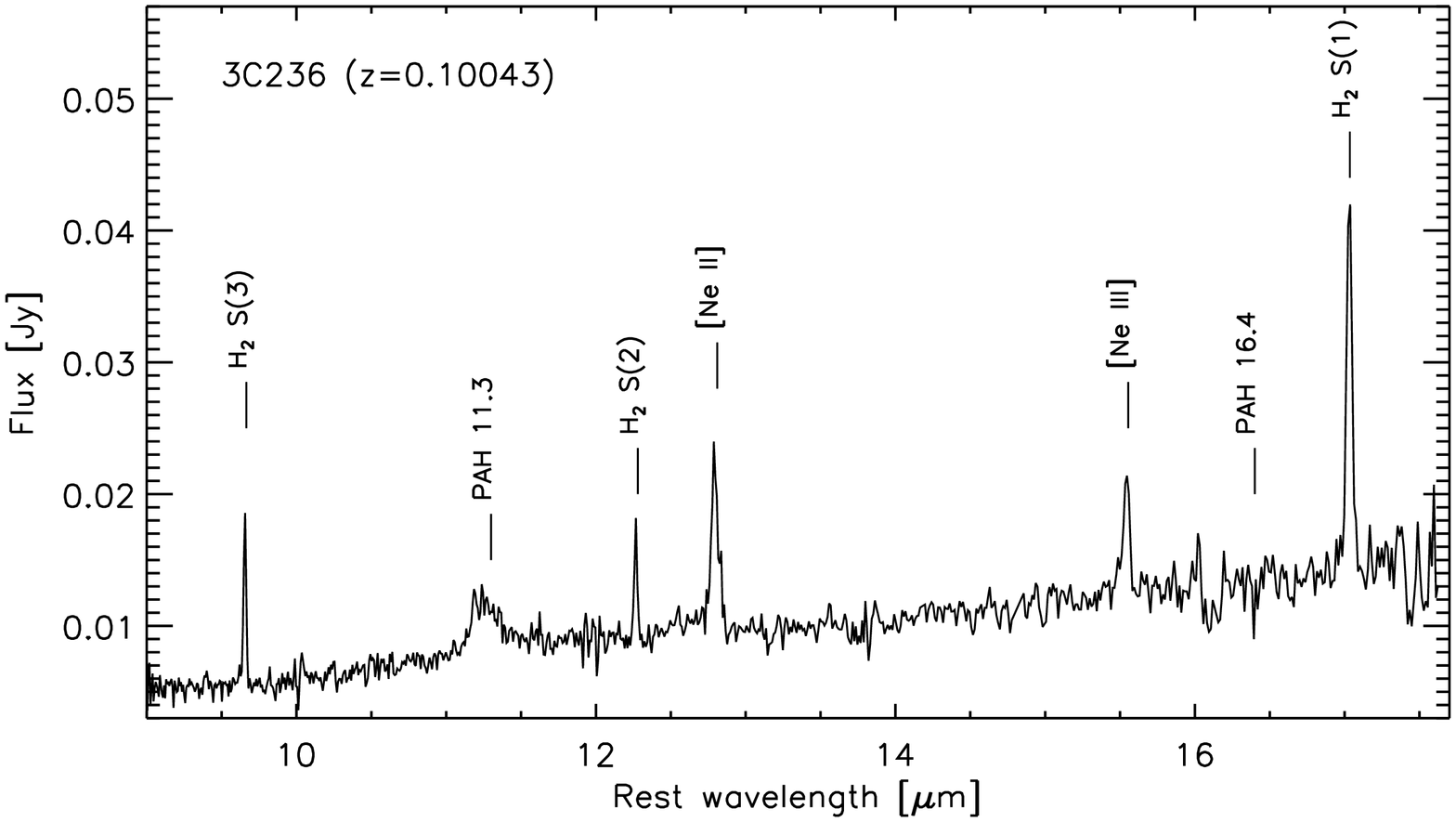}
    \includegraphics[height=0.49\textwidth, angle=90]{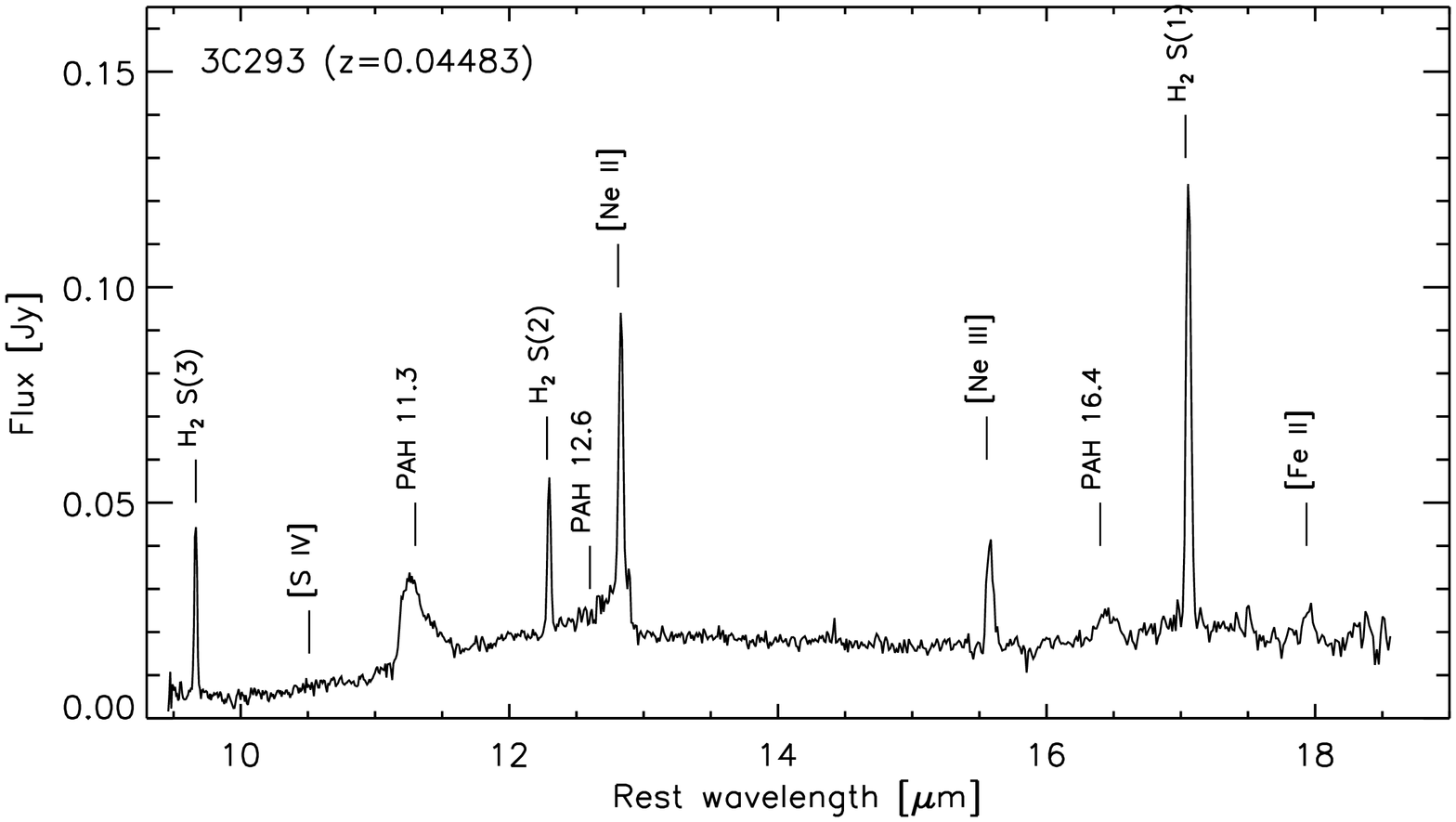}
    \includegraphics[height=0.49\textwidth, angle=90]{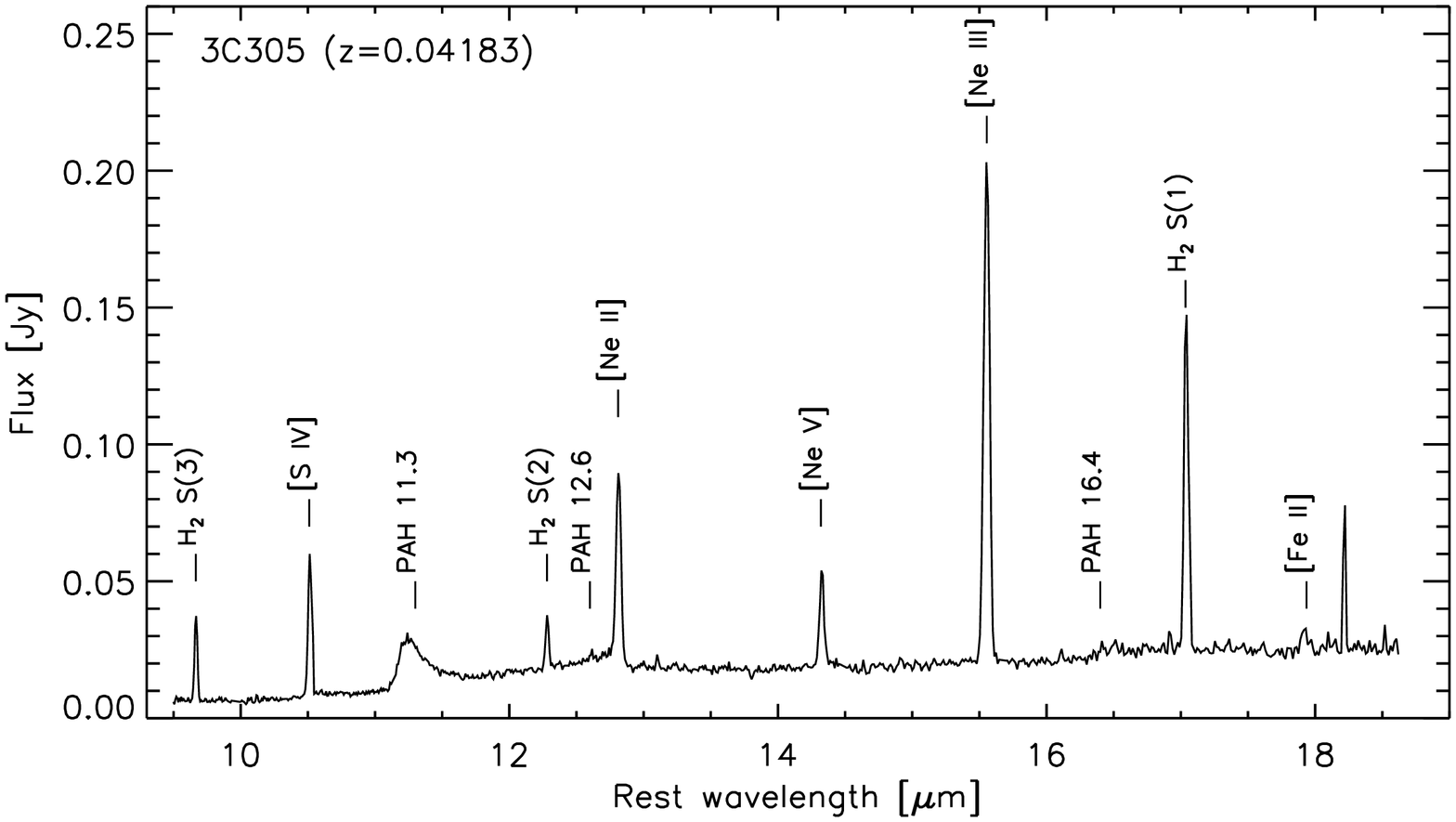}
    \includegraphics[height=0.49\textwidth, angle=90]{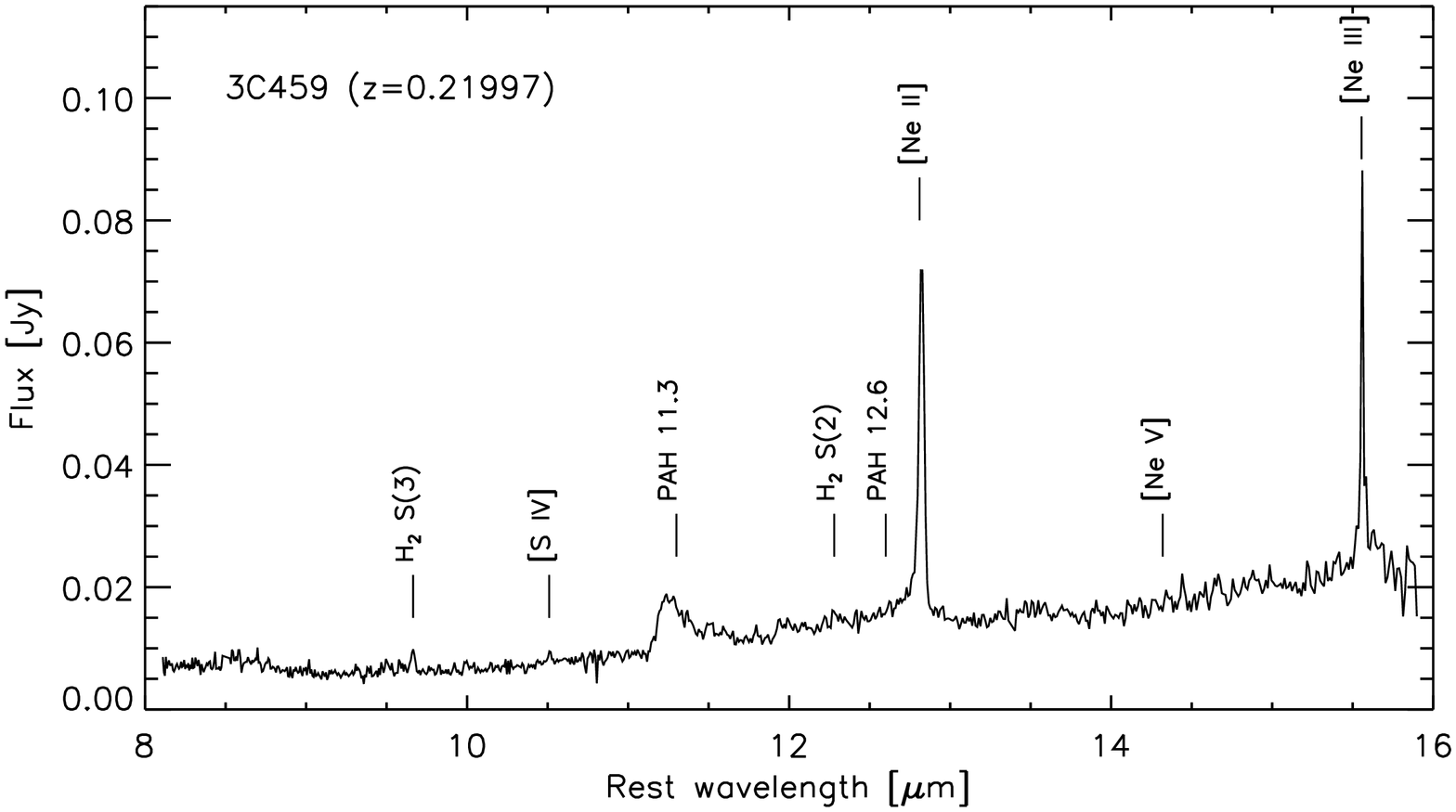}
    \includegraphics[height=0.49\textwidth, angle=90]{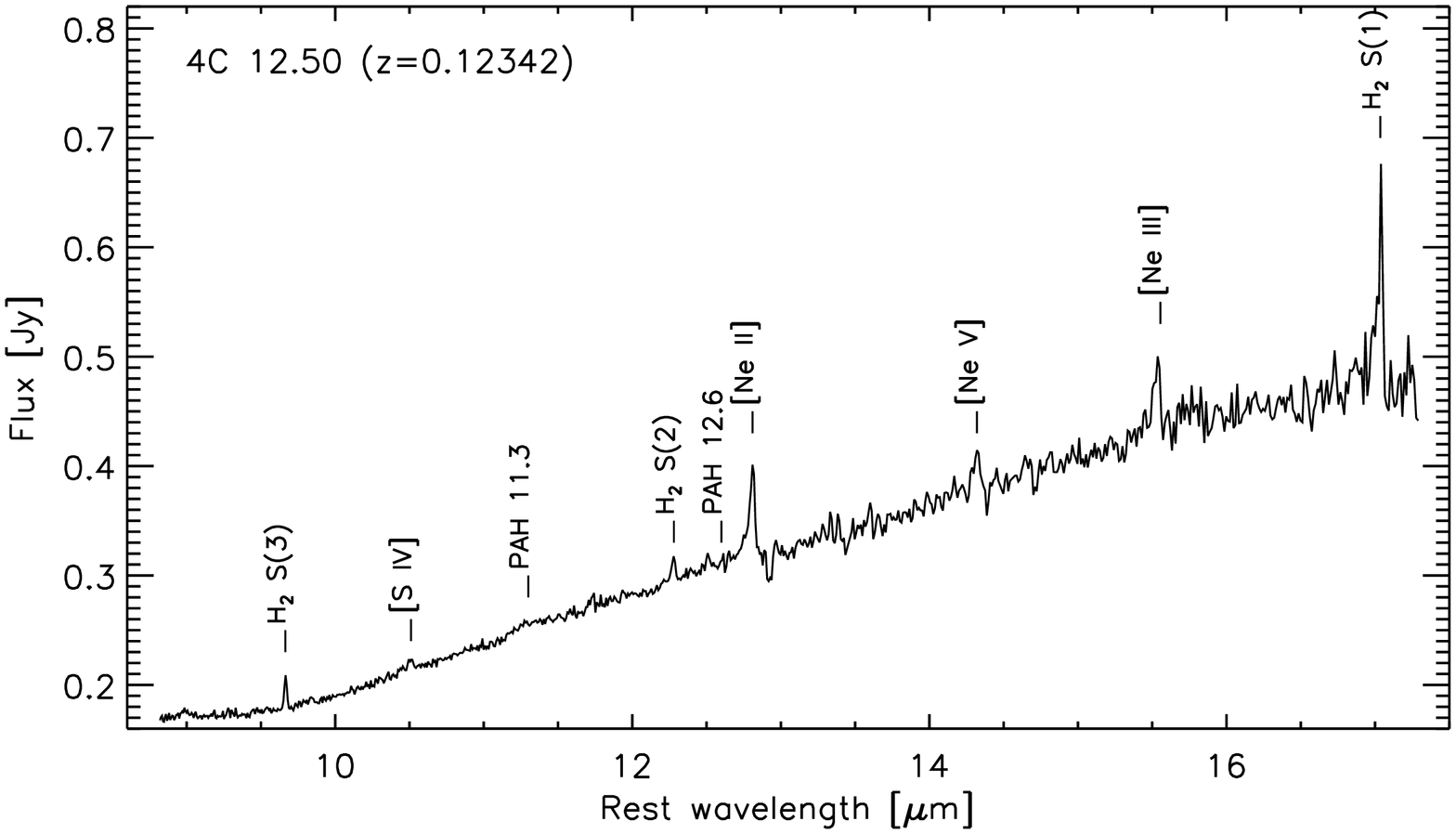}
    \includegraphics[height=0.49\textwidth, angle=90]{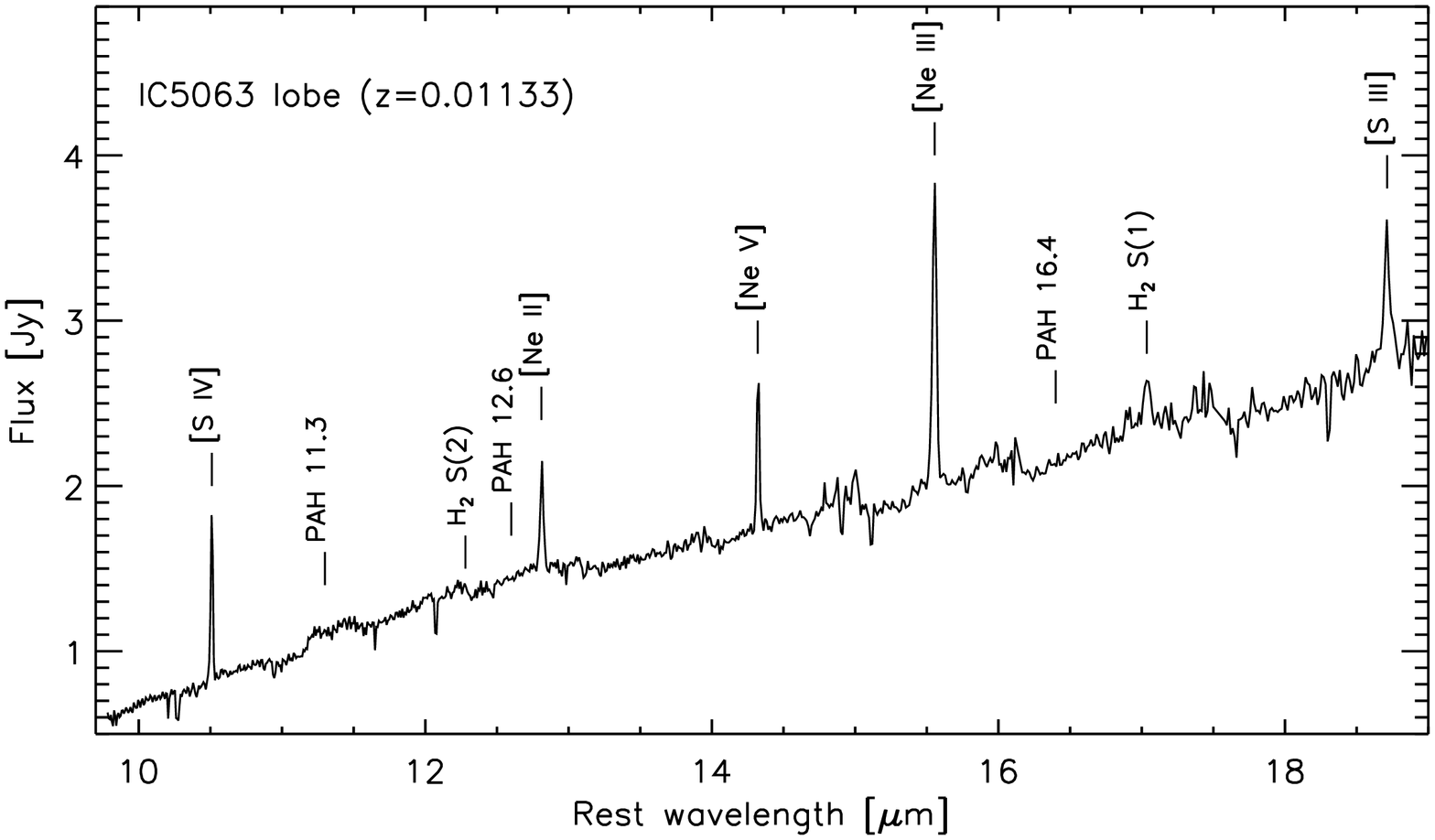}
    \includegraphics[height=0.49\textwidth, angle=90]{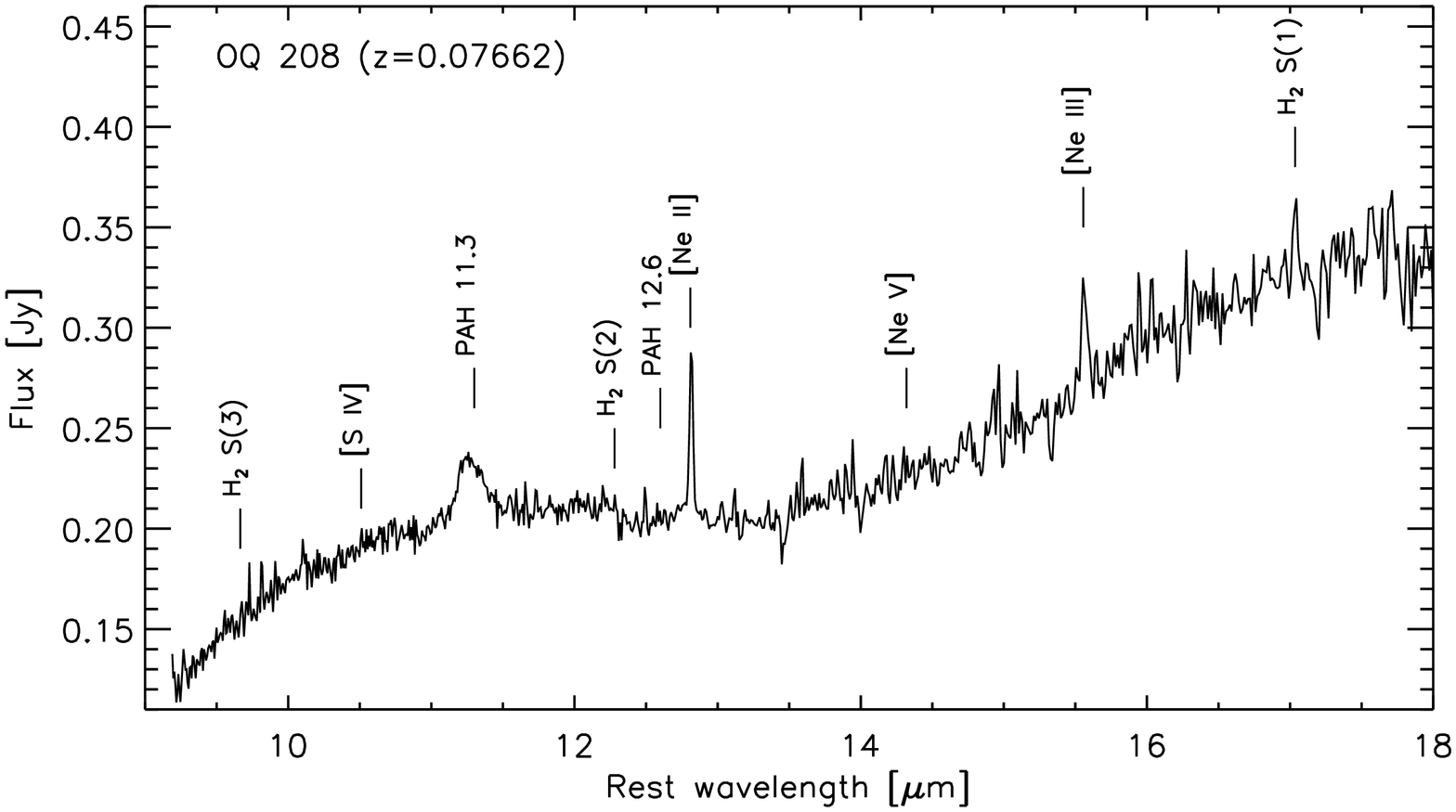}
    \includegraphics[height=0.49\textwidth, angle=90]{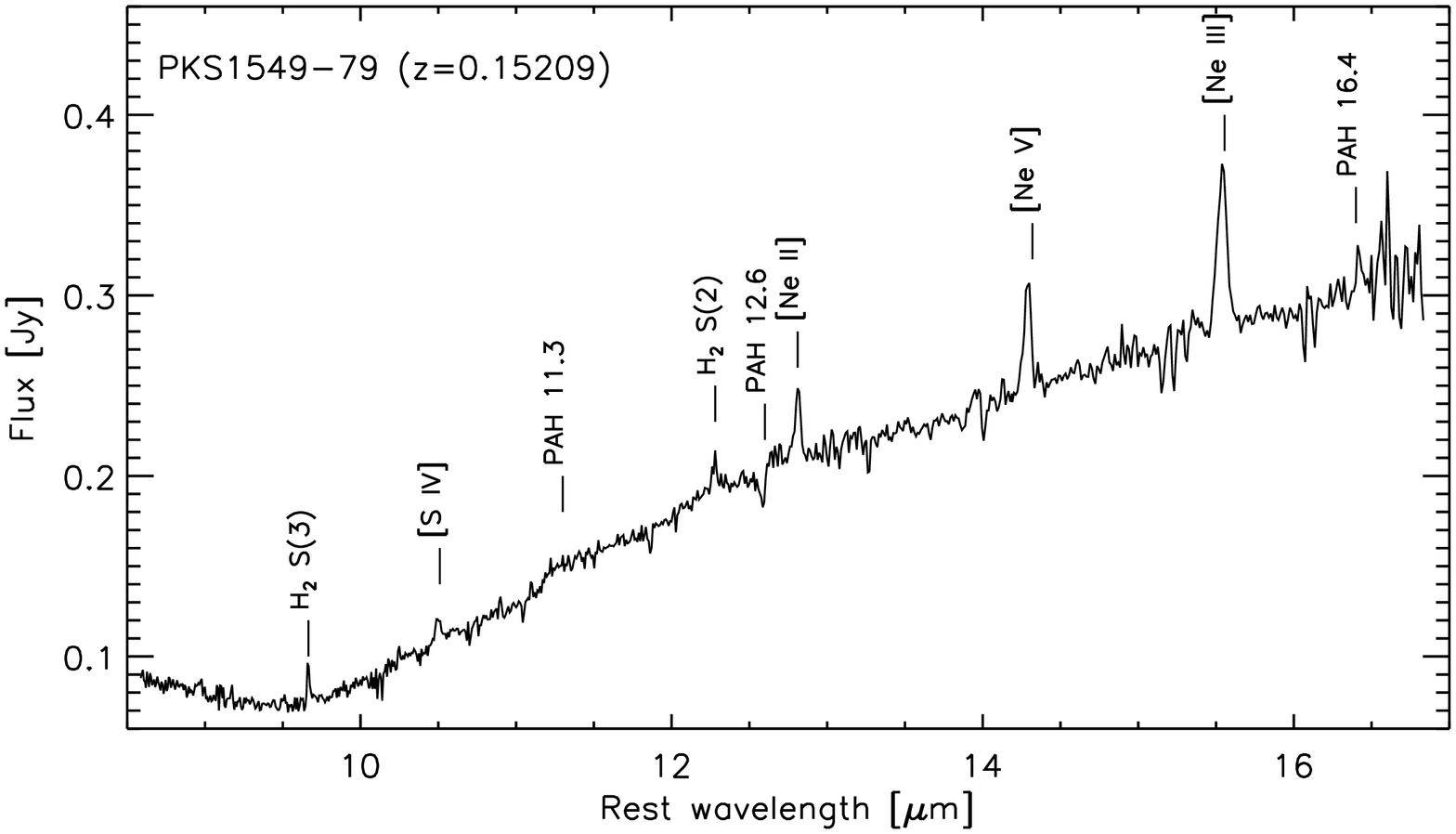}
      \caption{Spitzer IRS high-resolution spectra of the H{\sc i}-outflow radio galaxies, zoomed on the wavelength coverage of the SH (Short High) module.  }
       \label{fig:SH_spectra}
   \end{figure*}


\begin{figure*}
   \centering

   \framebox{\includegraphics[height=0.48\textwidth, angle=90]{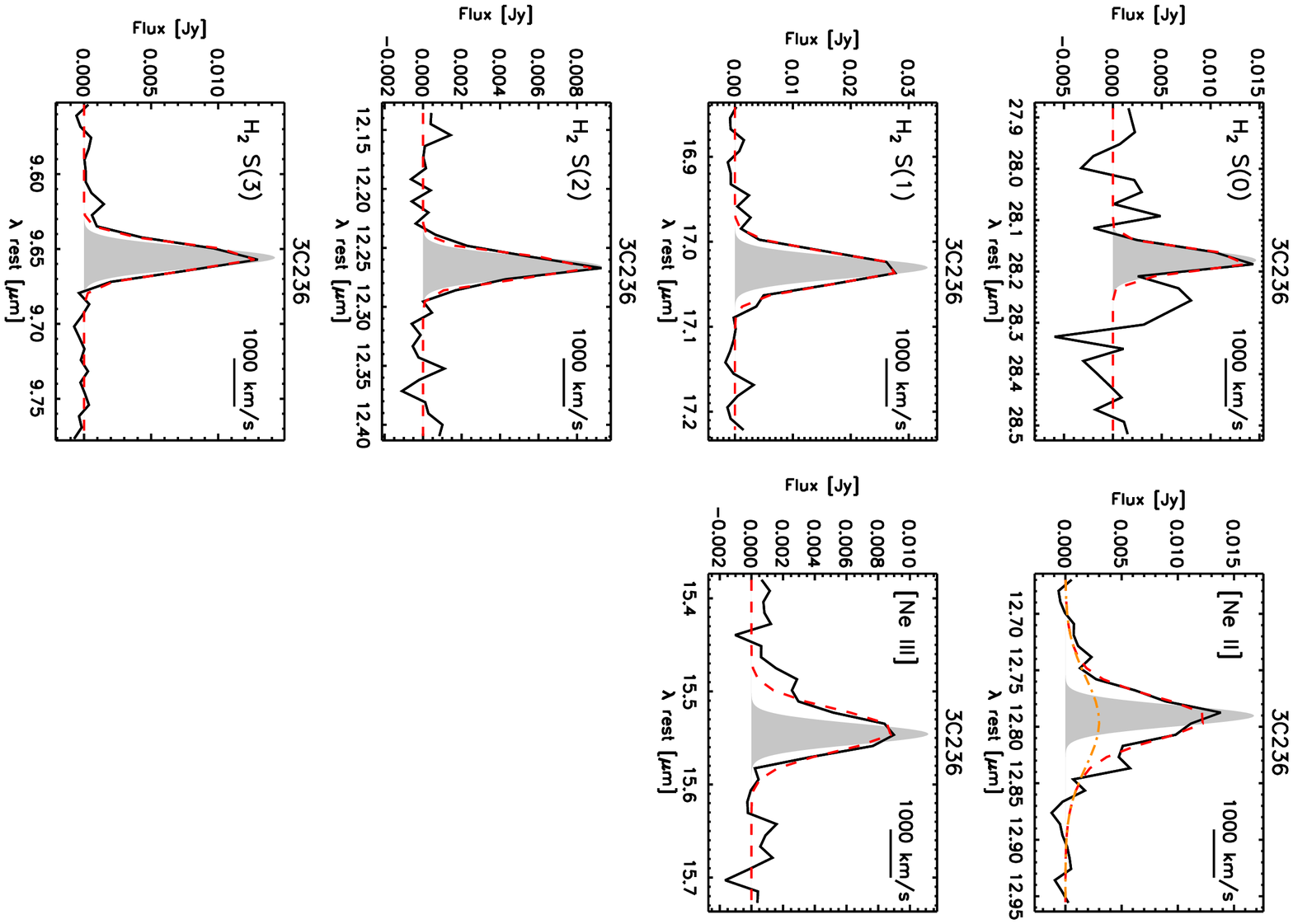}}
    \framebox{\includegraphics[height=0.48\textwidth, angle=90]{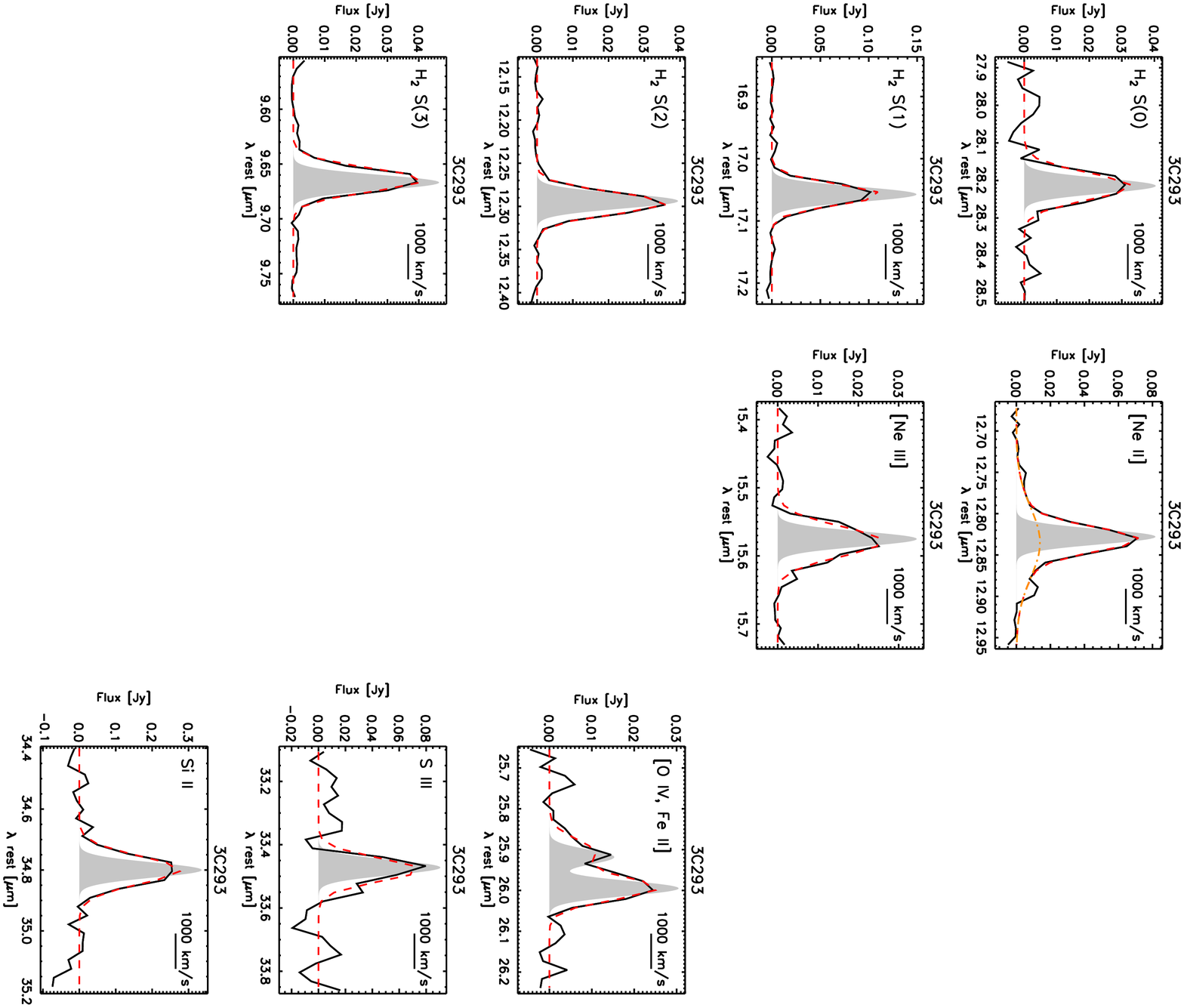}}
    \framebox{\includegraphics[height=0.48\textwidth, angle=90]{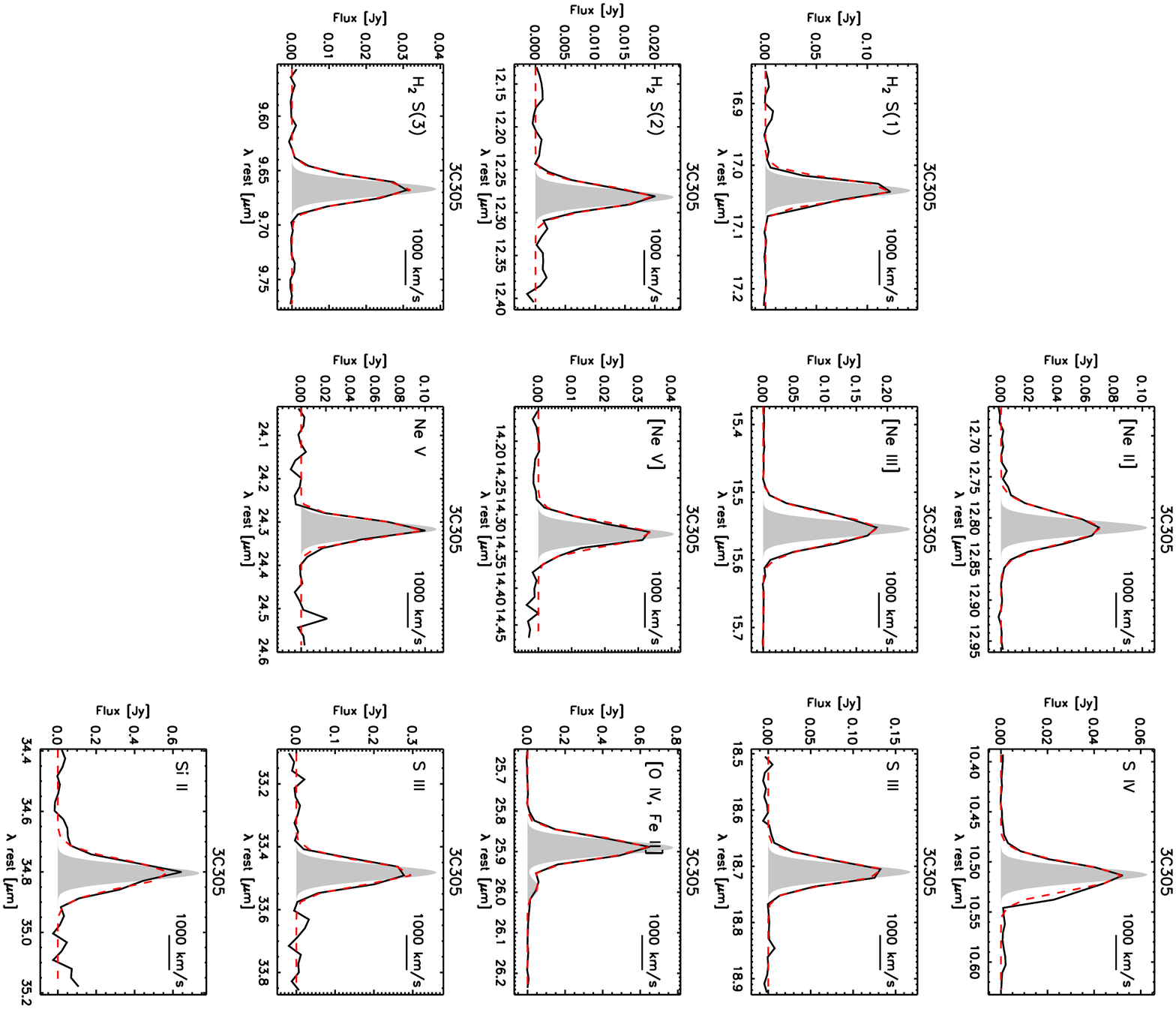}}
    \framebox{\includegraphics[height=0.48\textwidth, angle=90]{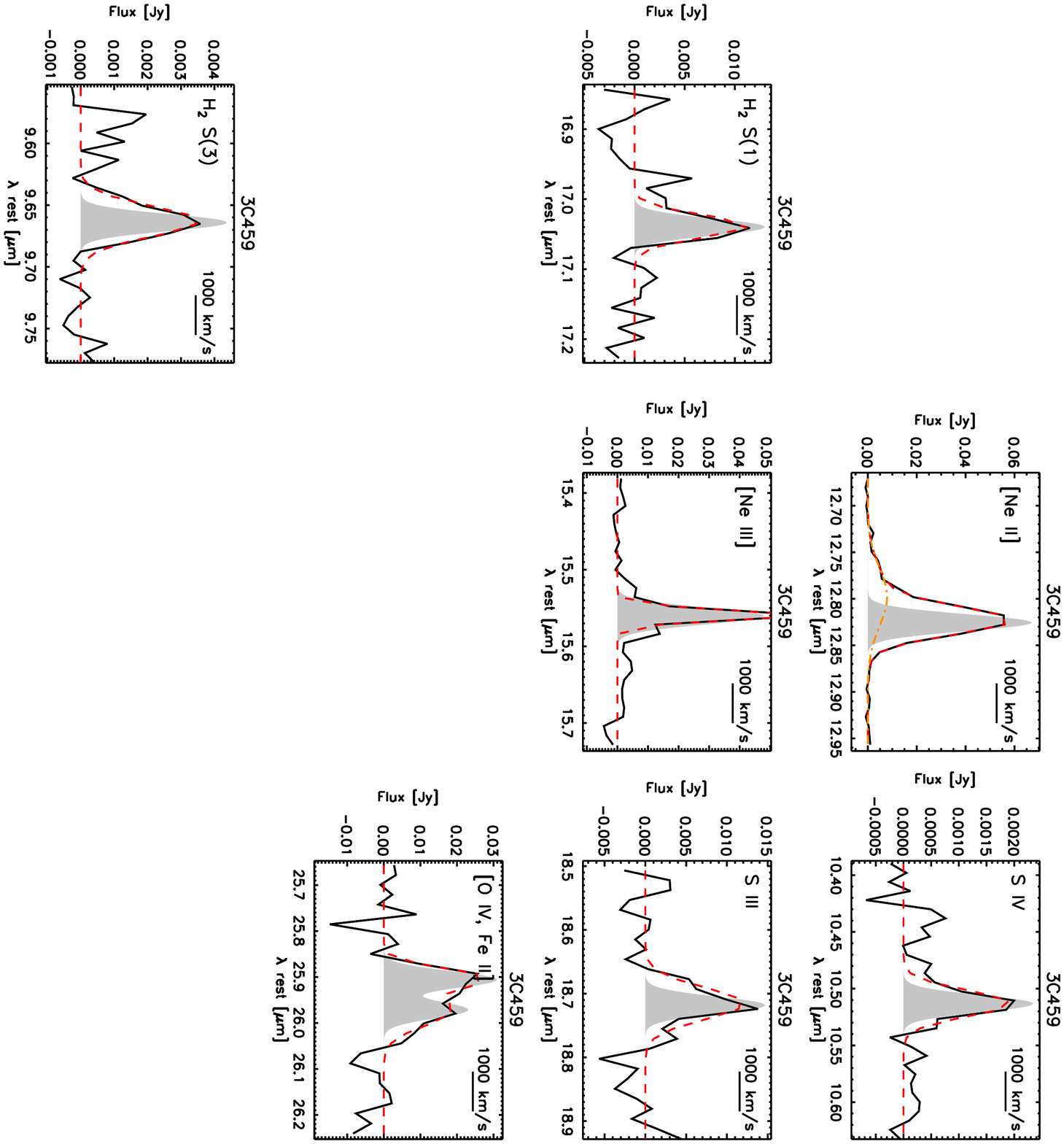}} 
            \caption{Detected spectral emission lines. The grey area is the \textit{Spitzer} IRS instrumental profile (a Gaussian with a FWHM corresponding to the resolution of the IRS), and the red dashed line is the result of the Gaussian fitting of the line. The [O{\sc iv}, Fe{\sc ii}] blend profile has been fitted with two Gaussian components when both lines are detected. For lines showing wings, the broad component of the Gaussian decomposition is overplotted (orange dashed-dotted line).}
       \label{fig:lines_1}
   \end{figure*}

    \begin{figure*}
   \centering
        \framebox{\includegraphics[height=0.48\textwidth, angle=90]{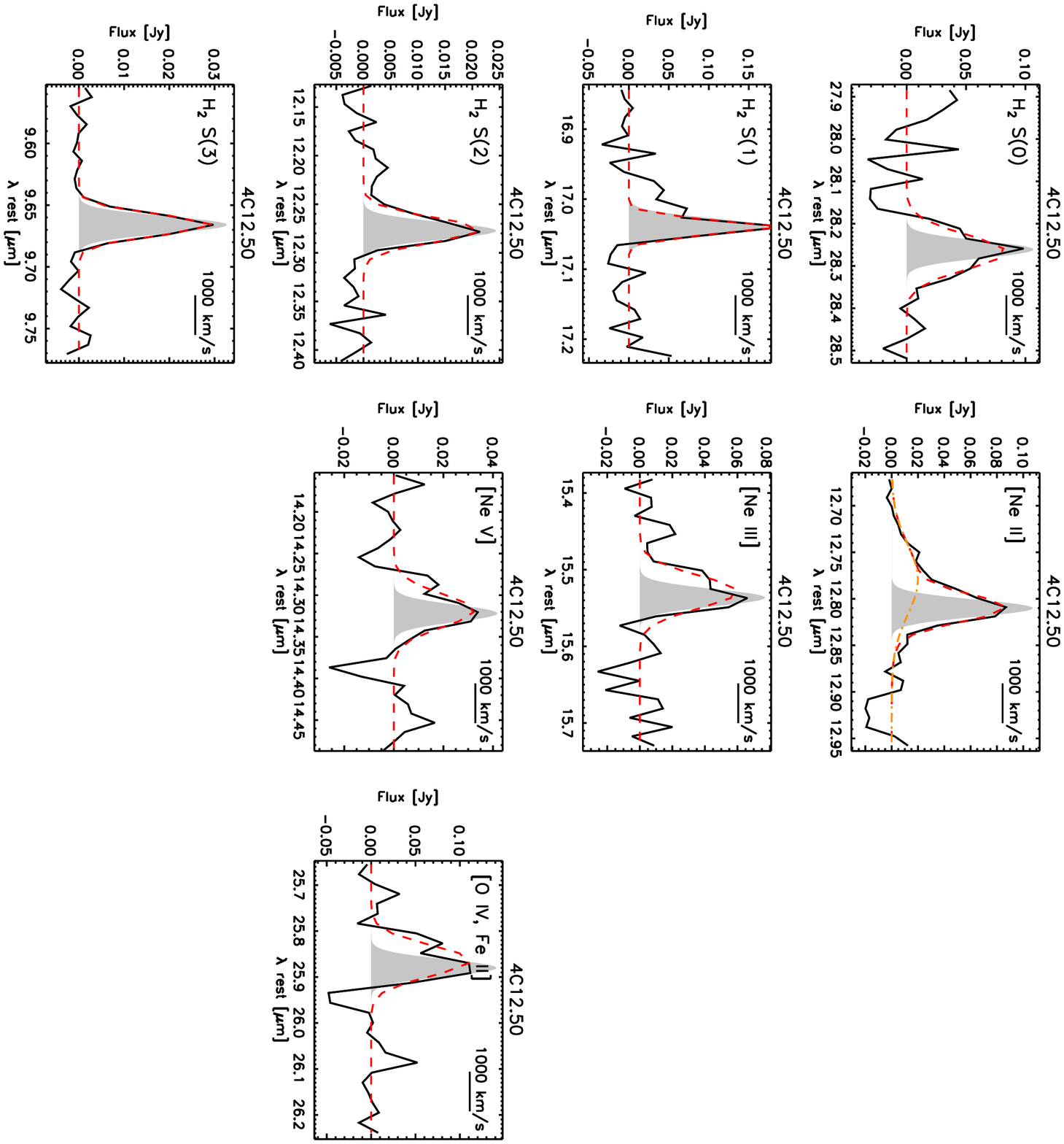}}
    \framebox{\includegraphics[height=0.48\textwidth, angle=90]{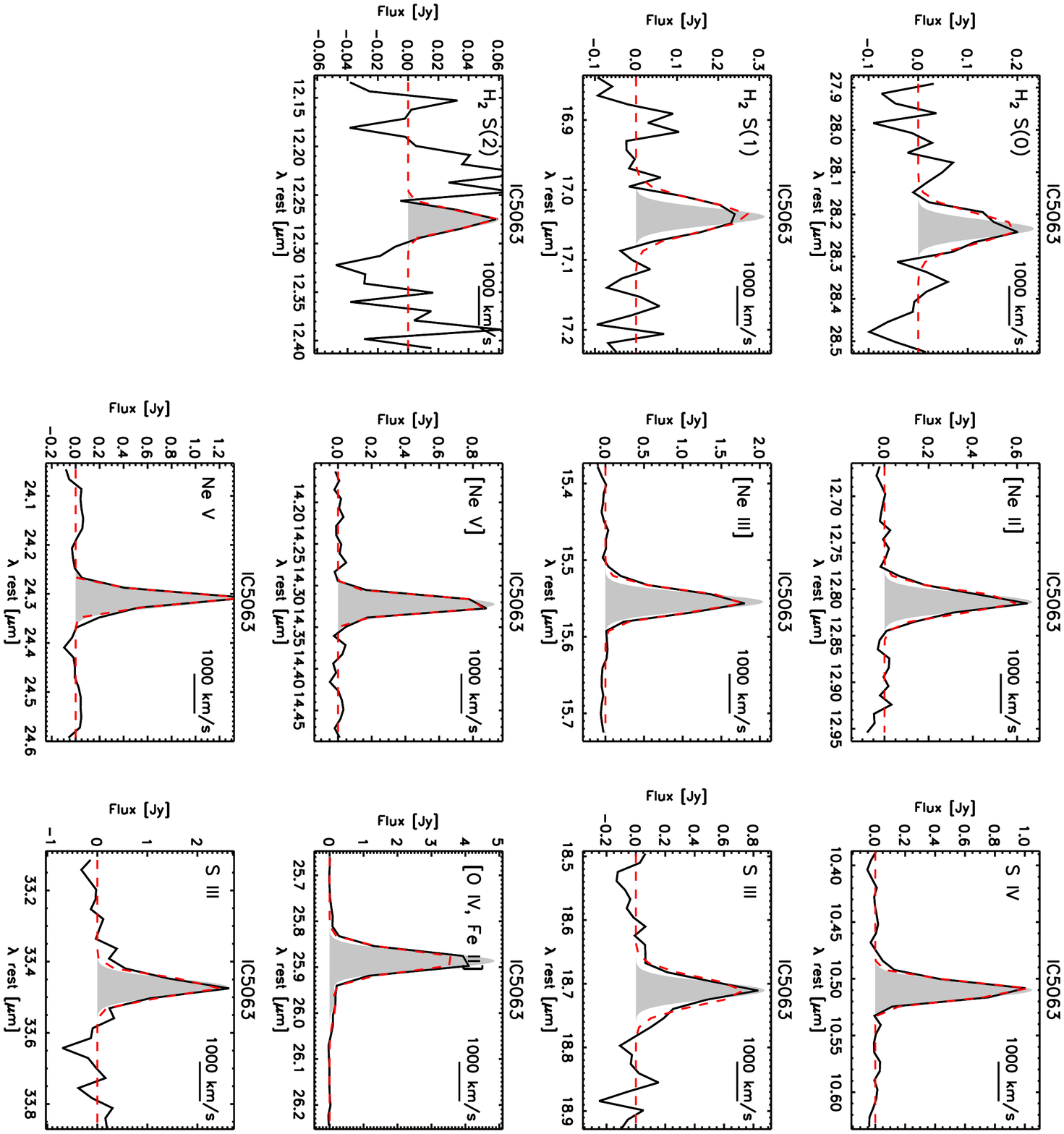}}
    \framebox{\includegraphics[height=0.48\textwidth, angle=90]{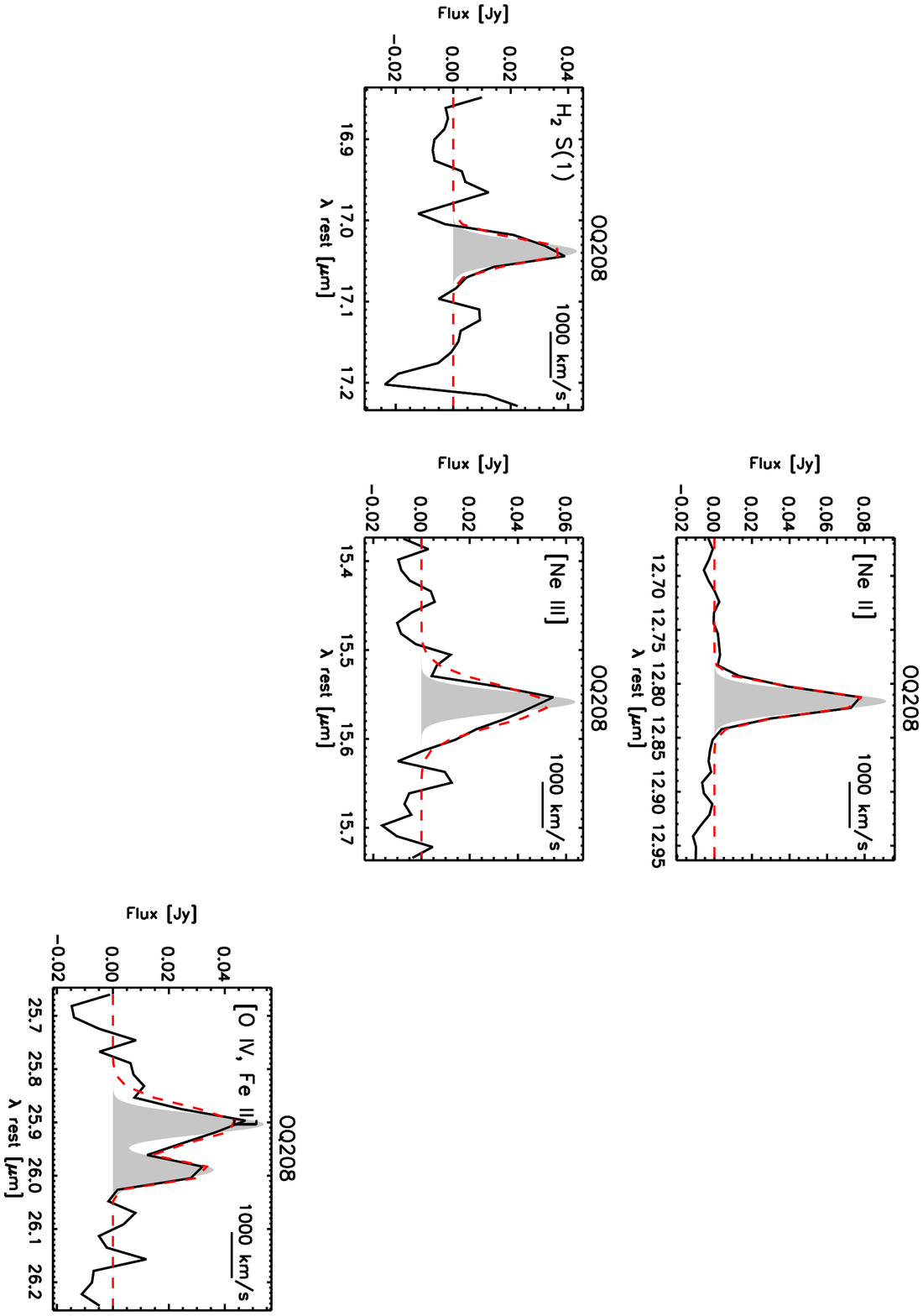}}
    \framebox{\includegraphics[height=0.48\textwidth, angle=90]{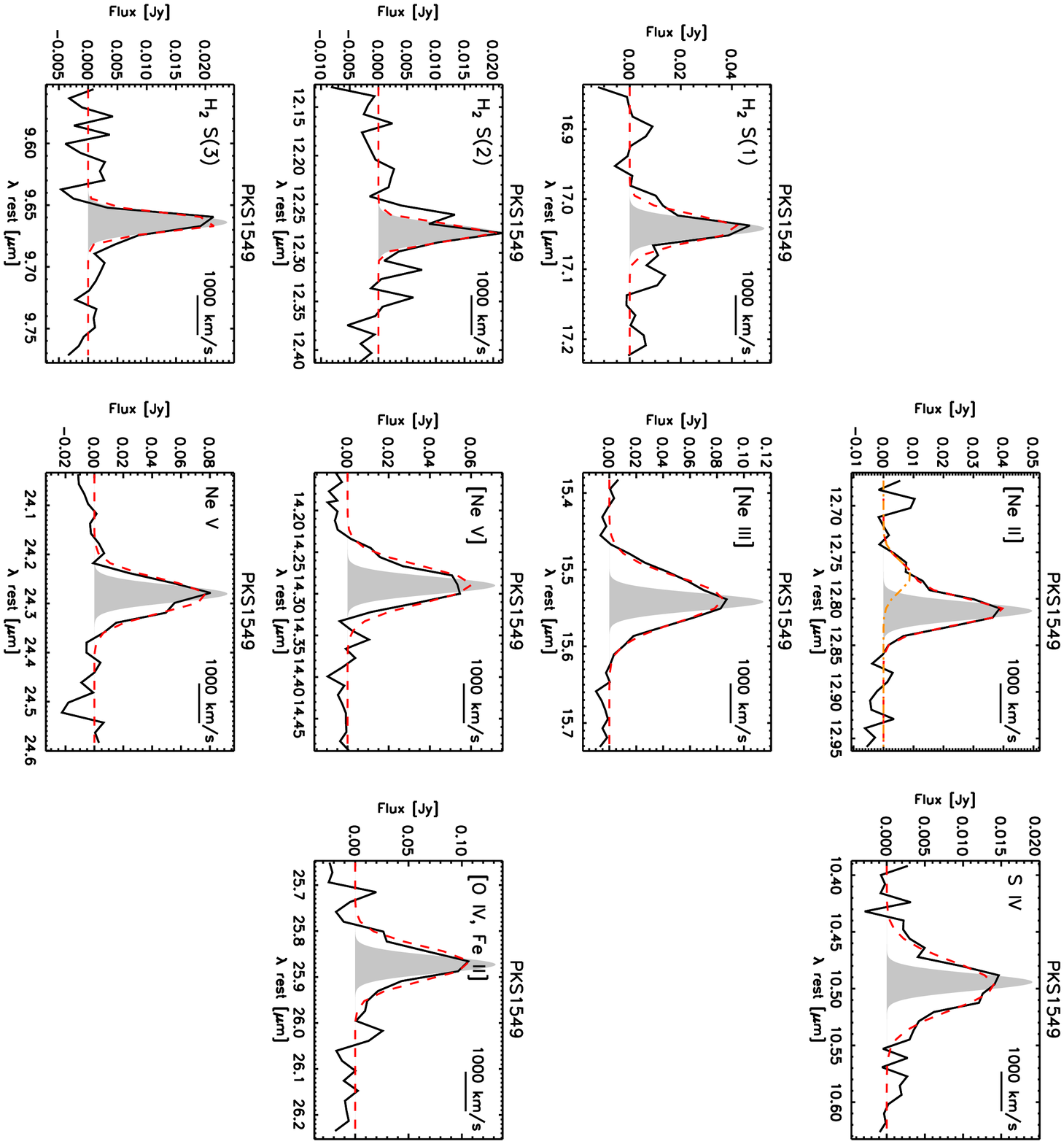}} 
      \caption{Detected spectral emission lines. The grey area is the \textit{Spitzer} IRS instrumental profile (a Gaussian with a FWHM corresponding to the resolution of the IRS), and the red dashed line is the result of the Gaussian fitting of the line. The [O{\sc iv}, Fe{\sc ii}] blend profile has been fitted with two Gaussian components when both lines are detected. For lines showing wings, the broad component of the Gaussian decomposition is overplotted (orange dashed-dotted line).}
       \label{fig:lines_2}
   \end{figure*}


\begin{figure*}
   \centering
    \includegraphics[width=\textwidth]{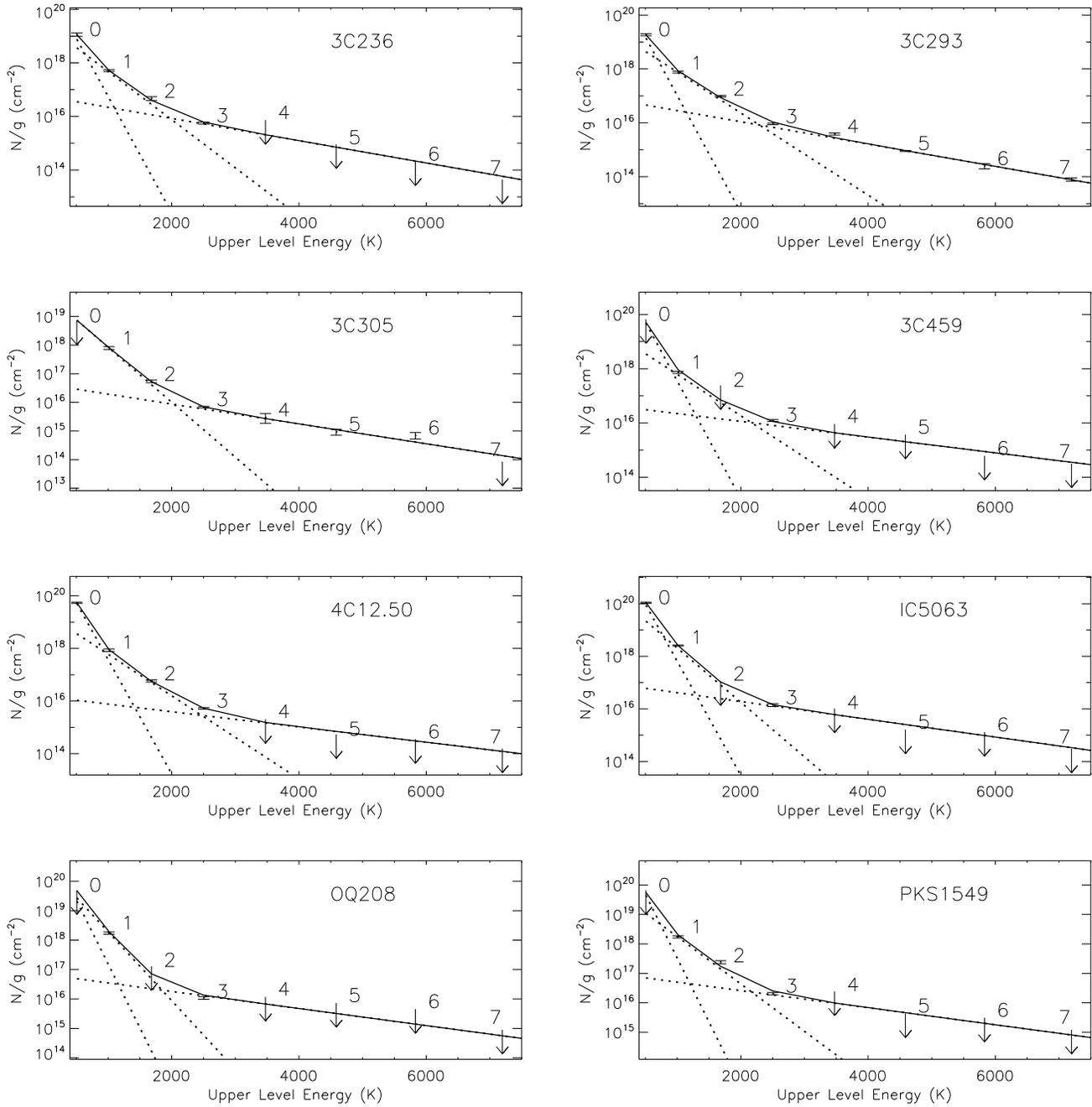}
      \caption{H$_2$ excitation diagrams with two or three temperature component fits overlaid. The H$_2$ temperatures, column densities, and masses are listed in Table~\ref{table_H2params_3T}.}
       \label{fig:H2exc_Tfit}
   \end{figure*}


\begin{figure*}
   \centering
    \includegraphics[width=0.45\textwidth]{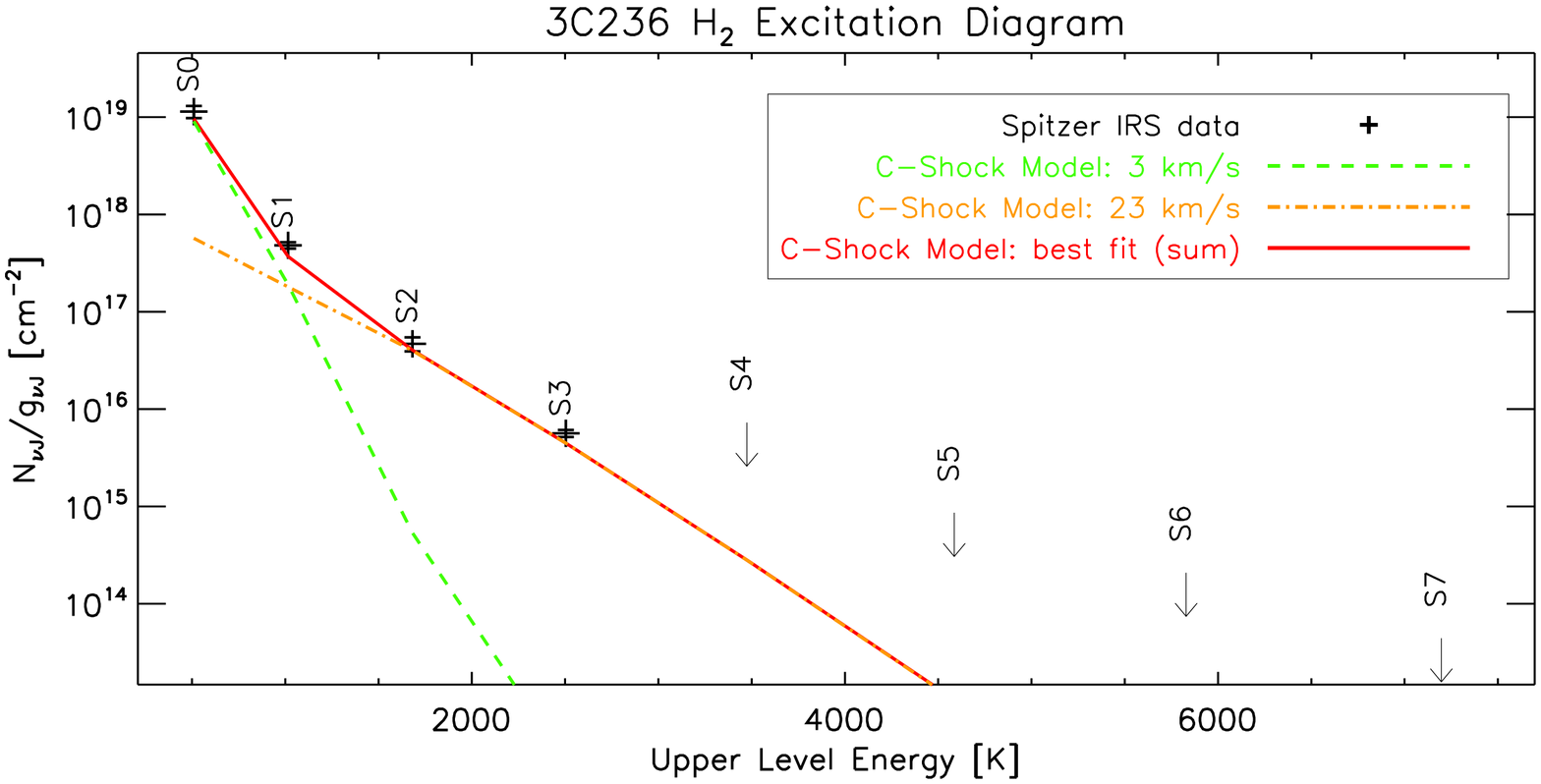}
    \includegraphics[width=0.45\textwidth]{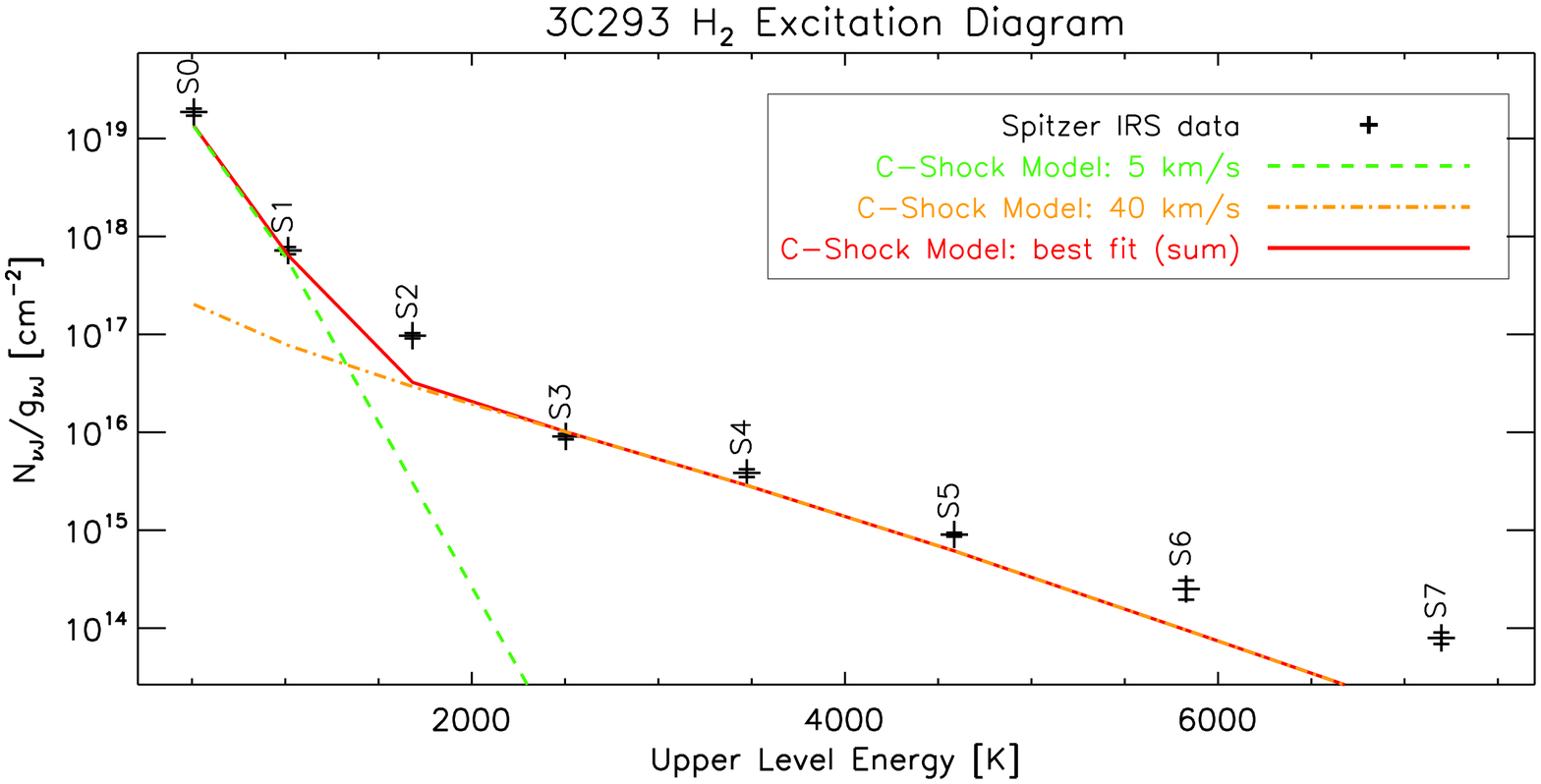}
    \includegraphics[width=0.45\textwidth]{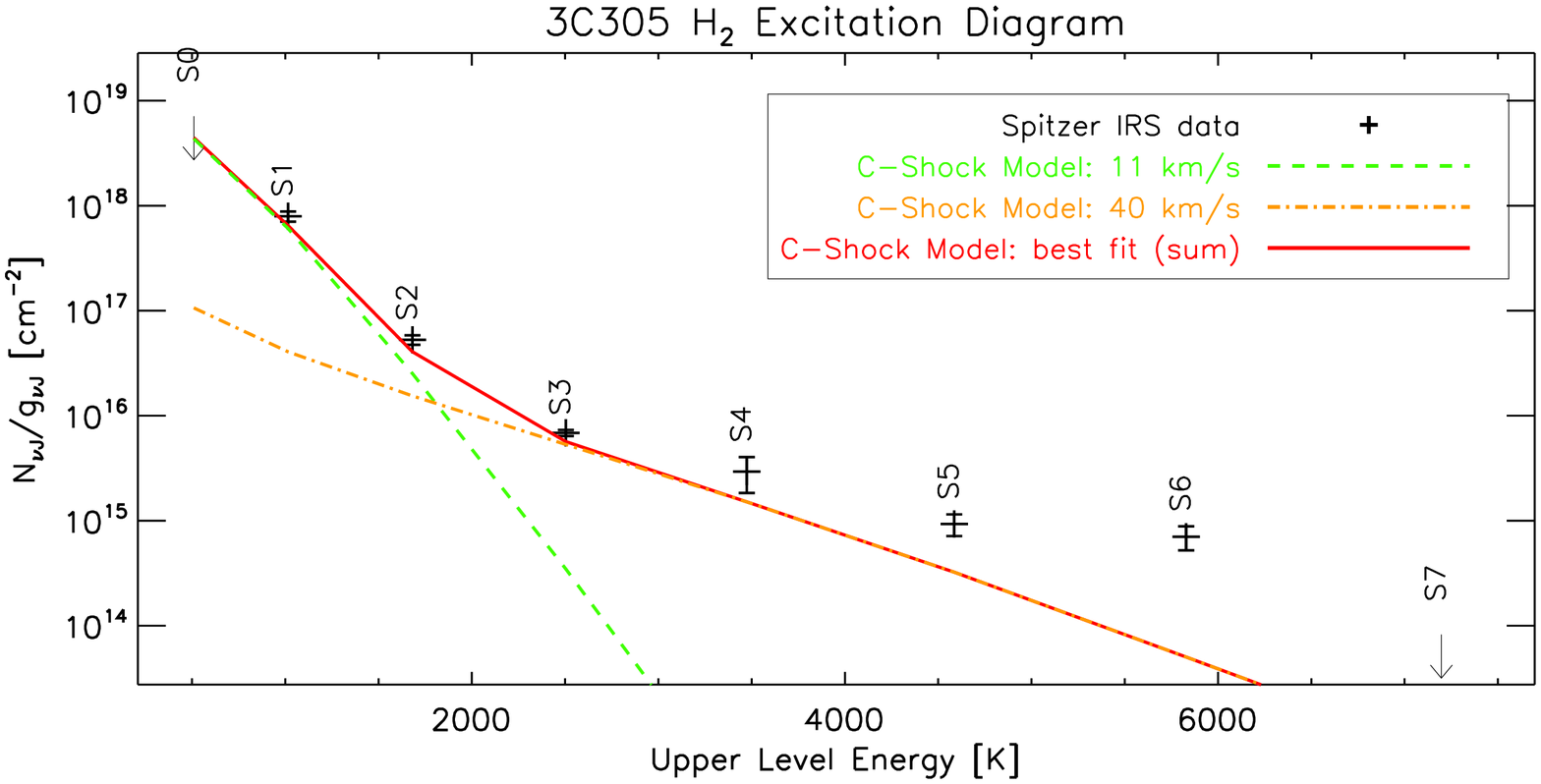}
    \includegraphics[width=0.45\textwidth]{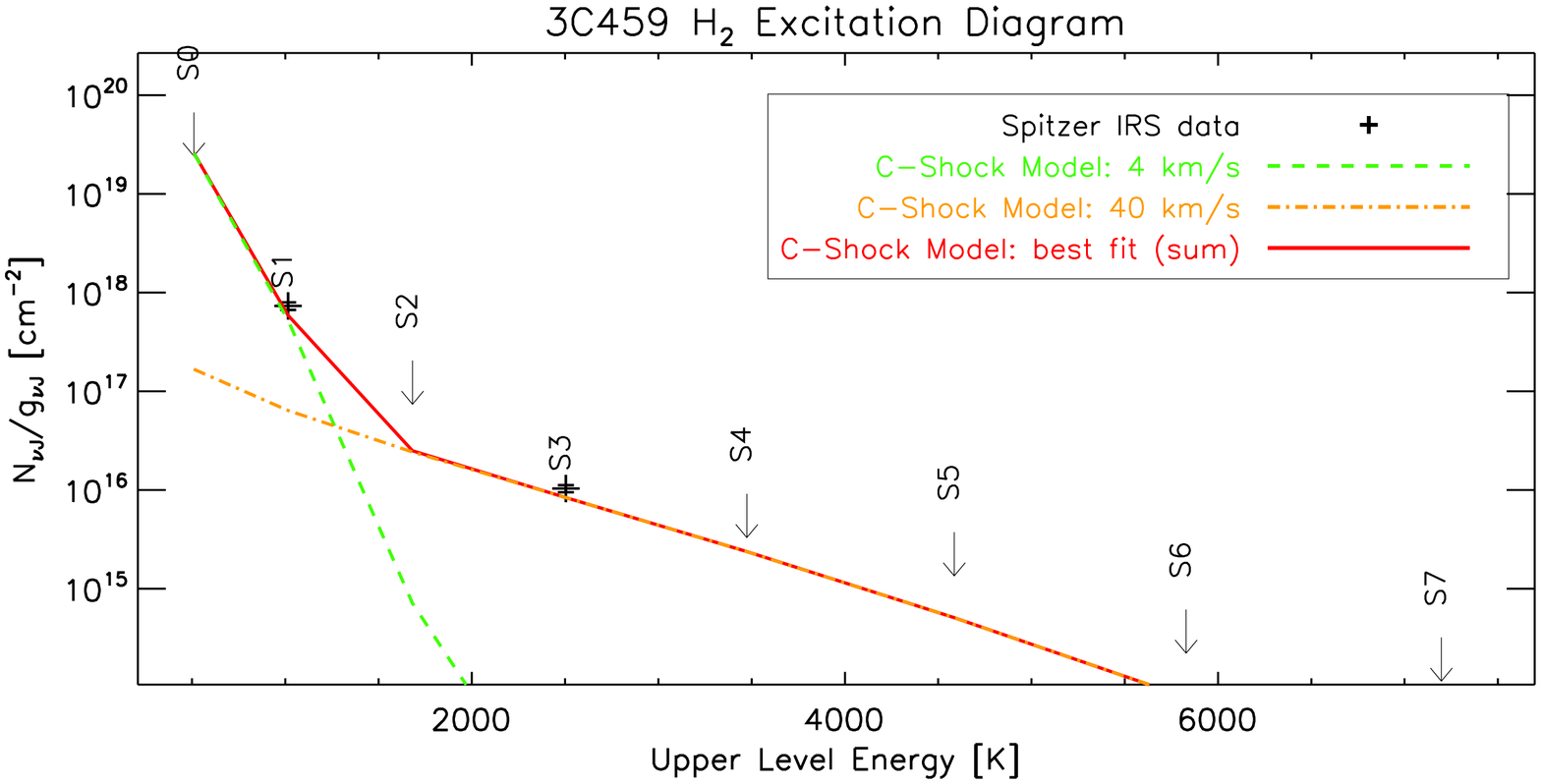}
    \includegraphics[width=0.45\textwidth]{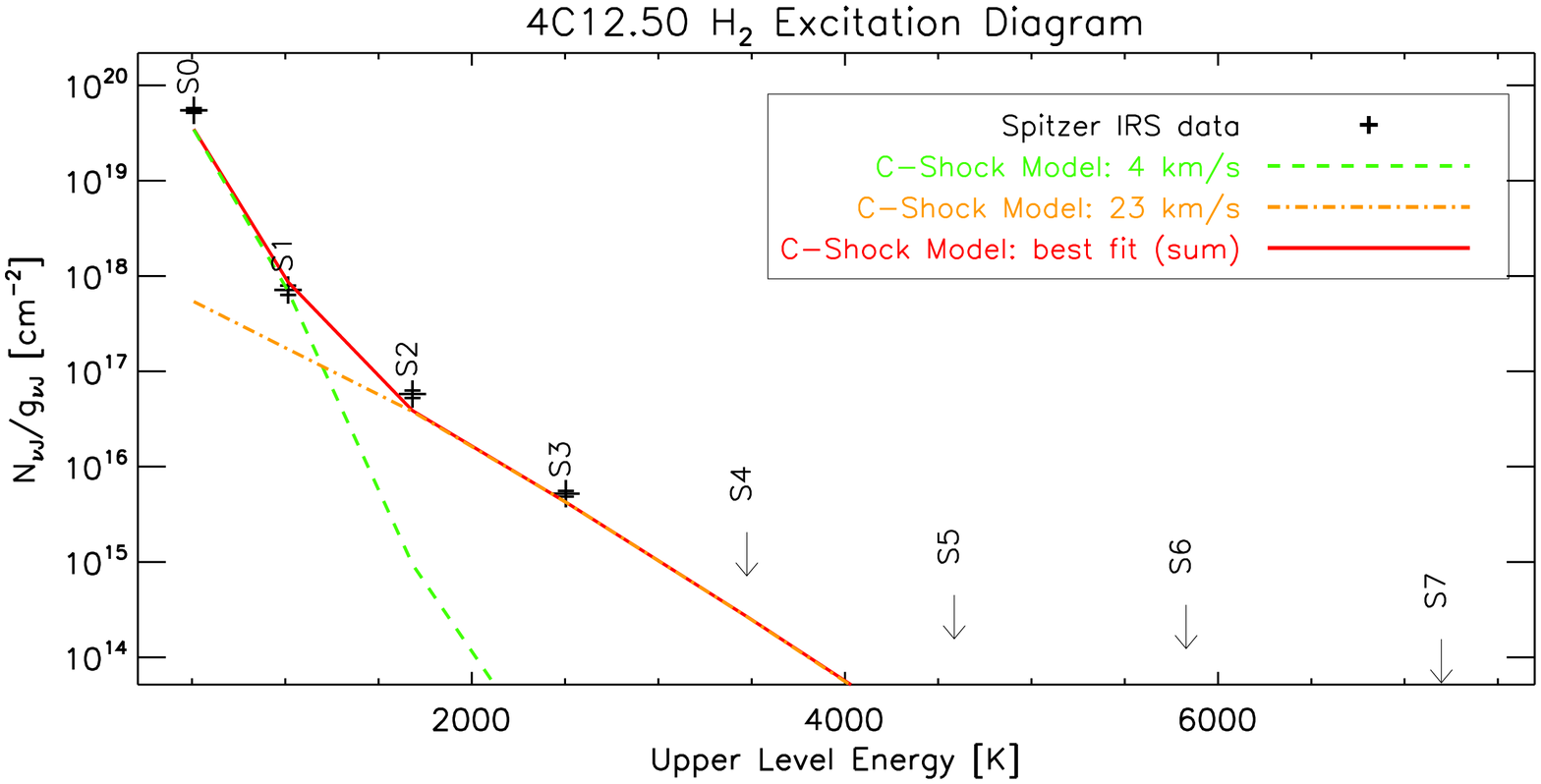}
    \includegraphics[width=0.45\textwidth]{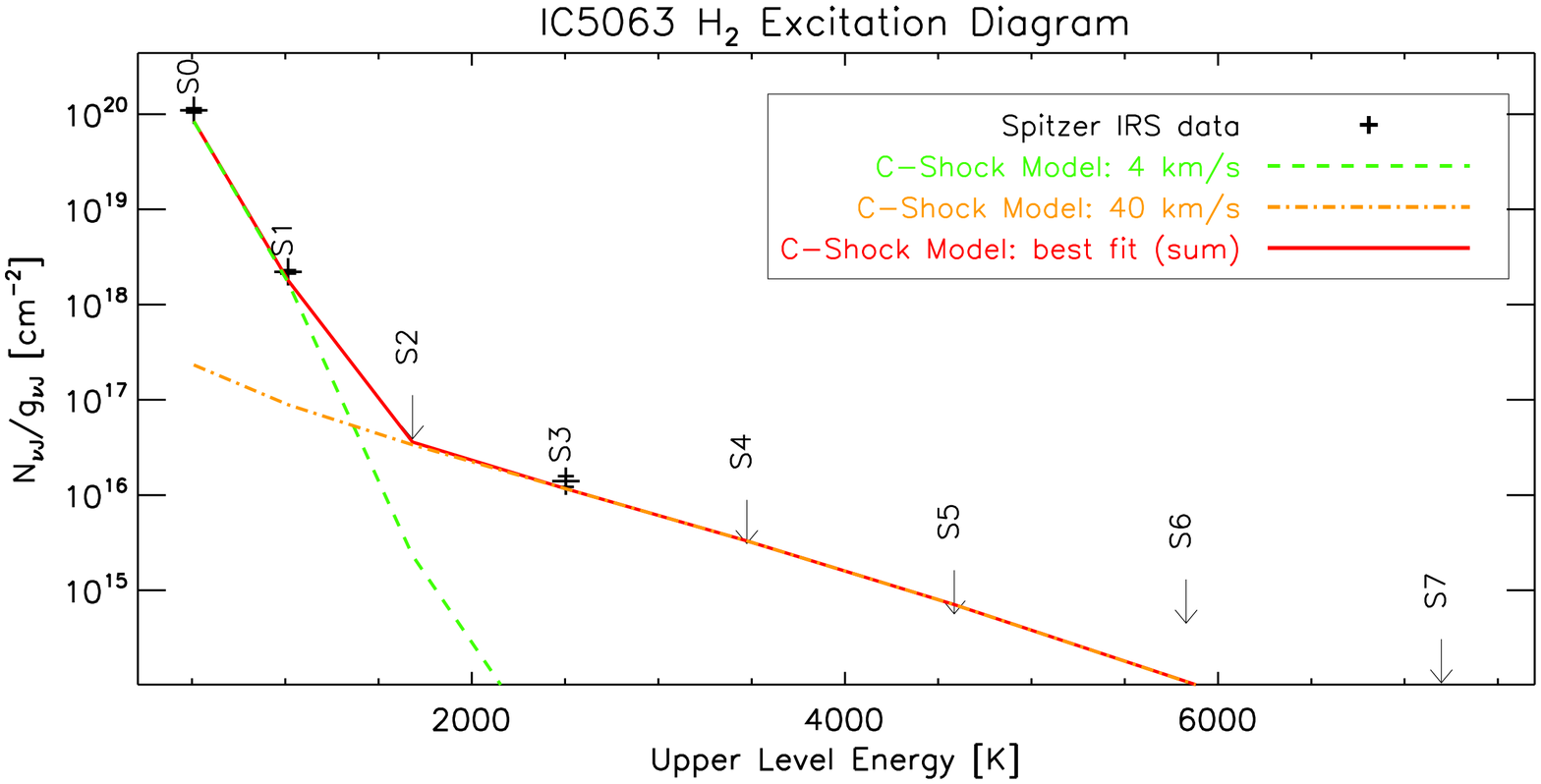}
    \includegraphics[width=0.45\textwidth,]{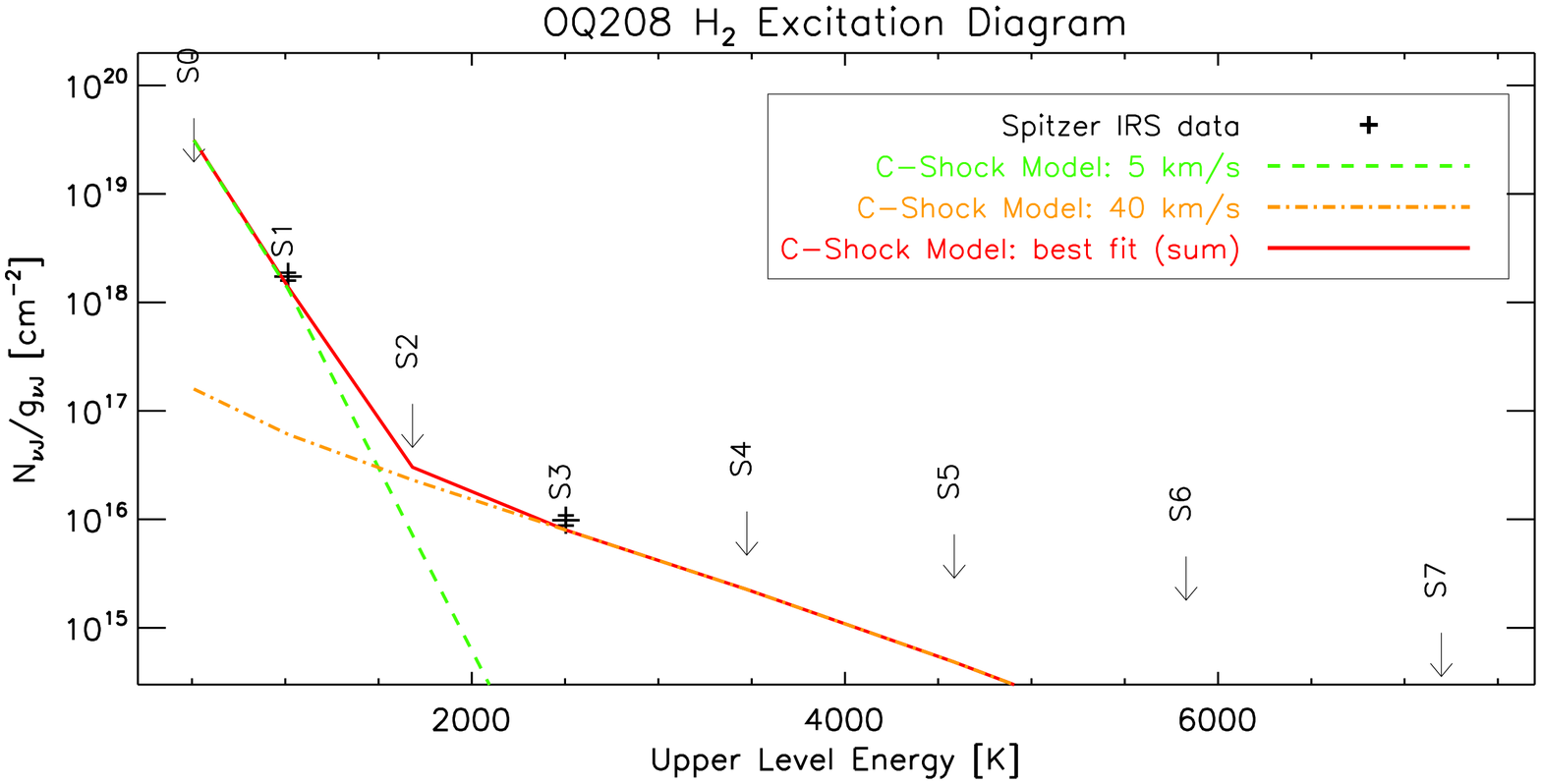}
    \includegraphics[width=0.45\textwidth]{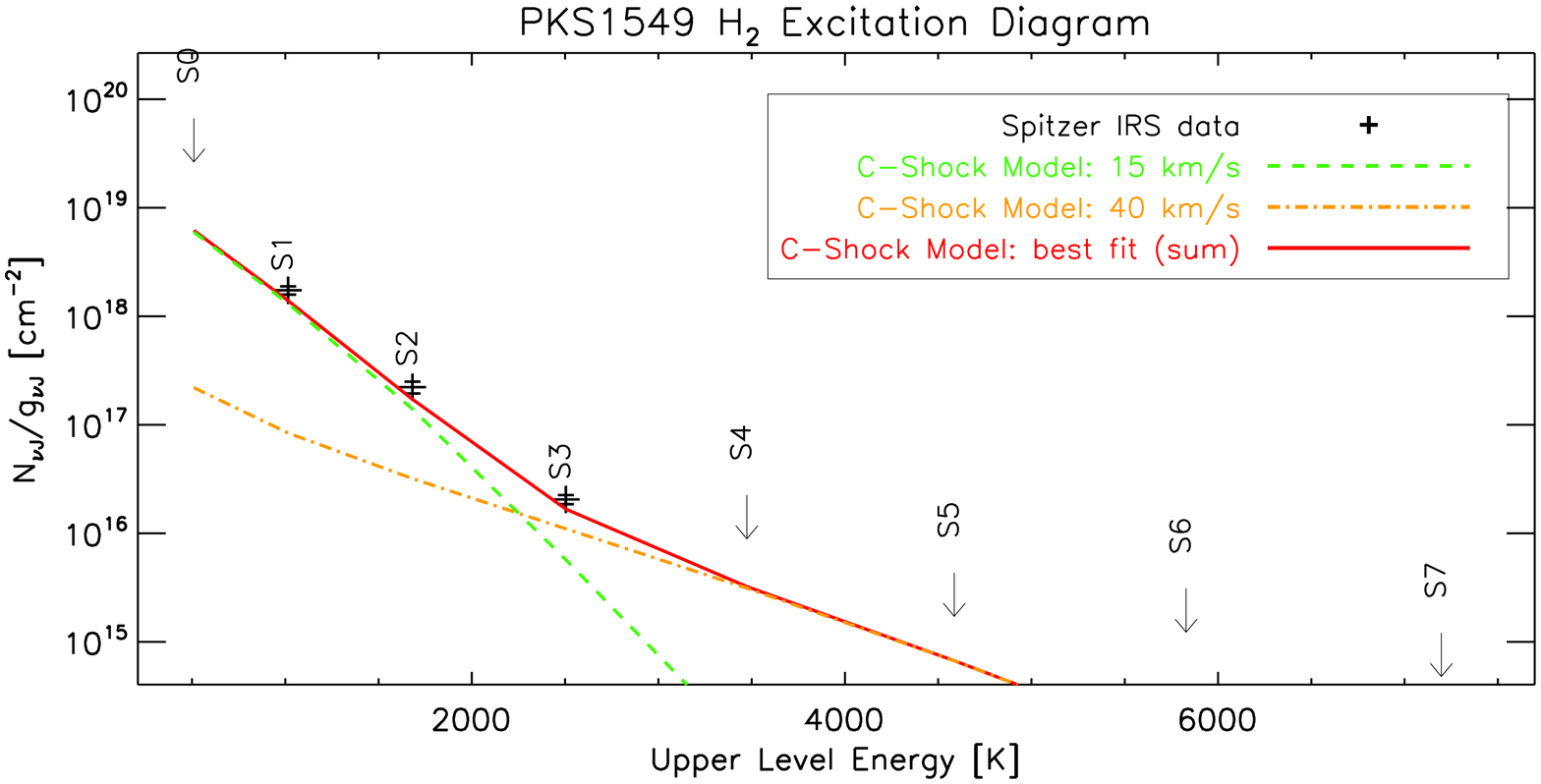}
      \caption{H$_2$ excitation diagrams fitted with a combination of two C-shock models, for a pre-shock density of $n_{\rm H} = 10^3$~cm$^{-3}$. The shock model parameters, and H$_2$ masses are listed in Table~\ref{table_H2params_shocks_nH1e3}.}
       \label{fig:H2exc_shocks_nH1e3}
   \end{figure*}

\begin{figure*}
   \centering
    \includegraphics[width=0.45\textwidth]{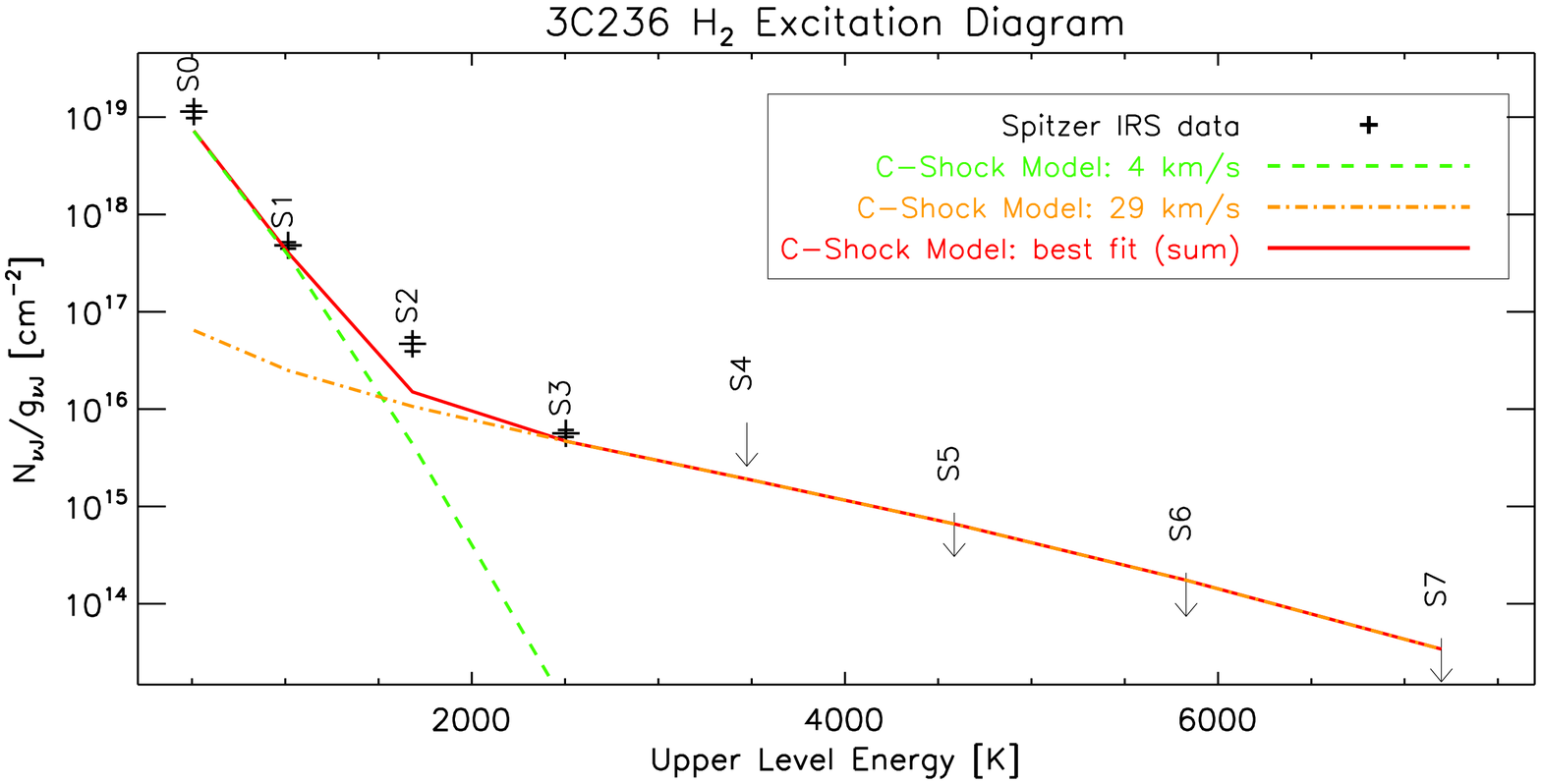}
    \includegraphics[width=0.45\textwidth]{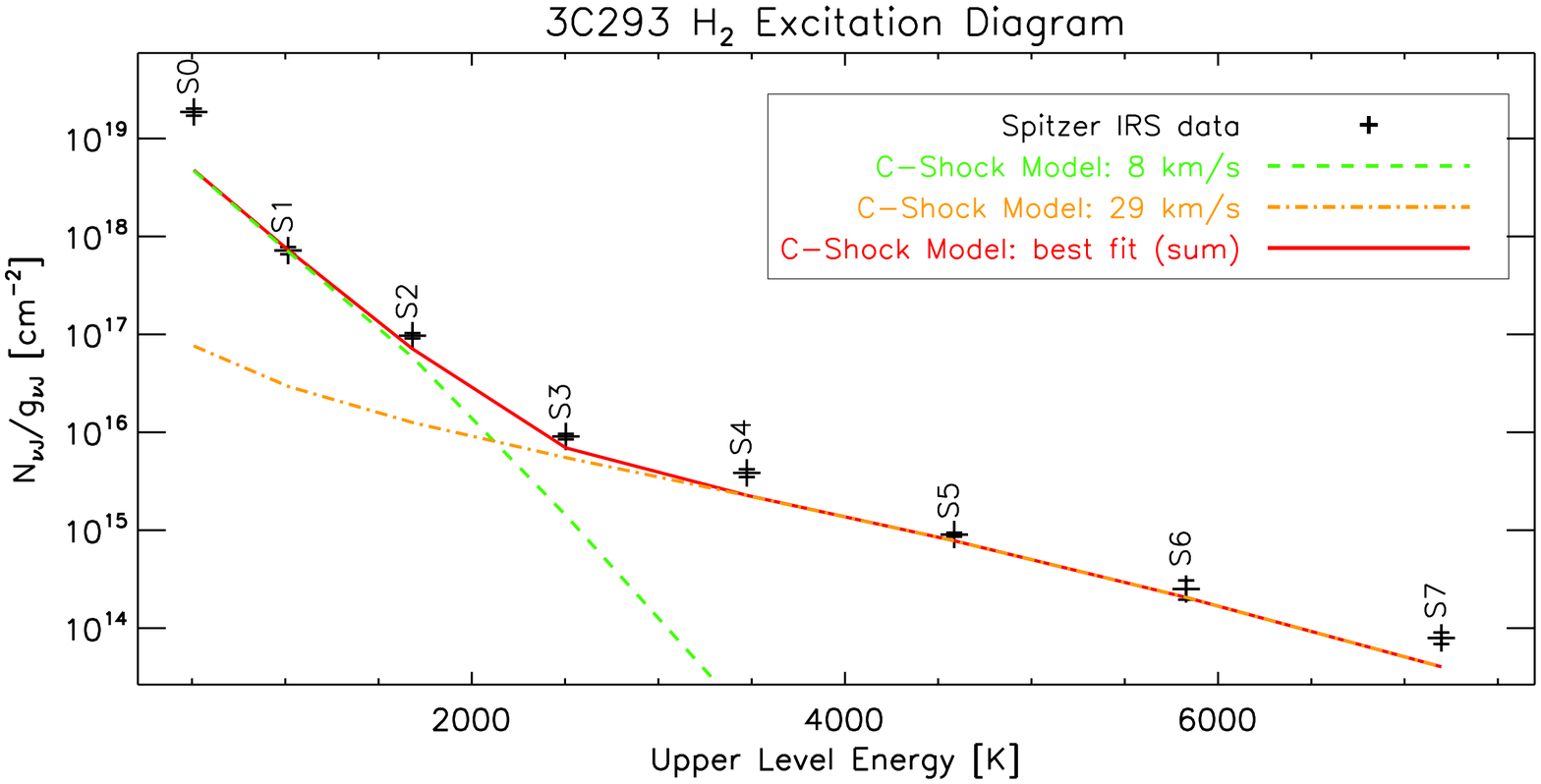}
    \includegraphics[width=0.45\textwidth]{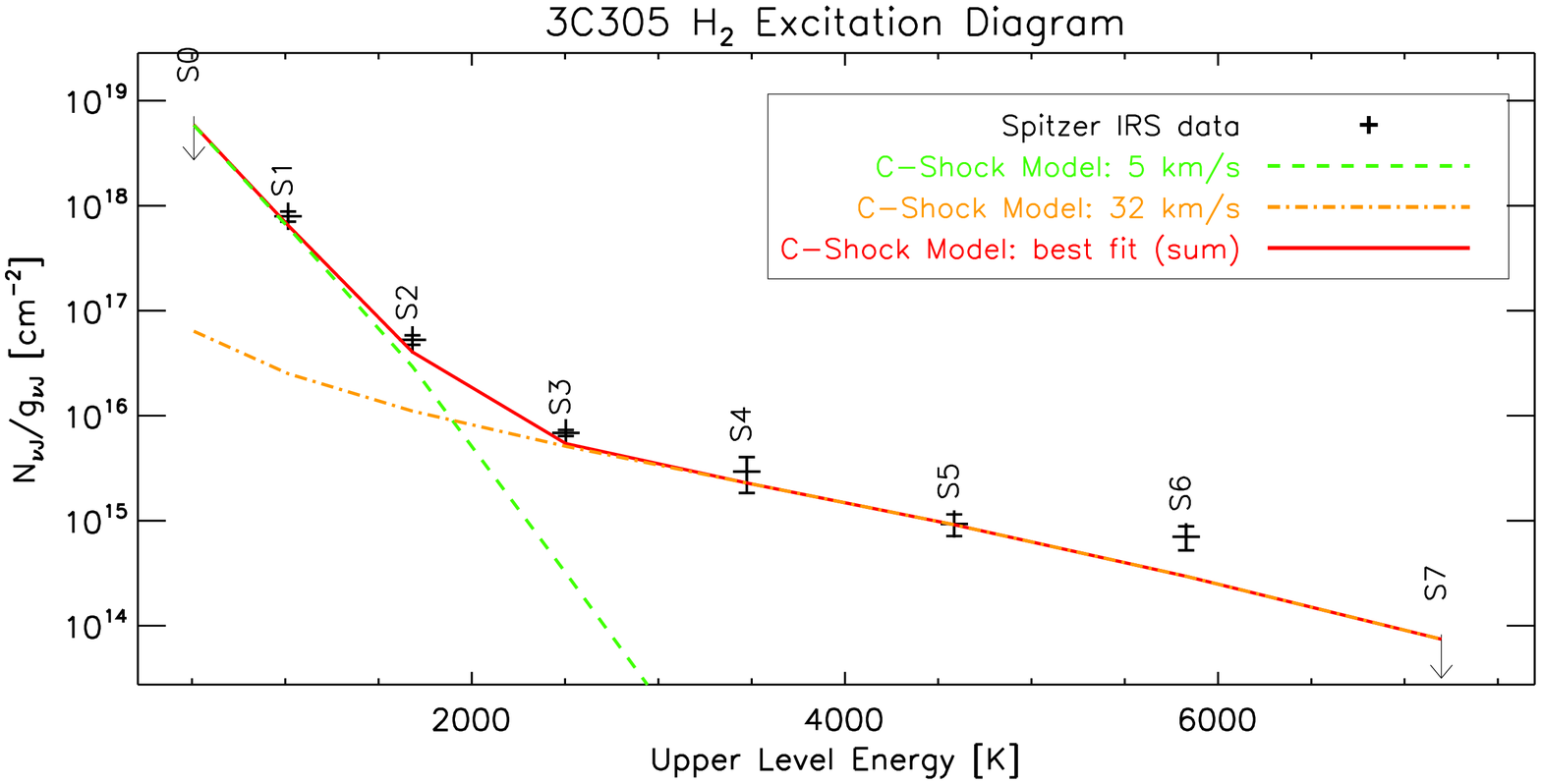}
    \includegraphics[width=0.45\textwidth]{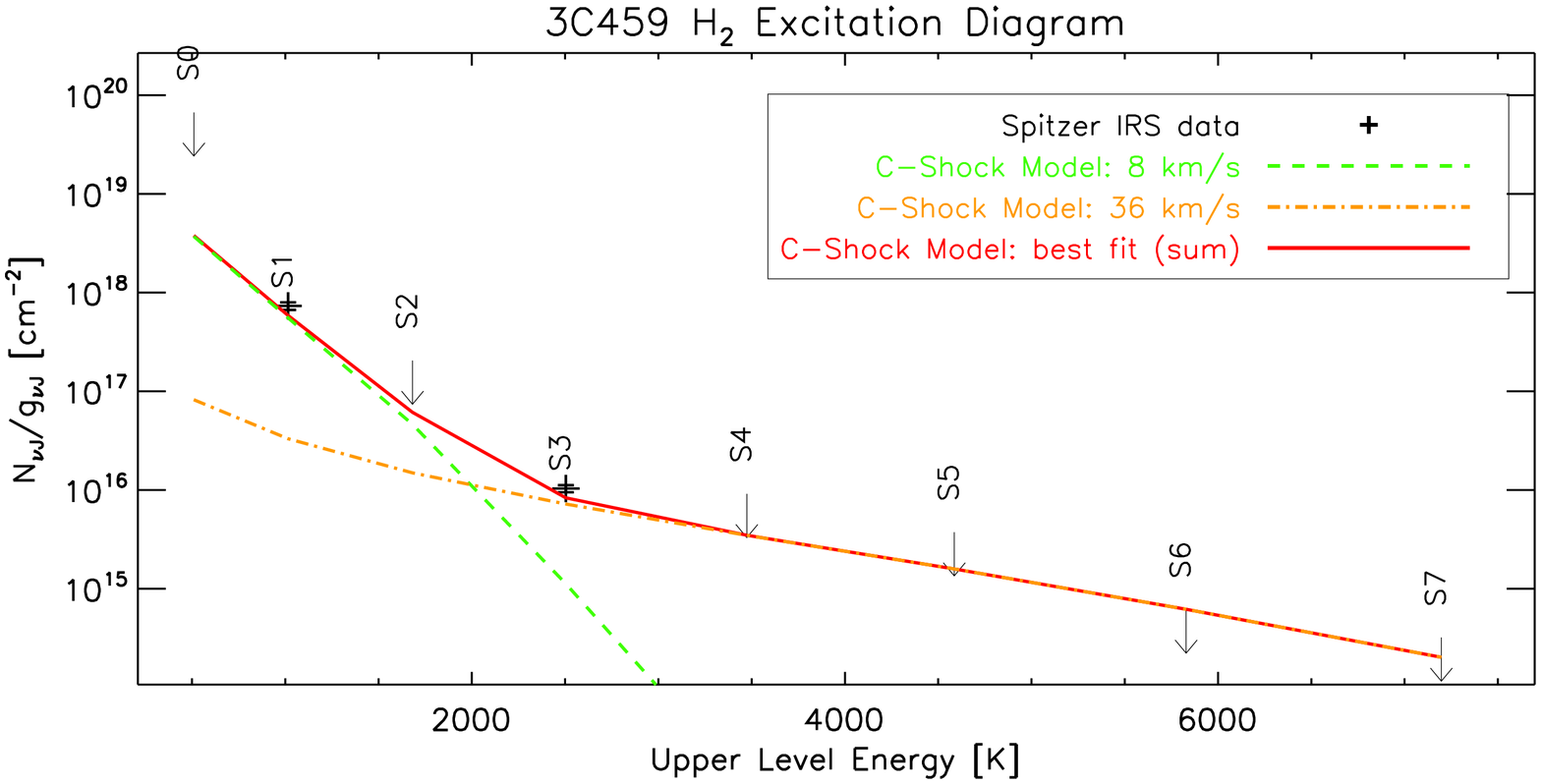}
    \includegraphics[width=0.45\textwidth]{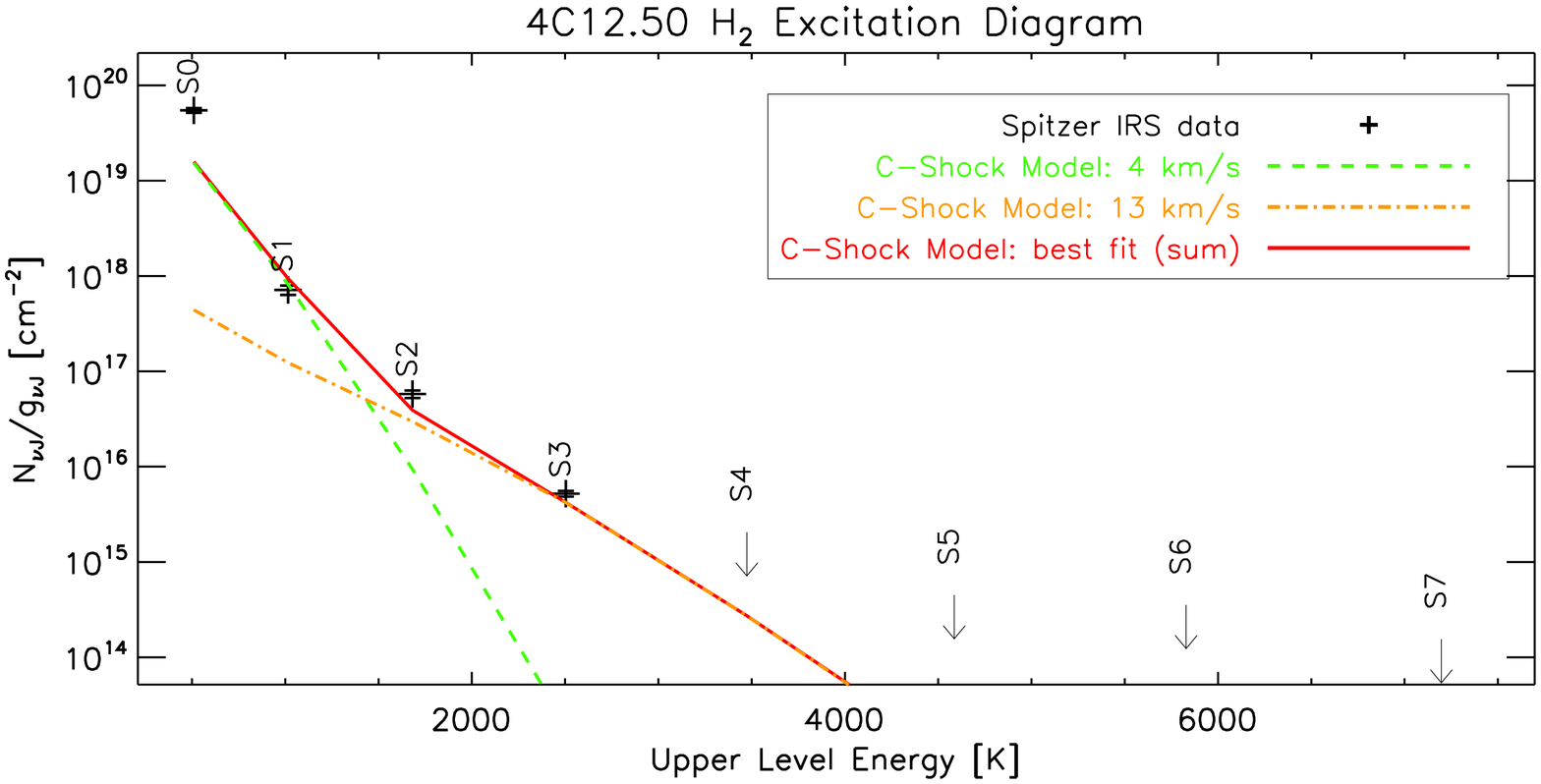}
    \includegraphics[width=0.45\textwidth]{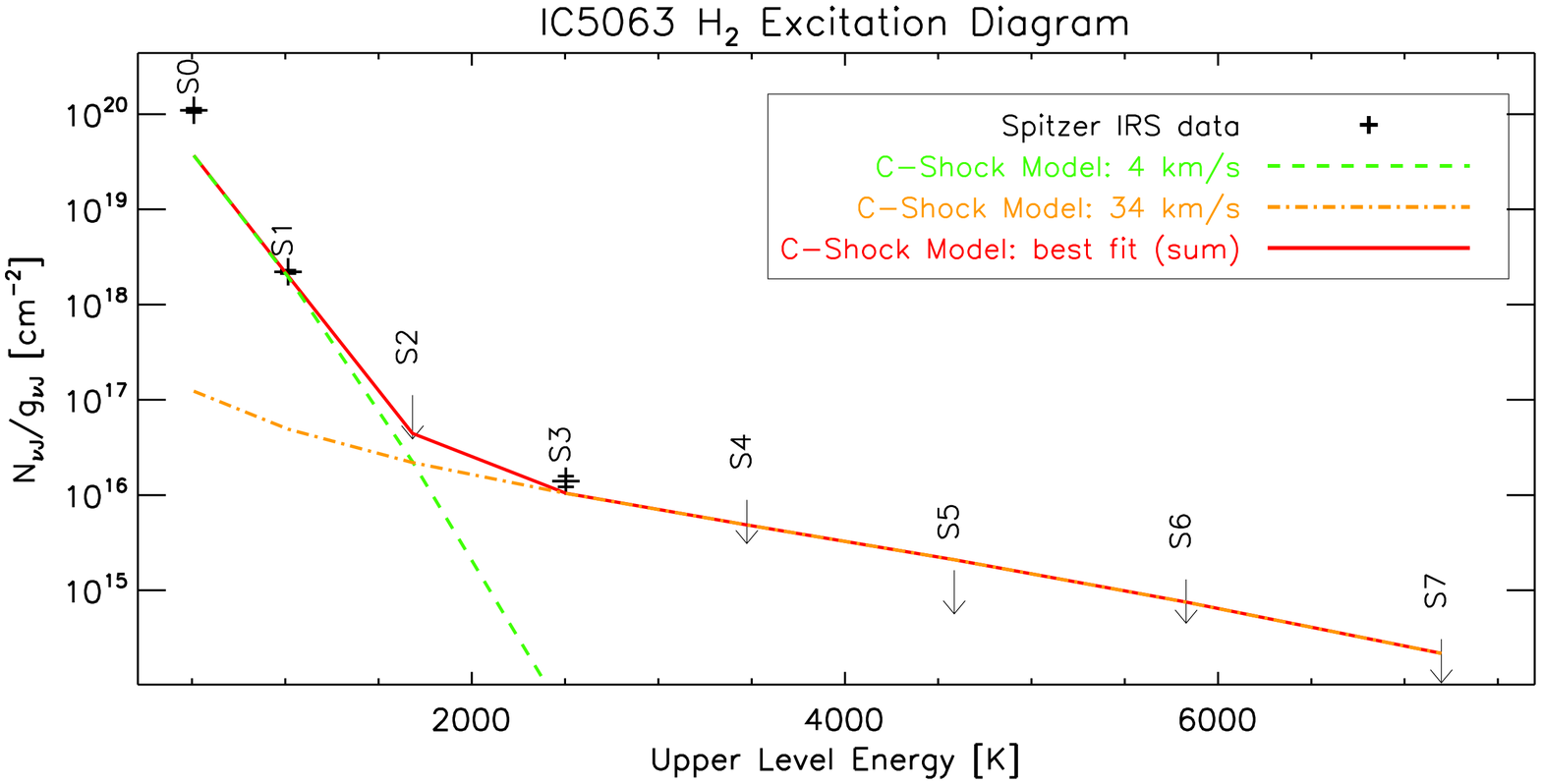}
    \includegraphics[width=0.45\textwidth]{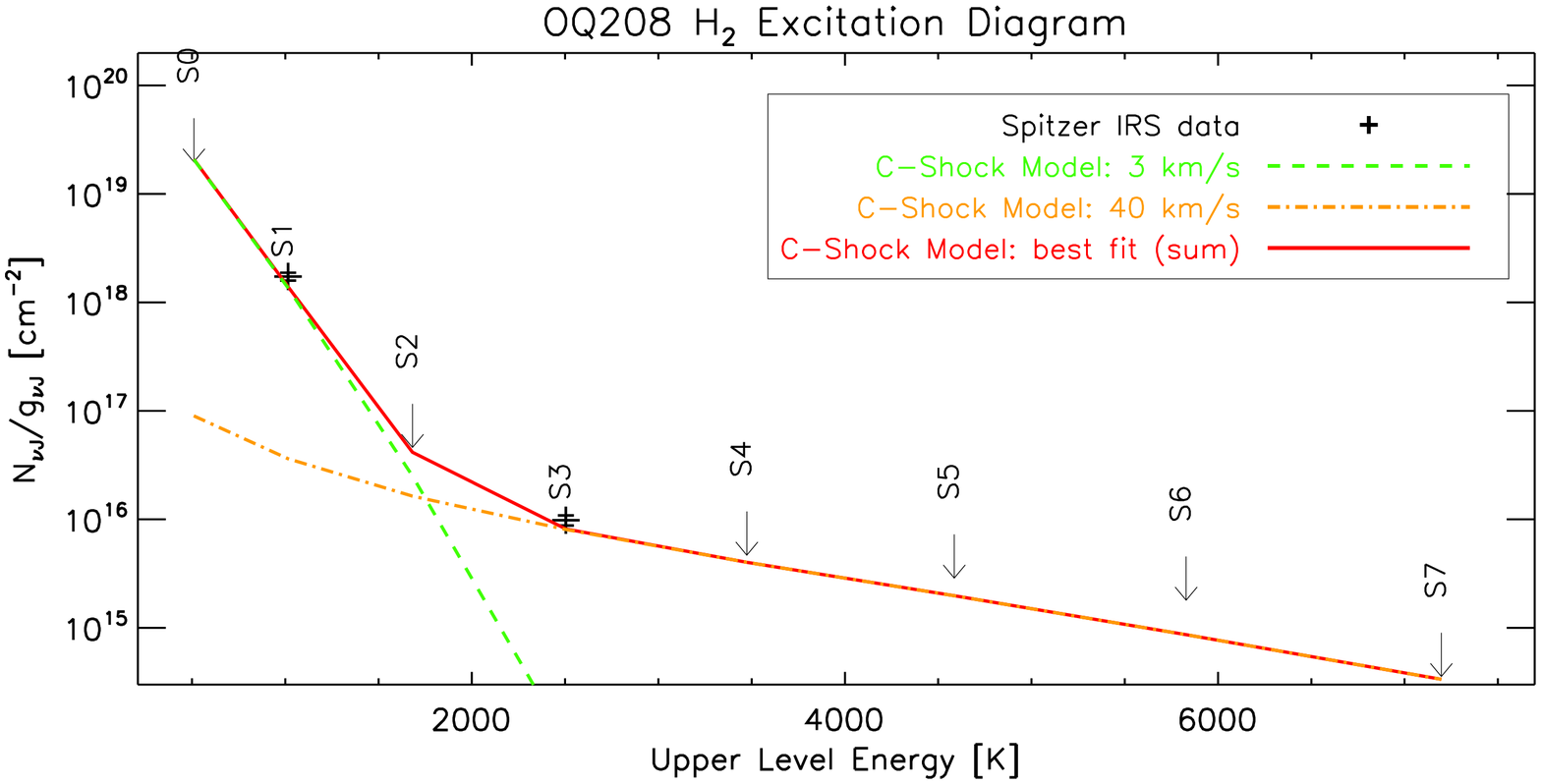}
    \includegraphics[width=0.45\textwidth]{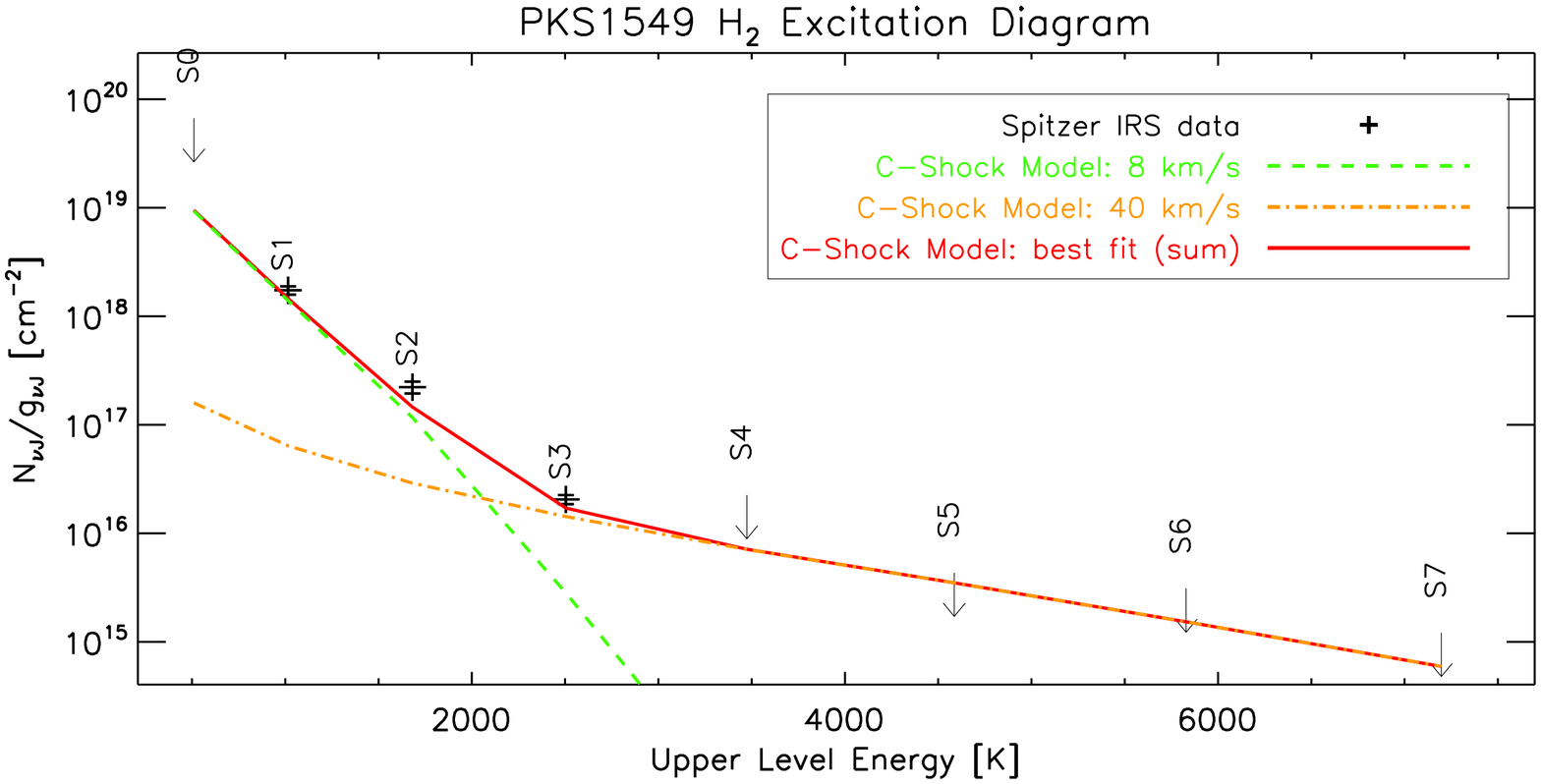}
      \caption{H$_2$ excitation diagrams fitted with a combination of two C-shock models, for a pre-shock density of $n_{\rm H} = 10^4$~cm$^{-3}$. The shock model parameters, and H$_2$ masses are listed in Table~\ref{table_H2params_shocks_nH1e4}.}
       \label{fig:H2exc_shocks_nH1e4}
   \end{figure*}
\begin{figure*}
   \centering
    \includegraphics[width=0.49\textwidth]{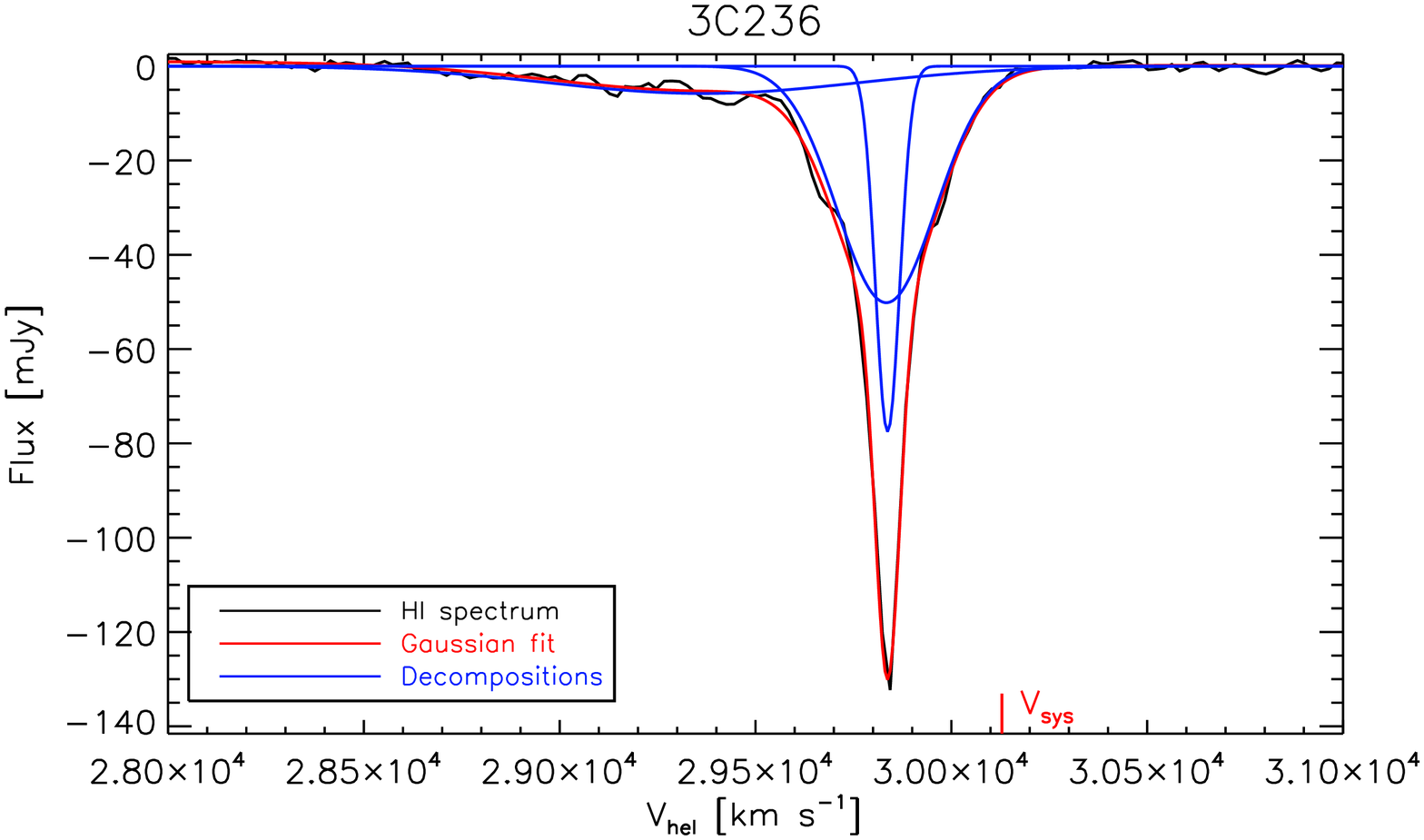}
    \includegraphics[width=0.49\textwidth]{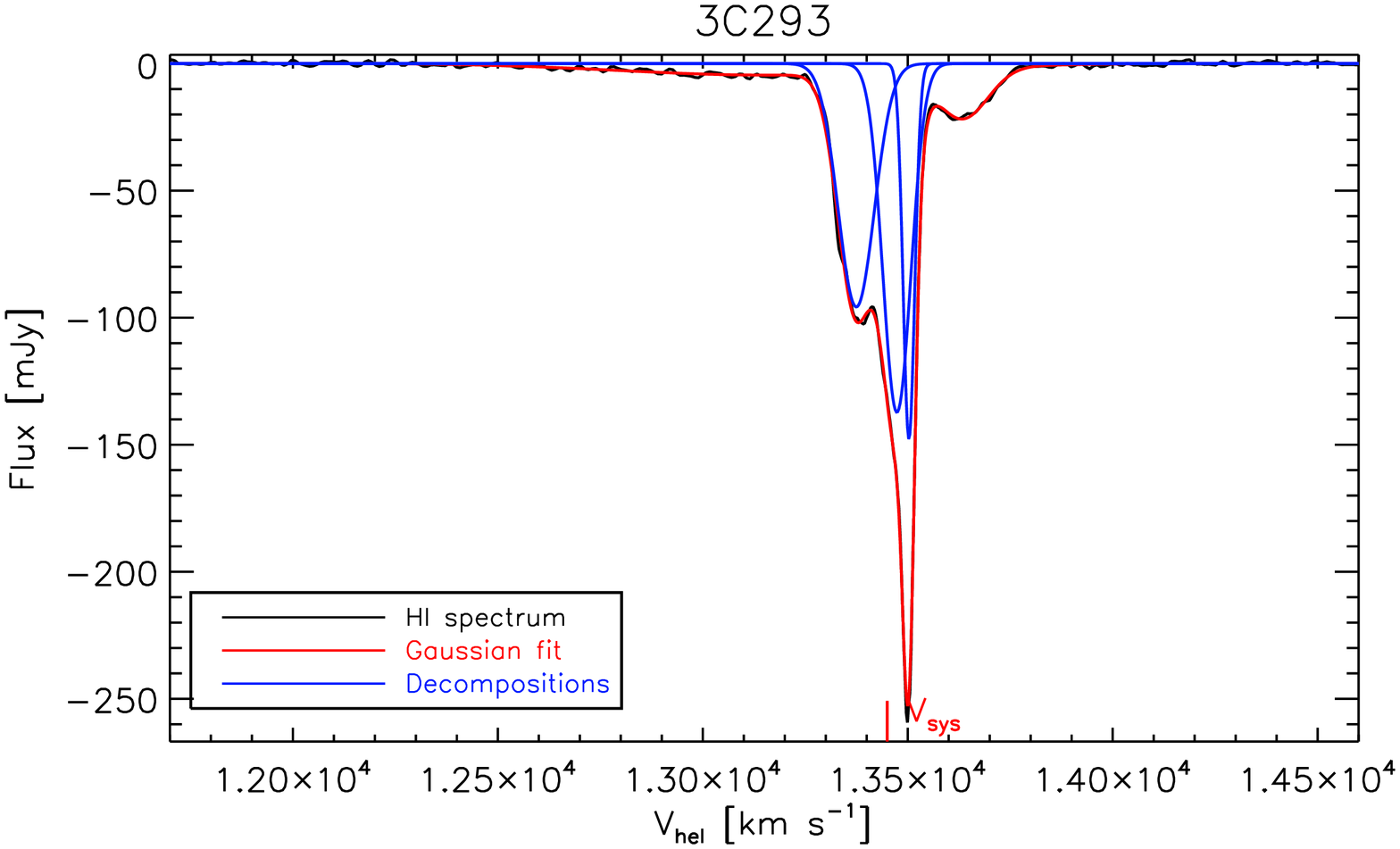}
    \includegraphics[width=0.49\textwidth]{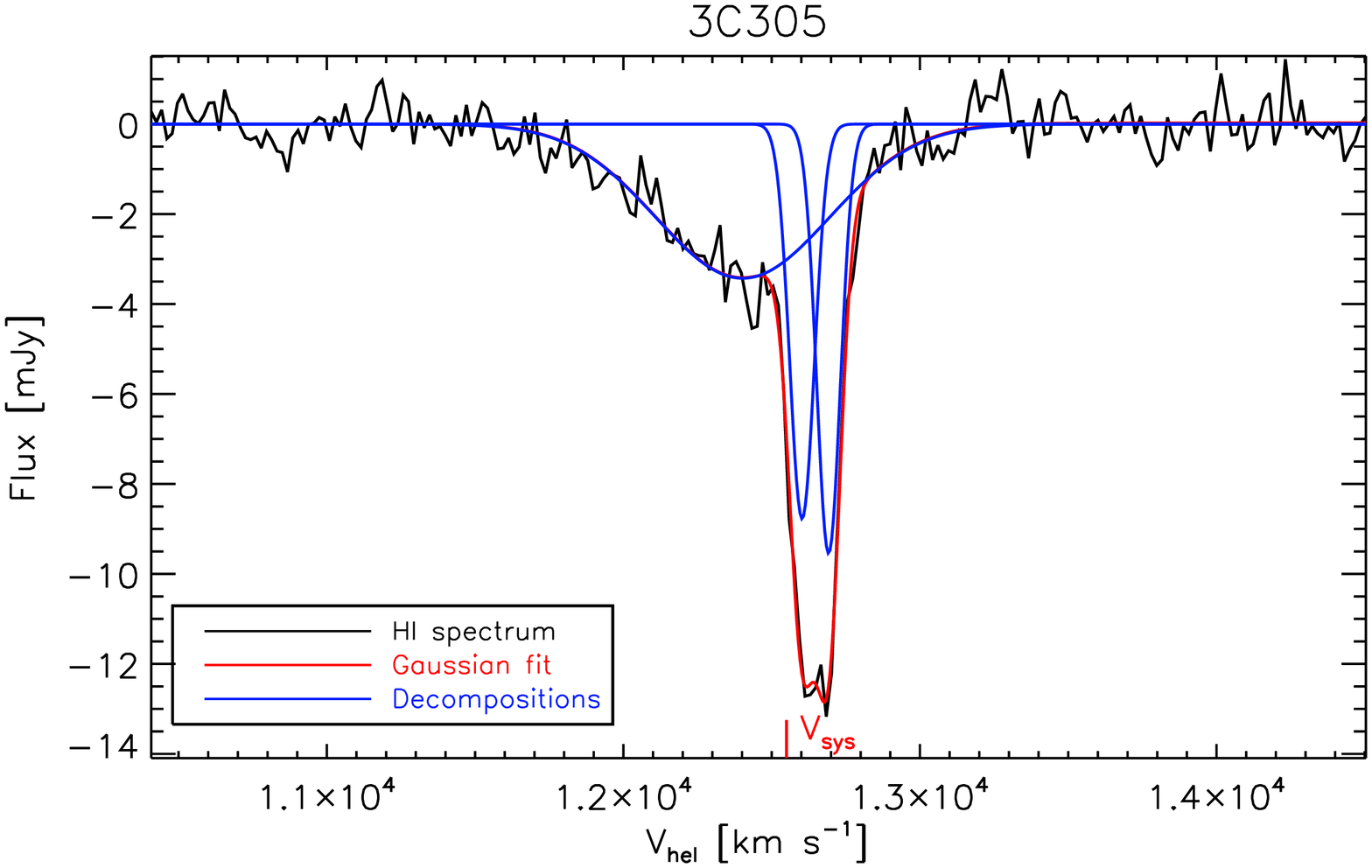}
    \includegraphics[width=0.49\textwidth]{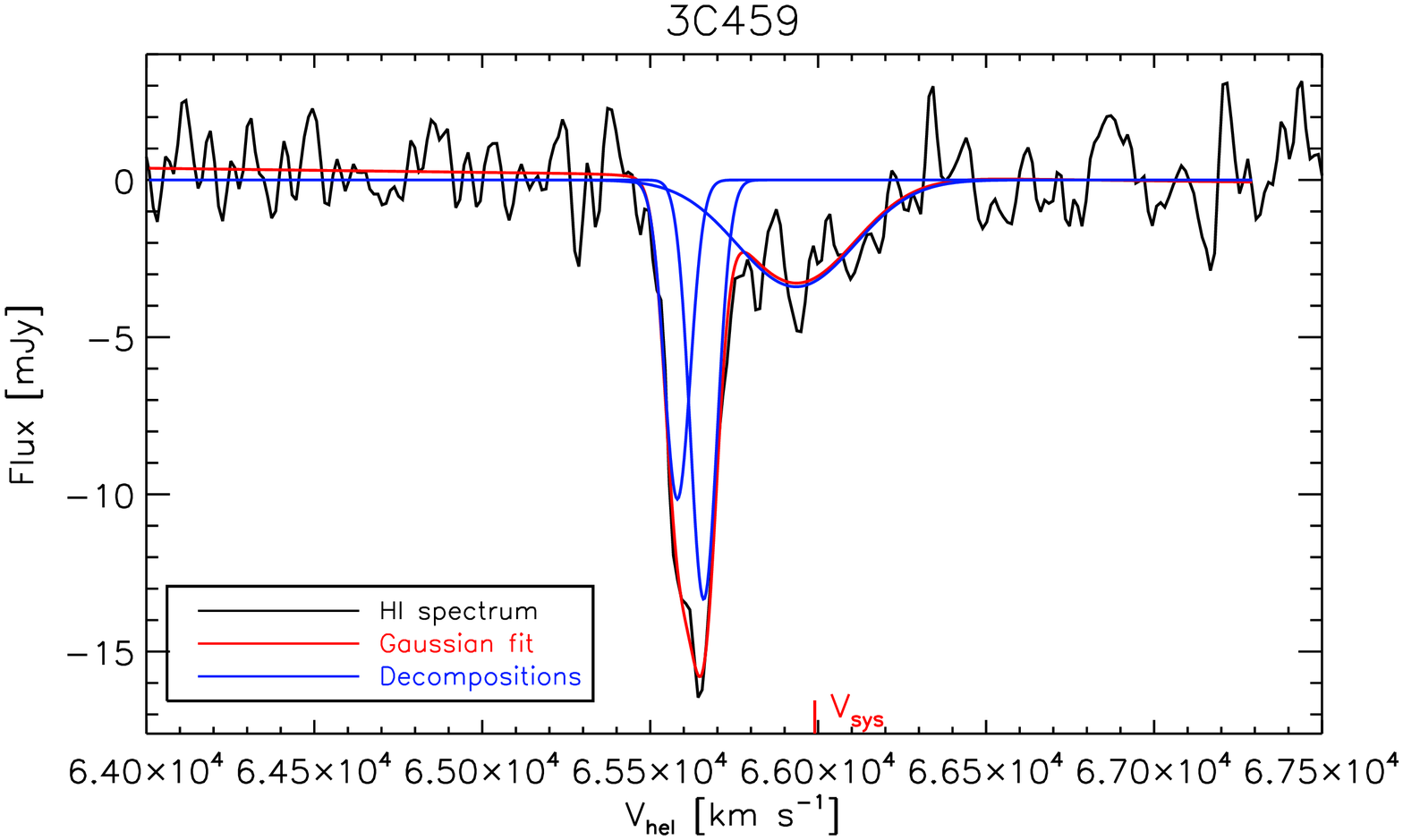}
    \includegraphics[width=0.49\textwidth]{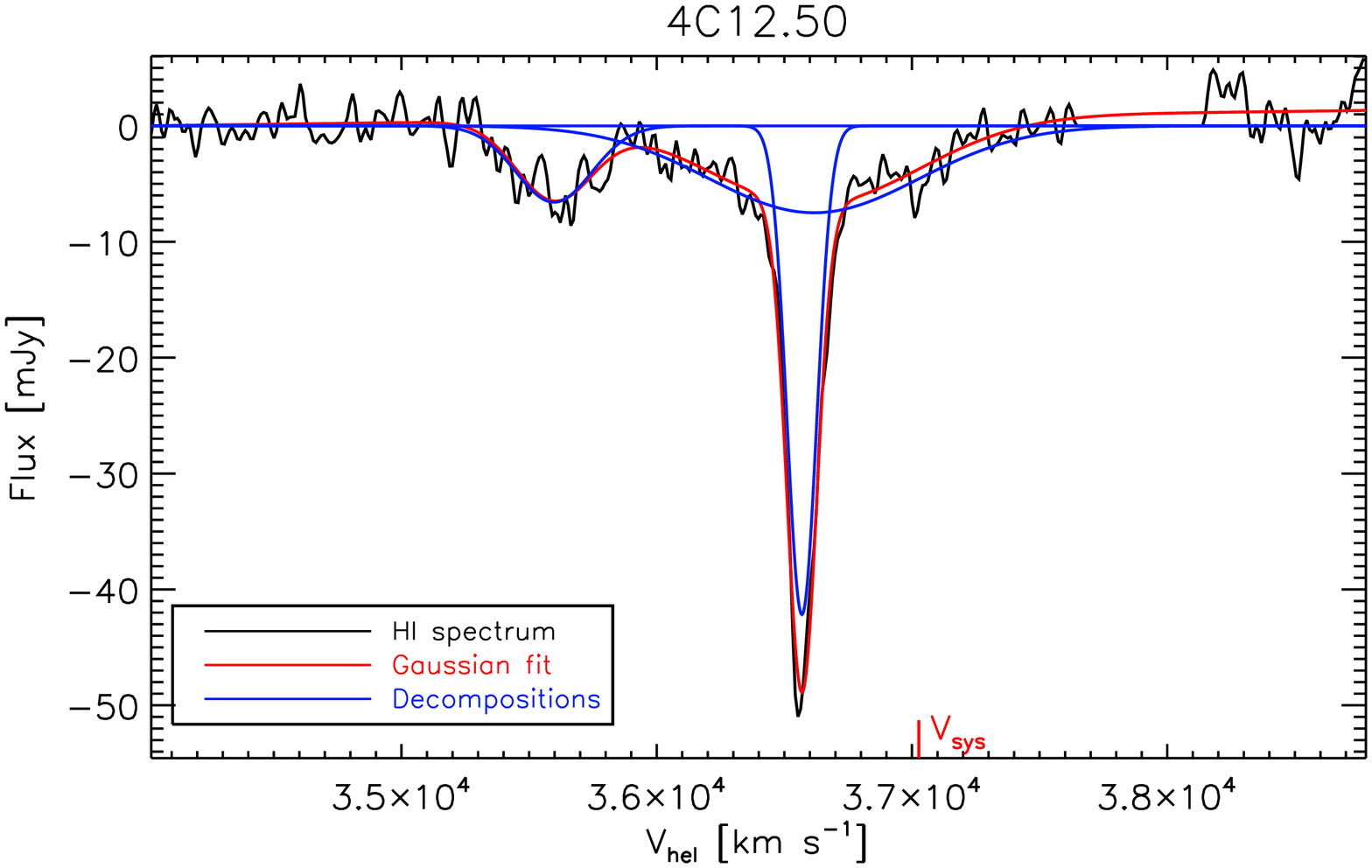}
    \includegraphics[width=0.49\textwidth]{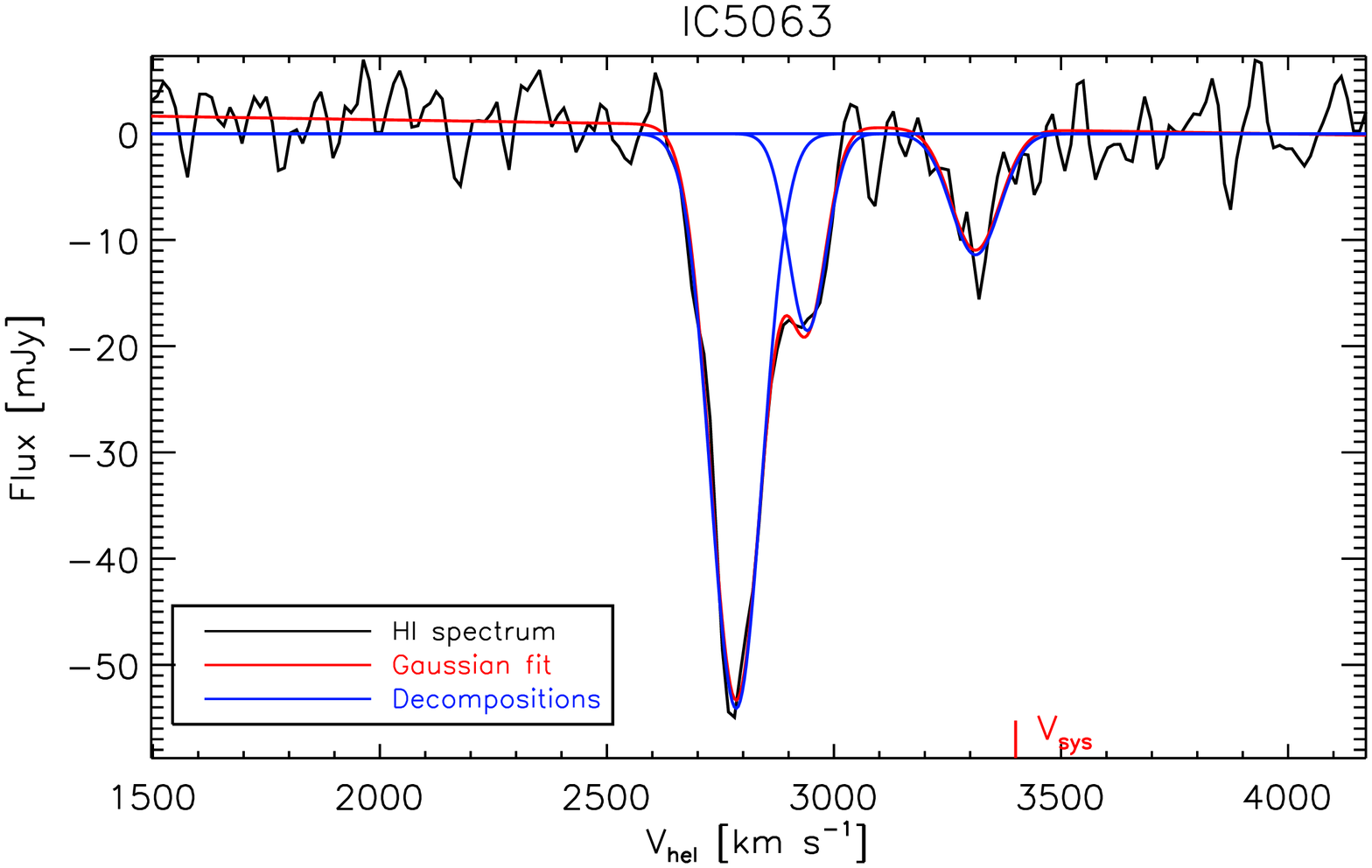}
    \includegraphics[width=0.49\textwidth]{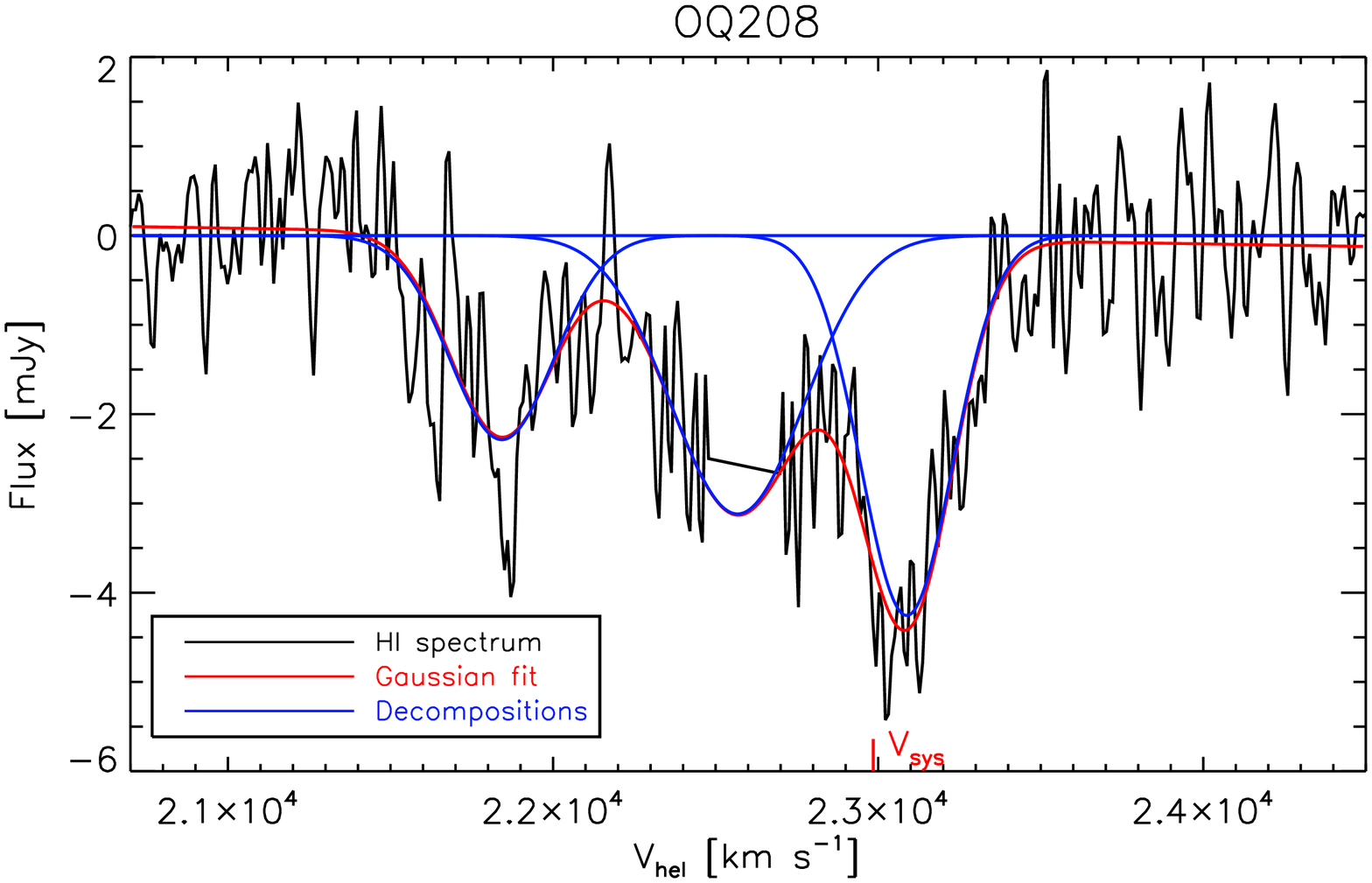}
    \includegraphics[width=0.49\textwidth]{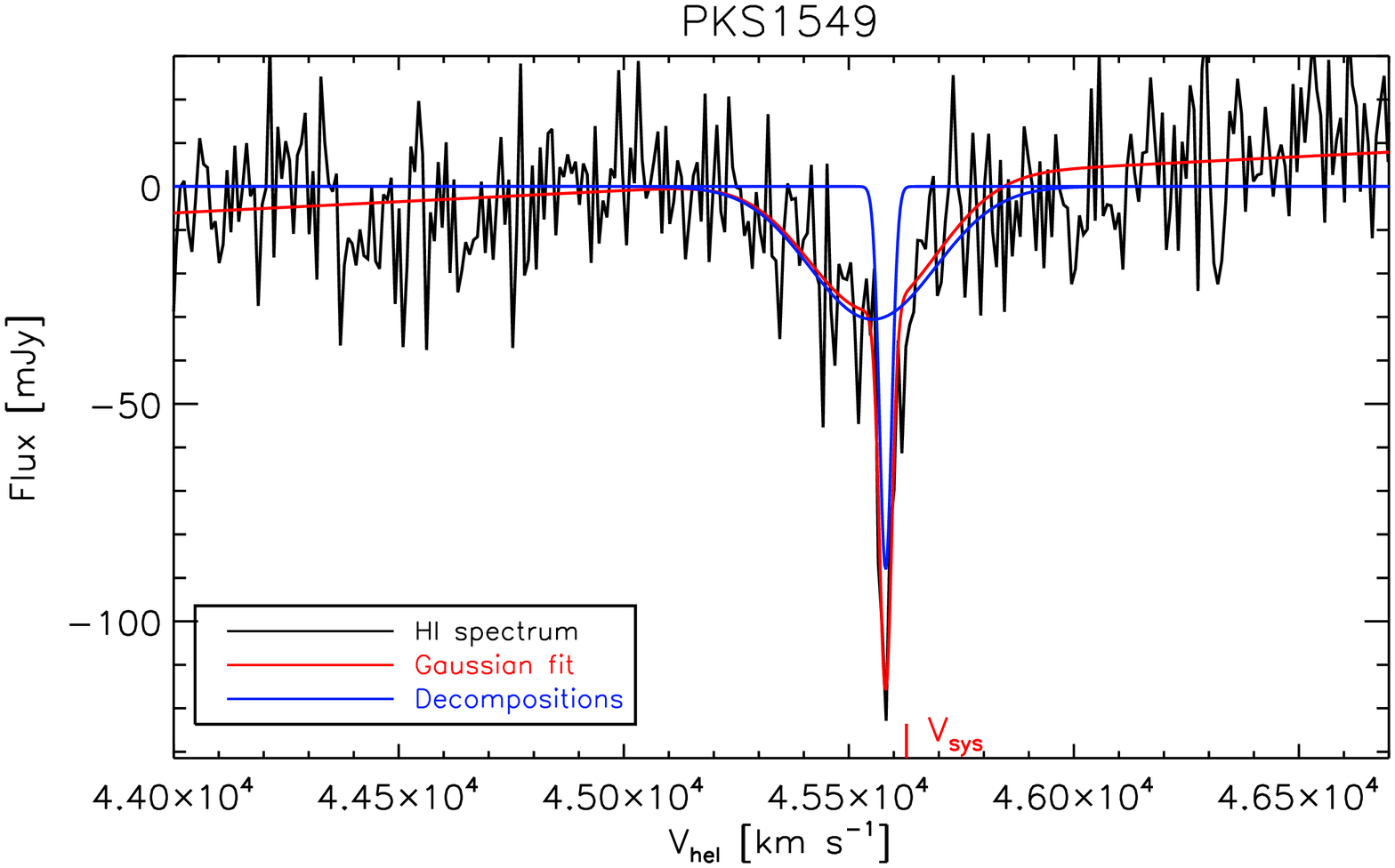}
    \caption{Gaussian decompositions of the H{\sc i} absorption profiles observed with the Westerbork telescope. The data is from \citet{Morganti2005}, except for  3C~293, where we used more recent observations taken by B. Emonts with the Westerbork telescope.}
       \label{fig:HIgauss_fit}
   \end{figure*}

\end{document}